\documentclass[a4paper, 11pt]{article}   
\usepackage{amsmath}
\usepackage{amsfonts}
\usepackage{latexsym}
\usepackage{graphicx}
\usepackage{multicol}
\usepackage{color}
\graphicspath{{./myplot/eps/}}
\usepackage{amssymb}
\usepackage{mathtools}
\usepackage{empheq}
\usepackage{caption}
\usepackage{subcaption}
\usepackage[toc,page]{appendix}
\usepackage{breqn} 
\usepackage{bm} 
\newcommand{\Sign}{\text{Sign}}
\begin{document}

\title{\textbf{The effect of flow on resonant absorption of slow MHD waves in magnetic flux tubes}}
\author{Mohammad Sadeghi$^{1}$, Karam Bahari$^{2}$,
	Kayoomars Karami$^{1}$\\
	\small{$^{1}$Department of Physics, University of Kurdistan, Pasdaran Street, P.O. Box 66177-15175, Sanandaj, Iran}\\
	\small{$^{2}$Physics Department, Faculty of Science, Razi University, Kermanshah, Iran}}

\maketitle

\begin{abstract}
In this paper, we study kink and sausage oscillations in the presence of longitudinal background flow. We study resonant absorption of the kink and sausage modes in the slow continuum under magnetic pore conditions in the presence of flow. we determine the dispersion relation then solve it numerically, and find the frequencies and damping rates of the slow kink and sausage surface modes. We also, obtain analytical solution for the
damping rate of the slow surface mode in the long wavelength limit. We show that in the presence of plasma flow, resonance absorption can result in strong damping for forward waves and can be considered as an efficient mechanism to justify the extremely rapid damping of slow surface sausage waves observed in magnetic pores. Also, the plasma flow reduces the efficiency of resonance absorption to damp backward waves. Furthermore, for the pore conditions, the resonance instability is avoided in our model.

\end{abstract}
\section{Introduction}
The mechanism of the heating of the solar corona (and the corona of the stars) is not
yet fully understood. Several non-thermal mechanisms have been proposed to explain this
phenomenon, and the problem of justifying this phenomenon remains. Surely the heating
must be tied to the magnetic field, because it is obvious that the heated areas have a
non-potential magnetic field. Plasma is bounded by magnetic field lines and can form
many types of visible structures. One of these is the propagation of magnetohydrodynamic (MHD) waves and their damping. Resonant absorption proposed as the damping
mechanism of MHD waves for the first time by Ionson \cite{Ionson}.  With the launch of space
satellites, the interest of theoretical physicists in studying waves in the solar atmosphere,
and especially the use of resonance absorption, increased. Nakariakov reported transverse oscillations in coronal loops with high damping rate \cite{Nakariakov} . Ruderman \& Roberts expressed the idea that the observed period of oscillation and their
damping time can be used to determine the transverse density distribution in a coronal
magnetic loop \cite{Ruderman2002}. This method was later used by many researchers (e.g., \cite{Goossens2002}- \cite{Raes2017}).\\

Because the source of the high-temperature energy of the corona originates from the
convection zone below the surface of the sun, it is important to study the dynamics of
MHD waves in the photosphere and chromosphere (e.g., \cite{Jess2015};
\cite{Jess2016}). In the photosphere, in addition to Alfv\'{e}n resonance, energy transfer by slow resonance absorption can be of particular importance.   Yu et al. showed that
 slow resonance absorption can affect the damping of waves in the photosphere \cite{Yu2017a} . They
 also found that the resonant damping of the fast surface kink mode is much stronger than
that of the slow surface kink mode. Yu et al. \cite{Yu2017b} considered linear profile for density and pressure in the transitional layers \cite{Yu2017b}. They showed in the cases where damping by
Alfv\'{e}n continuum is weak, the resonant absorption in slow continuum can be an effective mechanism for damping sausage and kink slow surface modes. Sadeghi \& Karami investigated resonance absorption in the presence of a weak magnetic twist in the
 photosphere condition \cite{Sadeghi}. They concluded that a magnetic twist could be effective on more
 intense damping. In this paper, we study effect of flow on the slow sausage and kink
MHD waves, which have been observed by Dunn Solar Telescope \cite{Grant2015}.\\

Observations by Brekke et al. and Tian et al. show that plasma flows in magnetic flux tubes are present everywhere in the solar atmosphere \cite{Brekke} and \cite{Tian2009}. Soler et al. reported that the flow velocities are usually less than $10\%$ of the plasma Alfv\'{e}n  speed \cite{Soler2011}. Grant et al. investigated wave damping observed in upwardly propagating sausage mode oscillations contained within a magnetic pore \cite{Grant2015}. They showed that the waves propagate only through 0.25 of it’s wavelength along the before they damp whereas theory would expect the wave to survive for the distance of a few wavelengths. They also showed that the average upflow speed in photosphere is about $1/3$ Alfv\'{e}n  speed. Although higher speeds have been observed up to about 1.15 Alfv\'{e}n speeds. MHD oscillations of flowing plasma have been investigated by a number of researchers \cite{Goossens} and \cite{Bahari2020b}. Joarder et al.
(1977) \cite{Joarder1997} investigated resonant instability of MHD waves in the presence of plasma flow.  They showed that if the plasma velocity is greater than a certain value, it will cause instability.  Soler et al. studied analytically and numerically the damping length of resonantly damped kink in static flux tubes including nonuniform transitional layer \cite{Soler2011}. They showed that flow affects the wavelength
and the damping length due to resonant absorption. Bahari considered propagating kink MHD waves  in the presence of magnetic twist and
plasma flow \cite{Bahari2018}. He showed that the damping of the waves depend on the direction of plasma flow and the wave number of the wave.
Bahari et al. studied the propagation and instability of kink waves in
a twisted magnetic tube in the presence of flow \cite{Bahari2020a}. They showed that for particular values
of flow speed in coronal flux tubes the kink MHD waves propagate without damping.
Ruderman \& Petrukhin investigated the effect of flow on the damping of standing
kink waves in the cold plasma approximation \cite{Ruderman2019}. They concluded that the effect of flow on
coronal seismography is weak but has a significant effect on prominences. Recently Geeraerts et al. studied the effect of electrical resistivity on the damping of slow surface sausage modes. They showed that electrical resistivity can play an important role in wave damping and greatly reduce the number of oscillations \cite{Geeraerts2020}.

Our aim in the present work is to investigate the effect of flow on the oscillation and damping of slow surface sausage and kink modes in the magnetic pore conditions. To study the effect of flow, we consider a model similar to the model of Yu et al. \cite{Yu2017b}, in which the plasma flow has been included too. In section \ref{model}, this model and the equations of motion governing the surface modes are presented. We find the dispersion relation in the case of no inhomogeneous layer in section \ref{disper}. Then in section \ref{conec}, we obtain the dispersion relation in the presence of the inhomogeneous layer using the connection formula for slow continuum. In Section \ref{num}, numerical calculations for magnetic pore conditions are shown. Finally, we conclude the paper in Section \ref{con}.
\section{Equations of Motion and Model}\label{model}
The linear perturbations of homogeneous flowing magnetized plasma are governed by the
 following equations \cite{kadomtsev1966hydromagnetic}
\begin{dgroup}
	\begin{dmath}\label{eqn:linearized:mhd1}
		\rho\left(\frac{\partial}{\partial t}+\bm{v}\cdot \nabla \right)^2 \bm{\xi}=-\nabla\delta p-\frac{1}{\mu_{0}}\Big(\delta \bm{B} \times (\nabla\times \bm{B})+\bm{B}\times (\nabla\times\delta \bm{B})\Big),
	\end{dmath}
	\begin{dmath}\label{eqn:linearized:mhd2}
		\delta p =- \bm{\xi} \cdot \nabla p-\gamma p \nabla \cdot \bm{\xi},
	\end{dmath}
	\begin{dmath}\label{eqn:linearized:mhd3}
		\delta \bm{B}=-\nabla \times (\bm{B}\times\bm{\xi}) ,
	\end{dmath}
\end{dgroup}
where $\rho, p$, $\bm{v}$ and $\bm{B}$ are the background density, kinetic pressure, plasma velocity and magnetic field, respectively. Also $\bm{\xi}$ is the Lagrangian displacement vector, $\delta p$
and $\delta \bm{B}$ are the Eulerian perturbations of the pressure and magnetic field, respectively. Here, $\gamma$ is the ratio of specific heats (taken to be $5/3$ in this work), and $\mu_0$ is the permeability of free space.

We consider a flux tube model with a unidirectional magnetic field
 which is in the direction of the tube axis. The model consists of interior and exterior
 regions in which the equilibrium and stationary quantities are constant and transitional
 layer in which the background quantities vary continuously. In the cylindrical coordinate
the magnetic field is\begin{dmath}
	\bm{B} = \Big(0,0,B_{z}(r)\Big).
\end{dmath}
Plasma pressure and magnetic field must be satisfied in the hydrostatic equilibrium equation
\begin{dmath}\label{eqn:pressure:r}
	\frac{{\rm d}}{{\rm d}r}\left(p + \frac{B_{z}^2}{2 \mu_0}\right)=0.
\end{dmath}
Here the background plasma density and magnetic field are assumed to be the same as those considered by Sadeghi \& Karami (2019) \cite{Sadeghi}
\begin{eqnarray}\label{rho}
	\rho(r)=\left\{\begin{array}{lll}
		\rho_{{\rm i}}, &r\leqslant r_i,&\\
		\rho_i+(\rho_e-\rho_i)\left(\frac{r-r_i}{r_e-r_i}\right), &r_i< r< r_e,&\\
		\rho_{{\rm e}}, &r\geqslant r_e,
	\end{array}\right.
\end{eqnarray}
where $r_i=R-l/2$ and $r_e=R+l/2$. Here, $R$ and $l$ are the tube radius and the
 thickness of the inhomogeneous layer, respectively,
\begin{eqnarray}\label{Bz}
	B^2_z(r)=\left\{\begin{array}{lll}
		B^2_{{z\rm i}}, &r\leqslant r_i,&\\
		B^2_{{z\rm i}}+\left(B^2_{{z\rm e}}-B^2_{{z\rm i}}\right)\left(\frac{r-r_i}{r_e-r_i}\right), &r_i< r< r_e,&\\
		B^2_{{z\rm e}}, &r\geqslant r_e,
	\end{array}\right.
\end{eqnarray}

where $\rho_i$ and $\rho_e$ are the constant densities of the interior and exterior
regions of the flux tube, respectively. Also $B_{z\rm i}$ and $B_{z\rm e}$ are the interior and exterior constant longitudinal magnetic fields, respectively. Putting Eqs. (\ref{Bz}) into the magnetohydrostatic equation (\ref{eqn:pressure:r}), we obtain the background gas pressure as follows

\begin{eqnarray}\label{p}
	p(r)=\left\{\begin{array}{lll}
		p_{{\rm i}}, &r\leqslant r_i,&\\
		p_{{\rm i}}+(p_{{\rm e}}-p_{{\rm i}})\left(\frac{r-r_i}{r_e-r_i}\right), &r_i< r< r_e,&\\
		p_{{\rm e}}, &r\geqslant r_e,
	\end{array}\right.
\end{eqnarray}
where
\begin{equation}
p_e=p_i+\frac{\left(B^2_{{z\rm e}}-B^2_{{z\rm i}}\right)}{2 \mu_{0}},
\end{equation}
and
$p_i$ is an arbitrary constant. The plasma flow is considered to be in the direction of the magnetic
field lines. as follows
\begin{eqnarray}\label{vz}
v_z(r)=\left\{\begin{array}{lll}
v_{{z\rm i}}, &r\leqslant r_i,&\\
v_{{z\rm i}}+\left(v_{{z\rm e}}-v_{{z\rm i}}\right)\left(\frac{r-r_i}{r_e-r_i}\right), &r_i< r< r_e,&\\
v_{{z\rm e}}, &r\geqslant r_e,
\end{array}\right.
\end{eqnarray}
where $v_{zi}$ and $v_{ze}$ are the constant flow of the interior and exterior
regions of the flux tube, respectively.
In addition, we define the following quantities
\begin{equation}
	v^2_{A(i,e)}\equiv\frac{B^2_{z(i,e)}}{\mu_0\rho_{(i,e)}},
\end{equation}

\begin{equation}\label{v2si}
	v^2_{s(i,e)}\equiv\gamma\frac{ p_{(i,e)}}{\rho_{(i,e)}},
\end{equation}

\begin{equation}
	v^2_{c(i,e)}\equiv\frac{v^2_{s(i,e)} v^2_{A(i,e)} }{v^2_{s(i,e)}+v^2_{A(i,e)}},
\end{equation}
where  $v_{A(i,e)}$, $v_{s(i,e)}$ and $v_{c(i,e)}$ are the interior/exterior Alfv\'{e}n, sound and cusp velocities, respectively.

Since the hydrostatic equilibrium is only a function of r, all
the perturbed quantities including $\bm{\xi}$ and $\delta P_{T}$  can be Fourier analyzed

\begin{dmath}\label{eqn:fourier:xi:pt}
	(\bm{\xi}, \delta P_{T}) \propto e^{i(m \phi + k_z z - \omega t)},
\end{dmath}
where $\omega$ is the oscillation frequency, $m$ is the azimuthal wavenumber for which only integer values are allowed and, $k_z$, is the longitudinal wavenumber in the $z$ direction. We study both forward and backward waves which propagate in the positive
and negative z directions respectively, for both the waves the longitudinal wavenumber is
restricted to positive values, the oscillation frequency is positive for forward waves and is
negative for backward wave. The perturbed quantity $\delta P_{T}= \delta p + {\bm B}. \delta{\bm B}/\mu_0$ is the Eulerian perturbation of total (gas and magnetic) pressure.  Putting Eq. (\ref{eqn:fourier:xi:pt}) into (\ref{eqn:linearized:mhd1})-(\ref{eqn:linearized:mhd3}), we obtain the two coupled first order differential equations
\begin{dgroup}\label{eqn:sakurai}
	\begin{dmath}\label{eqn:sakurai:xidel}
		\mathcal{D} \frac{ {\rm d}(r\bm{\xi})}{{\rm d}r} =- r C_2 \delta P_T,
	\end{dmath}
	\begin{dmath}\label{eqn:sakurai:ptdel}
		\mathcal{D} \frac{{\rm d}\delta P_T}{{\rm d}r} = C_3  \bm{\xi}.
	\end{dmath}
\end{dgroup}
The above equations derived earlier by Appert et al. \cite{Appert}
and later by Hain \& Lust \cite{hain1958normal}, Goedbloed \cite{goedbloed1971stabilization} and Sakurai et al. \cite{sakurai1991resonantI}. Here, the multiplicative factors are defined as
\begin{dgroup}\label{eqn:sakurai}
	\begin{dmath}\label{eqn:sakurai:xidel1}
		\mathcal{D}\equiv\rho \left(v^2_{s}+v^2_A\right)\left(\Omega^2-\omega^2_A\right)\left(\Omega^2-\omega^2_A\right),
	\end{dmath}
	
	\begin{dmath}\label{eqn:sakurai:ptdel1}
		C_2\equiv \Omega^4-\left(k_z^2+\frac{m^2}{r^2}\right)\left(v^2_{s}+v^2_A\right)\left(\Omega^2-\omega^2_A\right),
	\end{dmath}
	\begin{dmath}\label{eqn:sakurai:ptdel2}
		C_3\equiv\rho \mathcal{D} \left(\Omega^2-\omega^2_A\right),
	\end{dmath}
\end{dgroup}
in which
\begin{dgroup*}\label{eqn:sakurai}
	\begin{dmath*}\label{eqn:sakurai:ptdel3}
		f_B\equiv k_z B_z,
	\end{dmath*}
	\begin{dmath*}\label{eqn:sakurai:ptdel1}
		\omega^2_A\equiv\frac{f^2_B}{\mu_0 \rho}.
	\end{dmath*}
\end{dgroup*}
and
\begin{dmath*}\label{eqn:sakurai:xidel1}
	\omega^2_c\equiv\left(\frac{v^2_s}{v^2_A+v^2_s}\right)\omega^2_A,
\end{dmath*}
Here $\Omega=\omega-\omega_{f}$ is the Doppler shifted frequency which $\omega_{f}(=k_z v_z.)$ is the flow frequency, $\omega_A(=k_zv_A)$ is the Alfv\'{e}n oscillation frequency and $\omega_c(=k_zv_c)$ is the cusp oscillation frequency. Also $v_A=\left|\textbf{B}_z\right|/\sqrt{\mu_0 \rho}$ is the Alfv\'{e}n speed, $v_s=\sqrt{\gamma p/\rho}$ is the sound speed, and $v_c=\frac{v_s v_A}{(v^2_s+v^2_A)^{1/2}}$  is the cusp speed.

Combining Eqs. (\ref{eqn:sakurai:xidel}) and (\ref{eqn:sakurai:ptdel}), one can obtain a second-order ordinary differential equation for radial component of the  differential equation for $\delta P_{T}$ as \cite{Edwin1983Robertes}
\begin{dmath}\label{eqn:secon-order}
	\frac{{\rm d^2}\delta P_{T}}{{\rm d}r^2}+\frac{1}{r}\frac{{\rm d}\delta P_{T}}{{\rm d}r}-\left(k_r^2+\frac{m^2}{r^2}\right)\delta P_{T}=0,
\end{dmath}
where
\begin{equation}\label{k2}
k^2_{r}\equiv\frac{(\omega^2_{s}-\Omega^2)(\omega^2_{A}-\Omega^2)}{(v^2_{A}+v^2_{s})(\omega_c^2-\Omega^2)},
\end{equation}
solutions of Eq. (\ref{eqn:secon-order}) in the interior ($r\leqslant r_i$) and exterior ($r\geqslant r_e$) regions are given by
\begin{dgroup}
	\begin{dmath}\label{pi}
		\delta P_{Ti}(r) = A_{i } I_m(k_{ri} r),
	\end{dmath}
	\begin{dmath}\label{pe}
		\delta P_{Te}(r) = A_{e } K_m(k_{re} r),
	\end{dmath}
\end{dgroup}
where $A_i$ and $A_e$ are constant. Also $I(.)$ and $K(.)$ are the modified Bessel function of the second kind respectively. Replacing the solutions (\ref{pi}) and (\ref{pe}) into Eq. (\ref{eqn:sakurai:ptdel}) radial displacement can be determined as
\begin{dgroup}
	\begin{dmath}\label{xiri}
		\xi_{ri}(r) =\frac{A_{i }}{\rho_i (\Omega^2-\omega^2_{Ai})}  I^'_m(k_{ri} r),
	\end{dmath}
	\begin{dmath}\label{xire}
		\xi_{re}(r) =\frac{A_{e}}{\rho_i (\Omega^2-\omega^2_{Ae})}  K^'_m(k_{ri} r),
	\end{dmath}
\end{dgroup}
 in which prime denotes differentiation of the function with respect to its argument. These solutions are used in the next sections to determine the
 dispersion relation of the tube oscillations.
\section{Dispersion relation for the case of no inhomogeneous layer} \label{disper}
\begin{figure}
	\centering
	\begin{subfigure}[b]{0.45\textwidth}
		\centering
		\includegraphics[width=\textwidth]{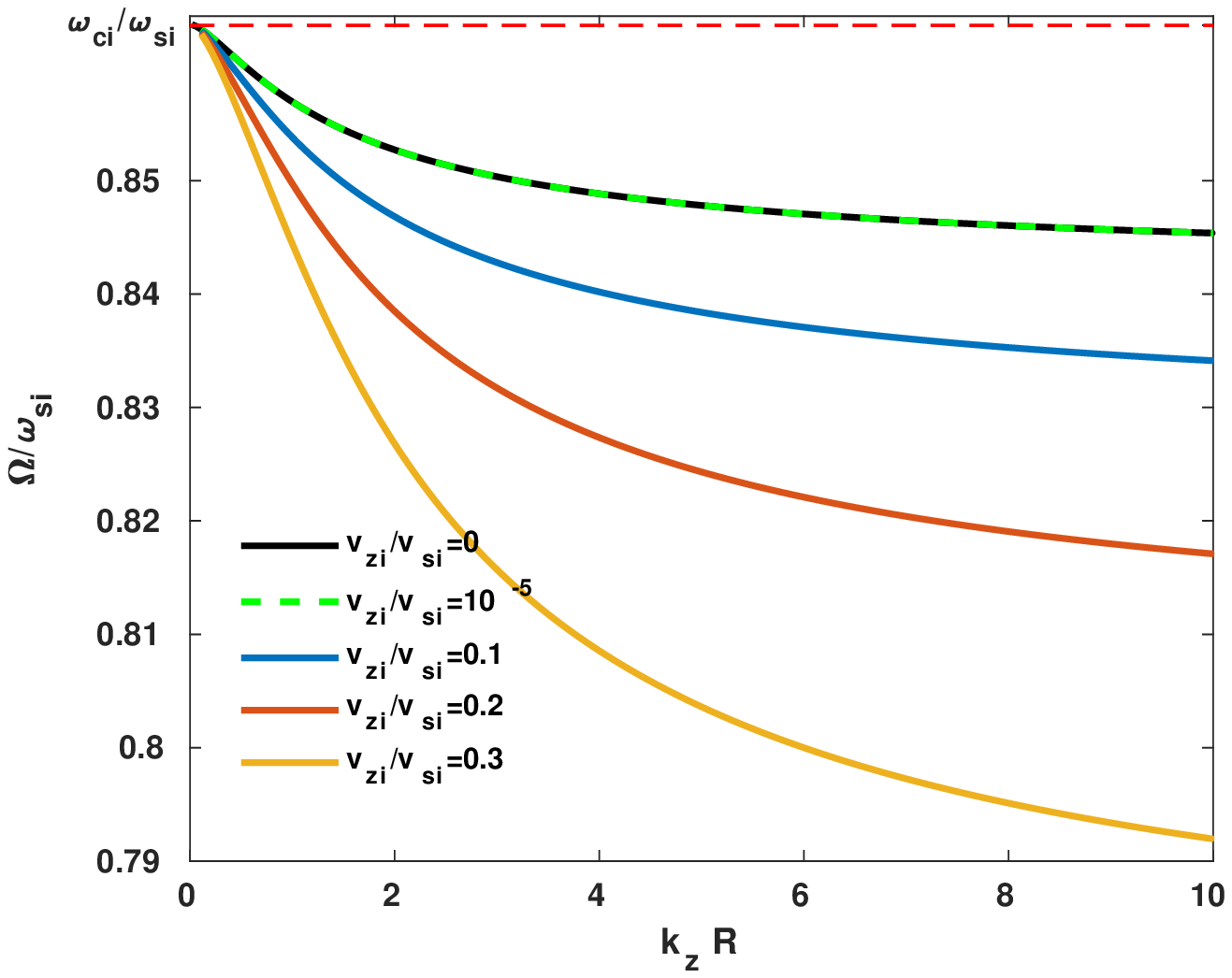}
		\caption{}
		\label{freqsausage}
	\end{subfigure}
	\hfill
	\begin{subfigure}[b]{0.45\textwidth}
		\centering
		\includegraphics[width=\textwidth]{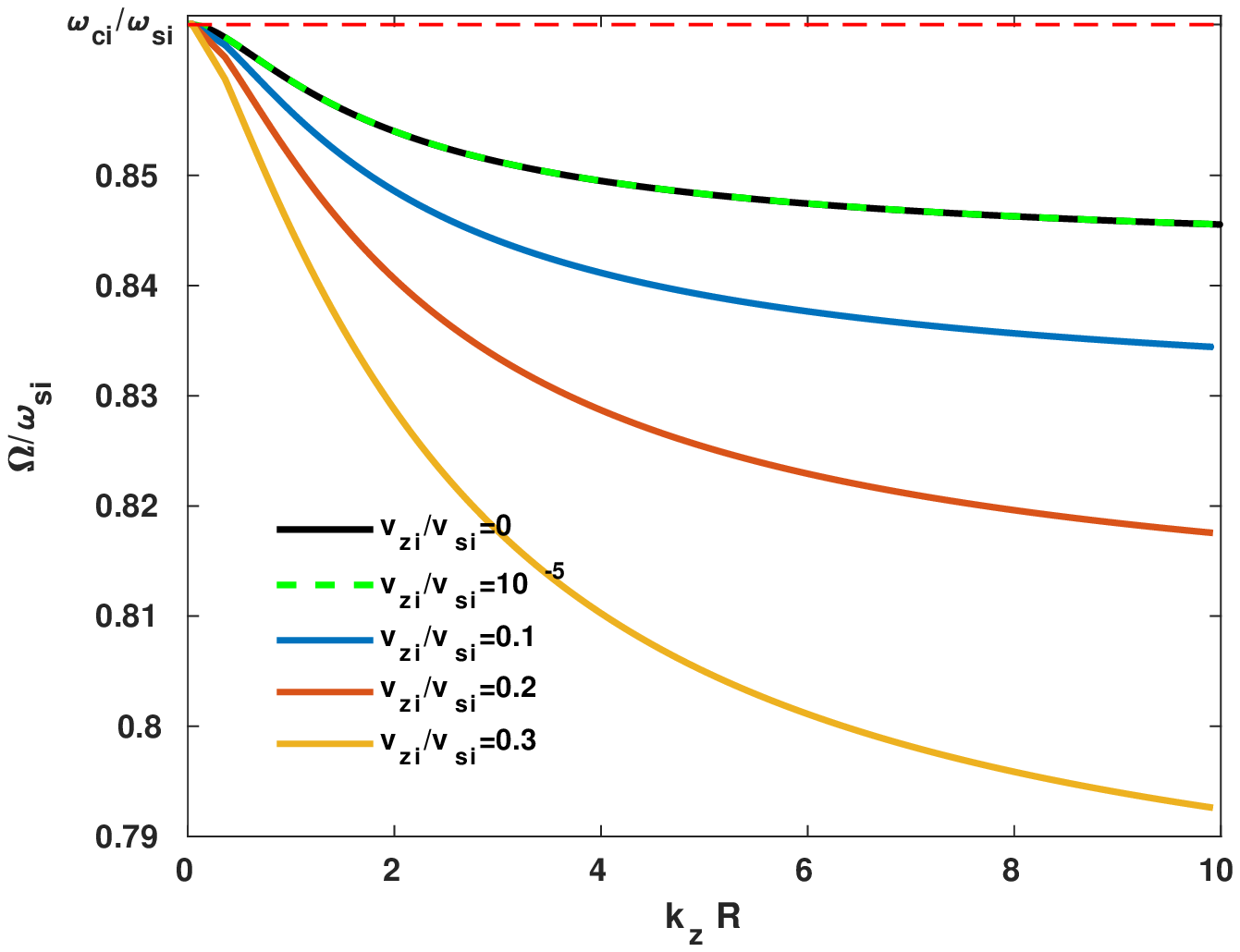}
		\caption{}
		\label{freqkink}
	\end{subfigure}
	\vfill
	\begin{subfigure}[b]{0.45\textwidth}
		\centering
		\includegraphics[width=\textwidth]{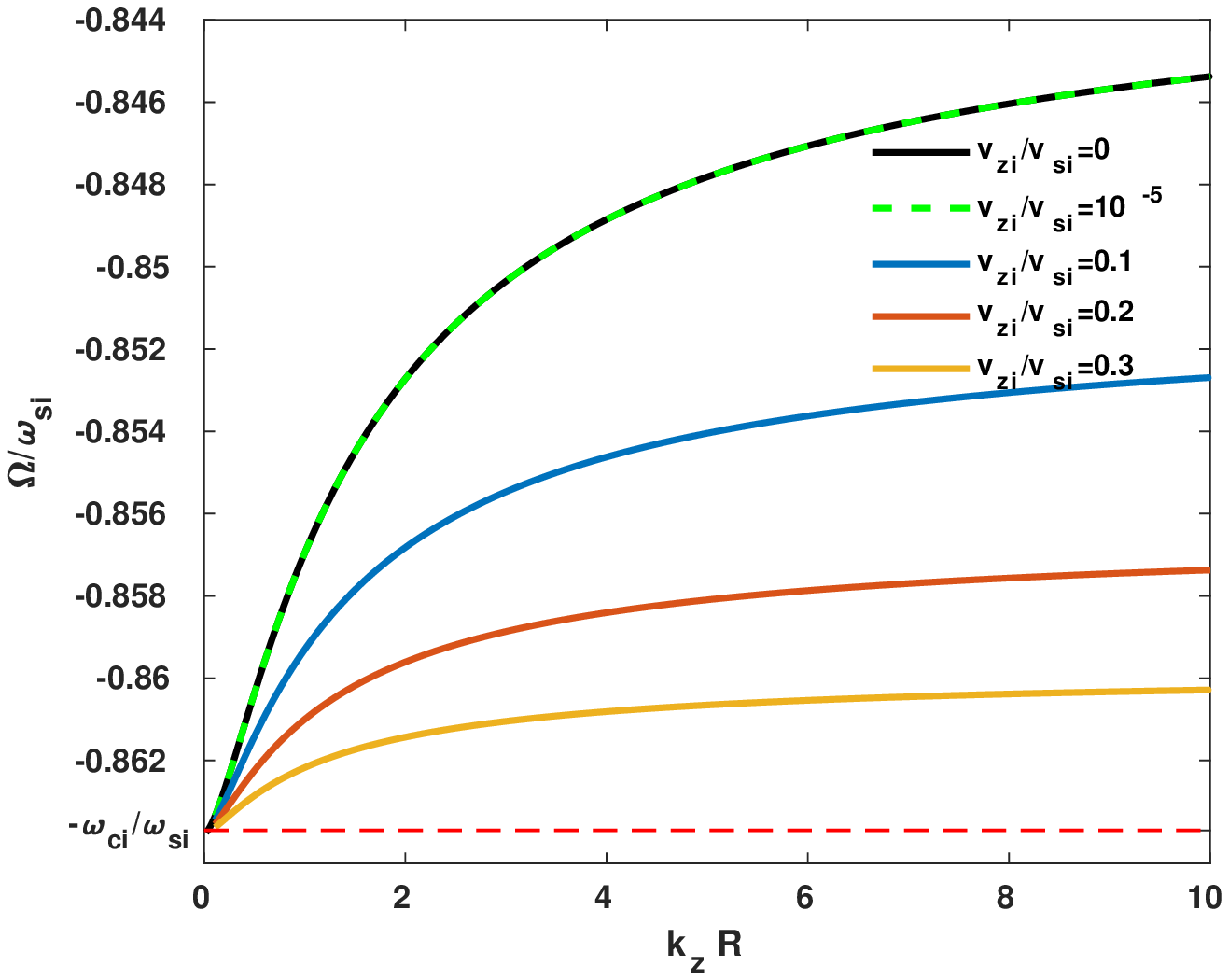}
		\caption{}
		\label{freqnsausage}
	\end{subfigure}
	\hfill
	\begin{subfigure}[b]{0.45\textwidth}
		\centering
		\includegraphics[width=\textwidth]{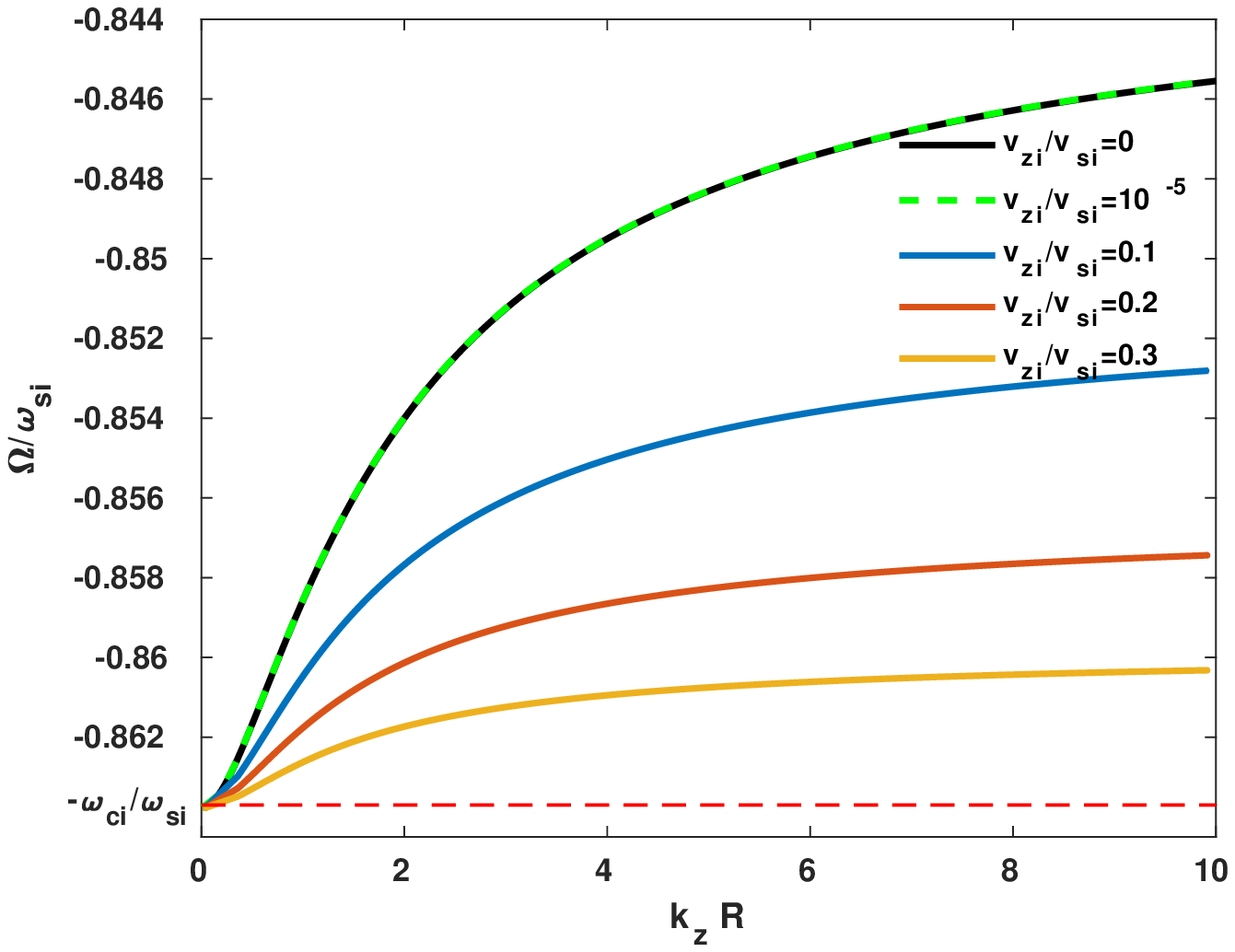}
		\caption{}
		\label{freqnkink}
	\end{subfigure}
	\vfill
	
	\caption{The Dopller shifted phase speed $\Omega/\omega_{si}$, Eq. (\ref{dispersionrelation}), of the slow surface sausage and kink modes versus $k_z R$ for various flow parameters $v_{zi}/v_{si}$ for  forward and backward waves. Panels (a) and (b) are for forward sausage and kink modes and panels (c) and (d) are for backward sausage and kink modes respectively. Under the magnetic pore conditions, following  \cite{Grant2015} the auxiliary parameters are taken as $v_{Ai}=12 ~{\rm km~s}^{-1}$, $v_{Ae}=0 ~{\rm km~s}^{-1}$ (i.e. $B_{ze}=0$), $v_{si}=7 ~{\rm km~s}^{-1}$, $v_{se}=11.5 ~{\rm km~s}^{-1}$, $v_{ci}=6.0464~{\rm km~s}^{-1}(\simeq 0.8638~ v_{si})$ and $v_{ce}=0 ~{\rm km~s}^{-1}$.}
	\label{freqsausagekink}
\end{figure}
In this section we consider a flux tube without the inhomogeneous layer and obtain the dispersion relation of oscillations. For this purpose, the
 solutions obtained for $\xi_{r}$ and $\delta P_{T}$ in the last section inside and outside the tube (i.e Eqs. (\ref{pi})-(\ref{xire})) must be satisfied in the following boundary conditions
\begin{dgroup}\label{boundaryconditions}
	\begin{dmath}\label{boundaryconditions1}
		\xi_{ri}\Big|_{r=R}=\xi_{re}\Big|_{r=R},
	\end{dmath}
	\begin{dmath}\label{boundaryconditions2}
		\delta P_{Ti}\Big|_{r=R}=\delta P_{Te}\Big|_{r=R},
	\end{dmath}
\end{dgroup}
where $R$ is the tube radius. Then the dispersion relation can be
 determined after some algebra as
\begin{equation}\label{dispersionrelation}
\rho_i \left(\Omega_{i}^2-\omega_{Ai}^2\right)-\frac{k_{ri}}{k_{re}}\rho_e \left(\Omega_{e}^2-\omega_{Ae}^2\right)Q_m=0,
\end{equation}
where
\begin{dmath*}
 Q_m=\frac{I_{m}^'\left(k_{ri} R\right)K_{m}\left(k_{re} R\right)}{I_{m}\left(k_{ri} R\right)K_{m}^'\left(k_{re} R\right)}.
\end{dmath*}

For the case with no flow $(\Omega_{i}=\Omega_{e}=\omega)$, the dispersion
 relation reduces to the result obtained  by Edwin \& Roberts \cite{Edwin1983Robertes} and Yu et al. \cite{Yu2017b}.

Here we solve the dispersion relation (\ref{dispersionrelation}) numerically and the
phase speed $\Omega/\omega_{si}$ of the slow surface sausage $(m = 0)$ and kink $(m = 1)$ modes versus
$k_zR$ for various values of the flow parameters $v_{zi}/v_{si}$ are displayed in Fig. \ref{freqsausagekink}. Panels (a) and (b) are for forward sausage and kink modes and panels (c) and (d) are for backward
 sausage and kink modes respectively. The figure shows that (i) for a given value of $k_zR$,
for forward waves when the flow speed increases the Doppler shifted phase speed decreases
and for backward waves the magnitude of the phase speed increases. (ii) For a given flow
speed $v_{zi}/v_{si}$ as $k_zR$ increases the Doppler shifted phase speed for forward decreases and
magnitude of the Doppler shifted phase speed for backward increases. (iii) For $k_zR\ll 1$, for both the forward and backward waves $\Omega/\omega_{si}$ tends to $\omega_{ci}/\omega_{si}$. (iv) These results show that for specific values of the flow speed, the Doppler shifted phase speed is between the
internal and external values of the cusp speed of the flux tube. (vi) For the case of no flow, the result of Yu et al. \cite{Yu2017b} is recovered.

\section{Dispersion relation in the presence of inhomogeneous layer and resonant absorption}\label{conec}
In this section we consider a flux tube with an inhomogeneous boundary layer. According to Equations (\ref{rho})-(\ref{p}), the density, magnetic field and pressure change continuously from the inside to the outside of the tube, so in this case, the Dopller shifted ($\Omega$) of the waves may be equal to the cusp $(\omega_{c})$ or Alfv\'{e}n $(\omega_{A})$ frequency. According to Yu et al. \cite{Yu2017b}, under photosphere conditions
the oscillation frequency will be equal to the cusp frequency at a point in the boundary
layer which causes a singularity in the equations of motion. This phenomenon is called
cusp resonant absorption.

 Sakurai et al. \cite{sakurai1991resonantI} showed that under the thin boundary approximation, the solutions inside and outside the tube can be connected using the connection formula
\begin{dgroup}\label{conection}
	\begin{dmath}\label{conection1}
		[\xi_r]\equiv\xi_{re}(r_e)-\xi_{ri}(r_i)=-i \pi \frac{\Sign~ \Omega}{|\Delta_c|}\frac{\mu \omega_{c}^4}{r B^2 \omega_{A}^2}\Big|_{r=r_c} \delta P_{Ti},
	\end{dmath}
	\begin{dmath}\label{conection2}
		[\delta P_T]\equiv\delta P_{Te}(r_e)-\delta P_{Ti}(r_i)=0,
	\end{dmath}
\end{dgroup}
where $[\xi_r]$ and $[\delta P_T]$ represent the jumps for the Lagrangian radial displacement and total pressure
perturbation across the inhomogeneous (resonant)  boundary, which connects the solutions inside and outside of the flux tube. The subscript $c$ in $\Delta_c$ shows that the quantity must be calculated in the surface where the cusp resonance occurs. We will determine the location of the cusp resonance, $r_c$ later. We obtain the dispersion relation in the presence of flow by substituting the solutions (\ref{pi})-(\ref{xire}) into the connection formula (\ref{conection1}) and (\ref{conection2}), the result is 
\begin{dmath}\label{dispersionrelationthin}
	\rho_i\left(\Omega_{i}^2-\omega_{Ai}^2\right)-\rho_e\left(\Omega_{e}^2-\omega_{Ae}^2\right)\dfrac{k_i}{k_e}Q_m+i \pi \frac{\Sign~  \Omega}{|\Delta_c|}\frac{k_{z}^2 }{\rho  }\left(\dfrac{v_{s}^2}{v_{s}^2+v_{A}^2}\right)^2\rho_i\rho_e\left(\Omega_{i}^2-\omega_{Ai}^2\right)\left(\Omega_{e}^2-\omega_{Ae}^2\right)\dfrac{G_m}{k_e}=0,~~
\end{dmath}
where $G_m=\frac{K_{m}(k_{re}r_e)}{K_{m}^'(k_{re}r_e)}$. It is clear that in the absence of plasma flow this equation
reduces to the dispersion relation obtained by Yu et al. \cite{Yu2017b}.

To display the background quantities in the boundary layer we define the variable
 $\delta\equiv\frac{r-r_i}{r_e-r_i}$ which varies from 0 to 1 in the boundary layer. Using Eqs. (\ref{rho}) to (\ref{p}), one can write the quantities  $v_s=\sqrt{\gamma p/\rho}$ , $v_A=\left|\textbf{B}_z\right|/\sqrt{\mu_0 \rho}$  in the inhomogeneous boundary layer
 as functions of $\delta$ as
\begin{dmath}\label{vs2}
	v_s^2=v_{si}^2\left[\frac{1+\delta(\chi v^2_{sei}-1)}{1+\delta(\chi -1)}\right],
\end{dmath}
\begin{dmath}\label{vA2}
	v_A^2=v_{Ai}^2\left[\frac{1+\delta(\chi v^2_{Aei}-1)}{1+\delta(\chi -1)}\right],
\end{dmath}
and the cusp velocity $v_c\equiv\frac{v_s v_A}{(v^2_s+v^2_A)^{1/2}}$ in the inhomogeneous layer ($r_i< r< r_e$) as
\begin{dmath}\label{v2C}
	v^2_c=\frac{v^2_{si}v^2_{Ai}\Big[1+\delta(\chi v^2_{sei}-1)\Big]\Big[1+\delta(\chi v^2_{Aei}-1)\Big]}{\Big[1+\delta(\chi -1)\Big]\Big[v^2_{si}\Big(1+\delta(\chi v^2_{sei}-1)\Big)+v^2_{Ai}\Big(1+\delta(\chi v^2_{Aei}-1)\Big)\Big]},
\end{dmath}

\begin{figure}
	\centering
	\begin{subfigure}[b]{0.47\textwidth}
		\centering
		\includegraphics[width=\textwidth]{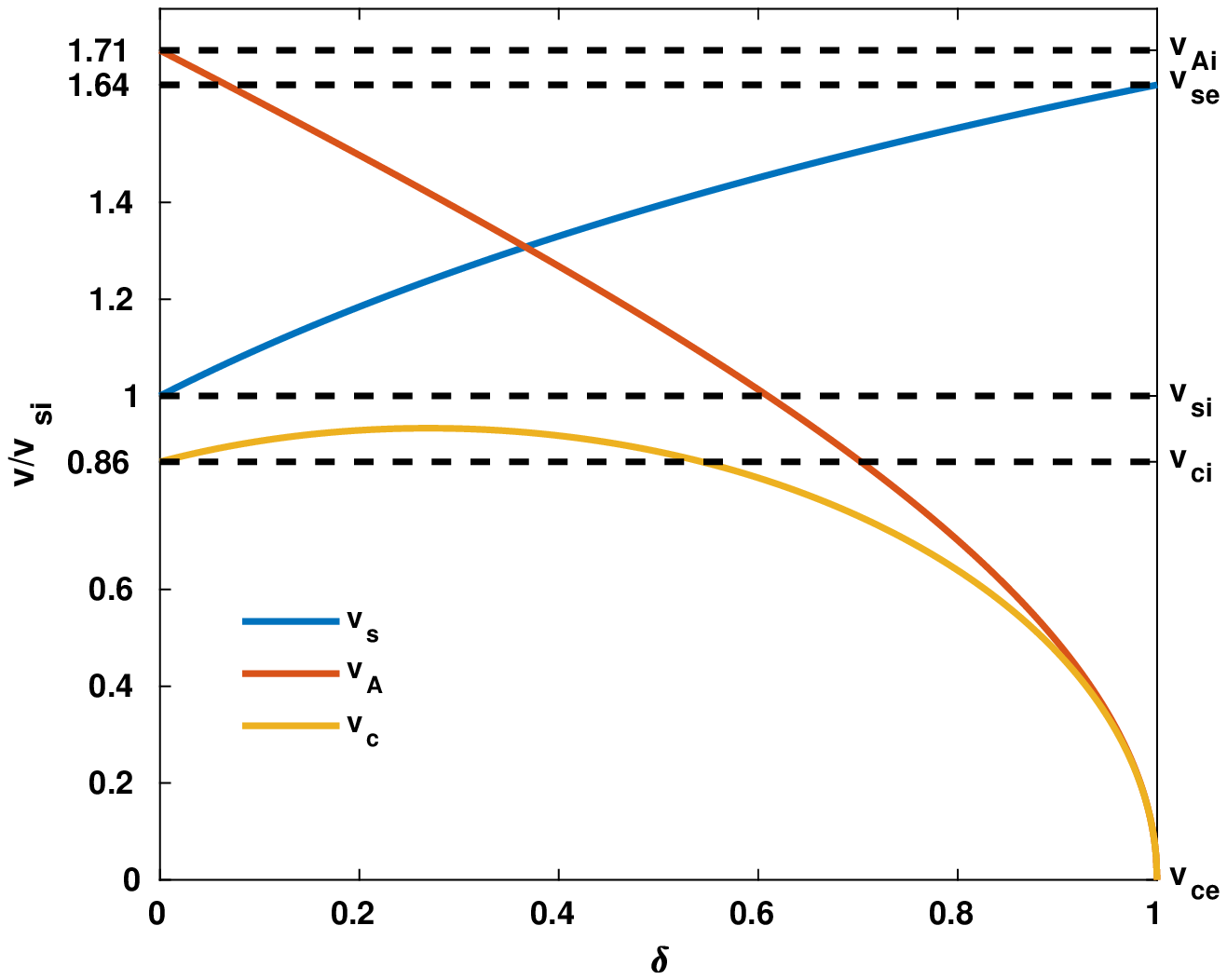}
		\caption{}
		\label{5al0.1tn}
	\end{subfigure}
	\hfill
	\begin{subfigure}[b]{0.47\textwidth}
		\centering
		\includegraphics[width=\textwidth]{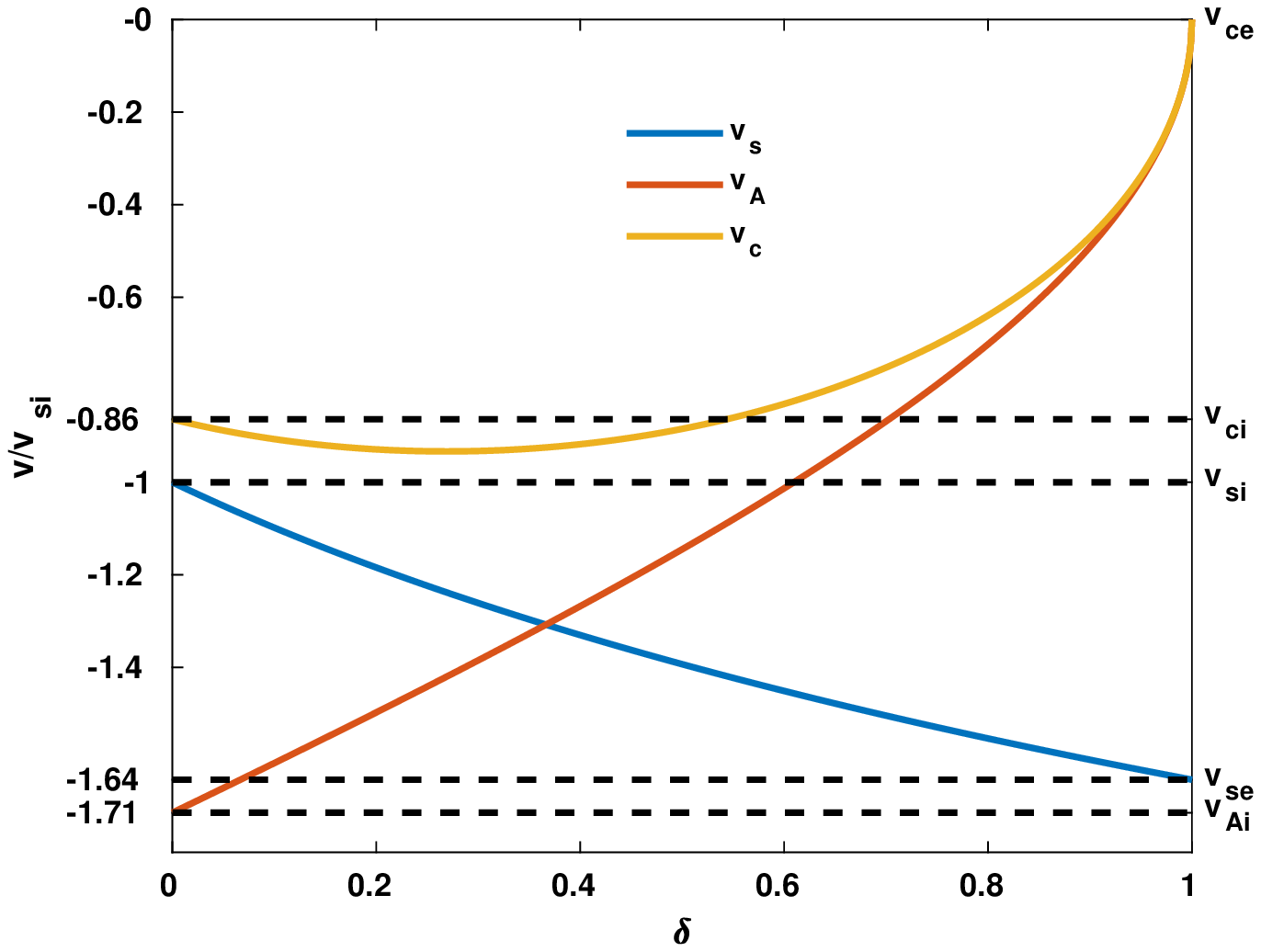}
		\caption{}
		\label{5al0.1tn}
	\end{subfigure}
	\caption{Variations the sound $v_s$ ($-v_s$), Alfv\'{e}n $v_A$ ($-v_A$) and cusp speeds $v_c$ ($-v_c$) versus $\delta$ in the annulus layer under magnetic pore conditions ( $v_{Ai}=12 ~{\rm km~s}^{-1}$, $v_{Ae}=0 ~{\rm km~s}^{-1}$, \textbf{$v_{ze}=0 ~{\rm km~s}^{-1}$}, $v_{si}=7 ~{\rm km~s}^{-1}$, $v_{se}=11.5 ~{\rm km~s}^{-1}$, $v_{ci}=6.0464~{\rm km~s}^{-1}(\simeq 0.8638~ v_{si})$ and $v_{ce}=0 ~{\rm km~s}^{-1}$ ).
    When $v_{ci} \leq v\leq v_{cm}$ resonance absorption occurs for the slow body modes and when $v < v_{ci}$  resonance absorption occurs for slow surface modes in the slow continuum.}
	\label{profile}
\end{figure}

where $\chi\equiv\rho_e/\rho_i$, $v_{sei}\equiv v_{se}/v_{si}$, $v_{Aei}\equiv v_{Ae}/v_{Ai}$.
Using Eqs. (\ref{vs2})-(\ref{v2C}) we plot the sound, Alfv\'{e}n and cusp velocities under magnetic pore conditions in Fig. \ref{profile}. The Figure shows that for $v_c< v_{ci}$ and $v_{ci}<v_c<v_{c_{\rm max}}$, the surface and body sausage modes can
resonantly damp in the slow continuum respectively. Here, $v_{c_{\rm max}}$ is the maximum value of the cusp speed in the transition layer.

Note that according to Yu et al. \cite{{Yu2017b}}, the position of the cusp resonance point $r_c$ is obtained by setting $\Omega^2=\omega_c^2\Big|_{r=r_c}\equiv k_z^2v^2_c\Big|_{r=r_c}$. Consequently, the resulting equation
in terms of the variable $\delta_c\equiv\delta\Big|_{r=r_c}=\frac{r_c-r_i}{r_e-r_i}$ yields the following second order equation
\begin{align}\label{rc}
A\delta_c^2+B\delta_c+C=0,
\end{align}
where $A$,$B$ and $C$ are similar to the constants defined in Eqs. (55)-(57) in Yu et al. 2017 \cite{Yu2017b}. The solutions for $\delta_c$  (see the curve $v_c$ in Fig. \ref{profile})

\begin{align}\label{rc1}
\delta_{c1}=-\frac{B}{2 A}+\frac{\sqrt{B^2-4AC}}{2 A},
\end{align}
\begin{align}\label{rc2}
\delta_{c2}=-\frac{B}{2 A}-\frac{\sqrt{B^2-4AC}}{2 A}.
\end{align}
For the slow surface sausage and kink  mode due to having resonance absorption,
$\Omega/\omega_{si}$ should be below $v_{ci}$, which means that only $\delta_{c2}$
satisfies this condition \cite{Yu2017b}.\\
Next, we turn to calculate the parameter $\Delta_c$ appeared in the dispersion relation (\ref{dispersionrelationthin}). To this aim, using Eq. (\ref{v2C}) and  $\omega_c^2(r_c)=k_z^2v^2_c\Big|_{r=r_c}$ we obtain
\begin{eqnarray}\label{delta}
\Delta_{c}&\equiv&\left[\frac{d}{dr}(\Omega^2-\omega_c^2)\right]_{r=r_c}=-2\left(\left(\omega-\omega_f\right)\frac{d\omega_{f}}{dr}+\omega_{c}\frac{d\omega_{c}}{dr}\right)_{r=r_c}\nonumber\\
&=&-2\left(\omega-\omega_f(r_c)\right)\frac{\omega_{fe}-\omega_{fi}}{l}-\Bigg(\frac{\omega_{c}^2(r_c)}{l}\Bigg)
\Bigg\{\frac{\left(\chi v_{sei}^2-1\right)}{1+\delta\left(\chi v_{sei}^2-1\right)}-\frac{(\chi-1)}{1+\delta(\chi-1)}\nonumber\\&&+\frac{(\chi v_{Aei}^2-1)}{1+\delta\left(\chi v_{Aei}^2-1\right)}\\&&-\frac{v_{si}^2\Big(\chi v_{sei}^2-1\Big)+v_{Ai}^2\Big(\chi v_{Aei}^2-1\Big)}{v_{si}^2\left[1+\delta\Big(\chi v_{sei}^2-1\right)\Big]+v_{Ai}^2\Big[1+\delta\left(\chi v_{Aei}^2-1\right)\Big]}\Bigg\}_{r=r_c},\nonumber
\end{eqnarray}
where $\omega_{f}=k_zv_z$.
\subsection{Weak Damping Limit—Slow Continuum}
Here, we study the dispersion relation (\ref{dispersionrelationthin}) in the weak damping limit. We first rewrite the dispersion relation as
\begin{equation}
D_{{\rm AR}}+iD_{\rm {AI}}=0,
\end{equation}
where $D_{{\rm AR}}$ and $D_{\rm {AI}}$ are the real and imaginary parts of Eq. (\ref{dispersionrelationthin}) respectively, given by
\begin{dmath}\label{limw1}
	D_{{\rm AR}}=\rho_i(\Omega_i^2-\omega_{Ai}^2) -\rho_e(\Omega_e^2-\omega_{Ae}^2) \frac{k_{ri}}{k_{re}}Q_m,
\end{dmath}
\begin{dmath}\label{limw2}
	D_{{\rm AI}}=\frac{ \pi \rho_i \rho_e k_z^2 }{k_{re} }\frac{\Sign~\Omega}{\rho_c|\Delta_c|}\Big|_{r=r_c}\Big(\frac{v_{sc}^2}{v_{Ac}^2+v_{sc}^2}\Big)^2 (\Omega_i^2-\omega_{Ai}^2)(\Omega_e^2-\omega_{Ae}^2) G_m.
\end{dmath}
Note that in Eqs. (\ref{limw1}) and (\ref{limw2}) we have the complex frequency $\omega=\omega_r+i\gamma$, in which $\omega_r$ and $\gamma$ are oscillation frequency and the damping rate, respectively.
In the limit of weak damping, i.e. $\gamma \ll \omega_r$, the damping rate $\gamma$ is given as \cite{Goossens}
\begin{equation}\label{gammac1}
\gamma_{mc}=-D_{{\rm AI}}(\omega_r)\left(\frac{\partial D_{{\rm AR}}}{\partial \omega}\Big|_{\omega_r}\right)^{-1}.
\end{equation}
Here, we want to simplify Eq. (\ref{gammac1}), to obtain the damping rate of surface sausage modes in the weak damping limit, i.e. $\gamma \ll \omega_r$. To this aim, we first calculate  $\frac{\partial D_{AR}}{\partial \omega}$ from Eq. (\ref{limw1})  as follows
\begin{align}\label{div}
&\frac{\partial D_{AR}}{\partial \omega}=2 \rho_i \Omega_i -2 \rho_e \Omega_e \frac{k_{ri}}{k_{re}}Q_m- \rho_e \left(\Omega_e-\omega_{Ae}^2\right)\left(\frac{1}{k_{re}}\frac{dk_{ri}}{d\omega}-\frac{k_{ri}}{k_{re}^2}\frac{dk_{re}}{d\omega}\right)Q_m-\rho_e \left(\Omega_e^2-\omega_{Ae}^2\right)\frac{k_{ri}}{k_{re}}\frac{dQ_m}{d\omega}.
\end{align}
Now from Eq. (\ref{k2}), one can obtain
\begin{equation}\label{kk}
\frac{dk_{ri}}{dw}=\frac{- \Omega_i^3(\Omega_i^2-2\omega_{ci}^2)}{(v_{si}^2+v_{Ai}^2)(\Omega_i^2-\omega_{ci}^2)^2 k_{ri}},
\end{equation}
\begin{equation}\label{kk2}
\frac{dk_{re}}{dw}=\frac{- \Omega_e^3(\Omega_e^2-2\omega_{ce}^2)}{(v_{se}^2+v_{Ae}^2)(\Omega_e^2-\omega_{ce}^2)^2 k_{re}}.
\end{equation}
With the help of Eqs. (\ref{kk}) and (\ref{kk2})
\begin{align}\label{div57}
&\frac{dQ_m}{d\omega}= xP_m\frac{\Omega_i^3(\Omega_i^2-2\omega_{ci}^2)}{(\omega_{si}^2-\Omega_i^2)(\omega_{Ai}^2-\Omega_i^2)(\Omega_i^2-\omega_{ci}^2)}+yS_m\frac{ \Omega_e^3(\Omega_e^2-2\omega_{ce}^2)}{(\omega_{se}^2-\Omega_e^2)(\omega_{Ae}^2-\Omega_e^2)(\Omega_e^2-\omega_{ce}^2) }.
\end{align}
Replacing this into Eq. (\ref{div}) yields
\begin{align}\label{di}
&\frac{\partial D_{AR}}{\partial \omega}=2 \rho_i \Omega_i -2 \rho_e \Omega_e \frac{k_{ri}}{k_{re}}Q_m-\rho_e  \left(\Omega_e^2-\omega_{Ae}^2\right)\frac{k_{ri}}{k_{re}}\nonumber\\
&\left(\frac{(Q_m+xP_m)(\Omega_i^2-2\omega_{ci}^2)\Omega_i^3}{(\omega_{si}^2-\Omega_i^2)(\omega_{Ai}^2-\Omega_i^2)(\Omega_i^2-\omega_{ci}^2)}-\frac{ (Q_m-yS_m)(\Omega_e^2-2\omega_{ce}^2)\Omega_e^3}{(\omega_{se}^2-\Omega_e^2)(\omega_{Ae}^2-\Omega_e^2)(\Omega_e^2-\omega_{ce}^2) }\right),
\end{align}
where
\begin{align}\label{div55}
&P_m\equiv\left(\frac{I''_m(x)}{I_m(x)}-\frac{{I'_m(x)}^2}{I_m(x)^2}\right)\frac{K_m(y)}{K_m'(y)},\nonumber\\
&S_m\equiv\left(1-\frac{K''_m(y) K_m(y)}{{K'_m(y)}^2}\right)\frac{I'_m(x)}{I_m(x)},
\end{align}
and $x=k_{ri}r_i$ and $y=k_{re}r_e$. Finally, substituting Eqs. (\ref{limw2}) and (\ref{di}) into Eq. (\ref{gammac1}) one can get the damping rate $\gamma$ in the limit of weak damping for the surface modes in the slow continuum as
\begin{align}\label{gammacwd}
\gamma_{mc}\Big|_{\omega=\omega_{r}}=-\frac{\frac{ \pi \rho_e k_z^2 }{k_{re}\rho_c }\frac{\Sign~\Omega}{|\Delta_c|}\Big|_{r=r_c}\left(\frac{v_{s}^2}{v_{A}^2+v_{s}^2}\right)^2 (\Omega_i^2-\omega_{Ai}^2)(\Omega_e^2-\omega_{Ae}^2) G_m}{2\left(\Omega_i-\chi\Omega_e\frac{k_{ri}}{k_{re}}Q_m\right)- \chi T_m},
\end{align}
where
\begin{align}\label{di11}
&T_m=\left(\Omega_e^2-\omega_{Ae}^2\right)\frac{k_{ri}}{k_{re}}\left(\frac{(Q_m+xP_m)(\Omega_i^2-2\omega_{ci}^2)\Omega_i^3}{(\omega_{si}^2-\Omega_i^2)(\omega_{Ai}^2-\Omega_i^2)(\Omega_i^2-\omega_{ci}^2)}-\frac{ (Q_m-yS_m)(\Omega_e^2-2\omega_{ce}^2)\Omega_e^3}{(\omega_{se}^2-\Omega_e^2)(\omega_{Ae}^2-\Omega_e^2)(\Omega_e^2-\omega_{ce}^2) }\right).
\end{align}
Equation (\ref{gammacwd}) can be more simplified in the long wavelength limit which we do in the next subsection.
\subsection{Weak damping rate in long wavelength limit - slow continuum}
In the limit $k_zR \ll 1$ i.e. $k_{ri}R(k_{re}R) \ll 1$ we can obtain a more simplified expansion for the damping rate $\gamma$, by using the asymptotic expansion of $Q_m, G_m, P_m$ and $S_m$.
For the sausage $(m=0)$ mode in the slow continuum we obtain (see Appendix \ref{A})
\begin{align}\label{gammacwd21121}
\gamma_{0c}=\frac{ 2\pi \chi^3 \Sign \Omega}{|\Delta_c| R}\left[\frac{ \omega_{ci}^7\omega_{si}^2\left(\Omega_{e}^2-\omega_{Ae}^2\right)^3 }{3\omega_{Ai}^{10} \omega_{ci}^2+8\chi \omega_{Ai}^8 \omega_{si}^2\left(\Omega_{e}^2-\omega_{Ae}^2\right)\ln(k_zR)}\right](k_zR)^4\ln^3(k_zR).
\end{align}
For the kink $(m=1)$ mode in the slow continuum we obtain (see Appendix \ref{B})
\begin{align}
\gamma_{1c}=-\frac{ \pi \chi^2 \Sign~\Omega}{8|\Delta_c| R}\frac{ \omega_{ci}^{11}\left(\Omega_e^2-\omega_{Ae}^2\right)^2}{\omega_{Ai}^{4}\left(\omega_{ci}^2\omega_{Ai}^2 -\chi\omega_{si}^2 \left(\Omega_e^2-\omega_{Ae}^2\right)\right)^2}(k_zR)^4.
\end{align}
Under magnetic pore condition $(v_{Ae}=0)$
\begin{align}\label{limitsu}
\gamma_{0c}=\frac{ 2\pi \chi^3 \Sign ~\Omega}{|\Delta_c| R}\left[\frac{ \omega_{ci}^7\omega_{si}^2 \Omega_{e}^6 }{3\omega_{Ai}^{10} \omega_{ci}^2+8\chi \omega_{Ai}^8 \omega_{si}^2 \Omega_{e}^2\ln(k_zR)}\right](k_zR)^4\ln^3(k_zR),
\end{align}
\begin{align}
\gamma_{1c}=-\frac{ \pi \chi^2 \Sign~\Omega}{8|\Delta_c| R}\frac{ \omega_{ci}^{11}\Omega_e^4}{\omega_{Ai}^{4}\left(\omega_{ci}^2\omega_{Ai}^2 -\chi\omega_{si}^2 \Omega_e^2\right)^2}(k_zR)^4.
\end{align}
In the absence of flow $(v_{zi}=v_{ze}=0)$, $\Omega_e=\omega_{ci}$ so
\begin{align}
\gamma_{0c}=\frac{ 2\pi \chi^3 \Sign~ \Omega}{|\Delta_c| R}\left[\frac{ \omega_{ci}^11\omega_{si}^2 }{3\omega_{Ai}^{10} +8\chi \omega_{Ai}^8 \omega_{si}^2 \ln(k_zR)}\right](k_zR)^4\ln^3(k_zR),
\end{align}
\begin{align}
\gamma_{1c}=-\frac{ \pi \chi^2 \Sign~\Omega}{8|\Delta_c| R}\frac{ \omega_{ci}^{11}}{\omega_{Ai}^{4}\left(\omega_{Ai}^2 -\chi\omega_{si}^2 \right)^2}(k_zR)^4,
\end{align}
where these relations are the same Eqs. $(79)$ in \cite{Sadeghi} and  $(38)$ in \cite{Yu2017b} respectively.
\section{Numerical results}\label{num}
 In this section we solve the dispersion relation (Eq. (\ref{dispersionrelationthin})) numerically to obtain the frequencies and damping rates of the slow surface sausage and kink modes and we compare the analytical results (Eq. \ref{gammacwd}) with the numerical results. Under the magnetic pore conditions, following  \cite{Grant2015} we set again the model parameters as $v_{Ai}=12 ~{\rm km~s}^{-1}$, $v_{Ae}=0 ~{\rm km~s}^{-1}$ (i.e. $B_{ze}=0$), $v_{si}=7 ~{\rm km~s}^{-1}$, $v_{se}=11.5 ~{\rm km~s}^{-1}$, $v_{ci}=6.0464~{\rm km~s}^{-1}(\simeq 0.8638~ v_{si})$ and $v_{ce}=0 ~{\rm km~s}^{-1}$. We have assumed the flow outside the tube to be zero $(v_{ze}=0 ~{\rm km~s}^{-1})$. Note that the dispersion relations, Eqs. (\ref{dispersionrelation}) and (\ref{dispersionrelationthin}), are symmetric under the exchange $(\omega,v_z)$ with $(-\omega,-v_z)$. Therefore, it is sufficient to consider only the positive values of flow velocity with both positive and negative values of oscillation frequency, i.e. forward and backward waves in the presence of upward plasma flow. Our numerical results are shown in Figs. \ref{5sl0.1} to \ref{5kkz4}.\\
Figures \ref{5sl0.1} and \ref{5sl0.4} represent variations of the phase speed (or normalized frequency) $v/v_{si}\equiv\omega_r/\omega_{si}$, Doppler shifted phase speed $\Omega/\omega_{si}$ and the damping rate $-\gamma_{0c}/\omega_r$ ($\gamma_{0c}/\omega_r$) of the slow surface sausage modes for forward and backward waves versus $k_z R$ for various flow parameters and various thickness of the inhomogeneous layer $l/R=(0.1,0.2)$. The left panels of these figures clear that for forward wave  and various flow parameters $v_{zi}/v_{si}=(10^{-5}, 0.2, 0.4, 0.6, 0.8)$ (i) The value of the phase speed $v/v_{si}$ increases with increasing the flow parameter $v_{zi}/v_{si}$. (ii) The minimum value of the Doppler shifted phase speed decreases with increasing the flow. (iii) The maximum value of $-\gamma_{0c}/\omega_r$ increases, and for low flow parameter correspond to smaller $k_zR$ when $v_{zi}/v_{si}$ increases but for high flow parameter correspond to larger $k_zR$ when $v_{zi}/v_{si}$ increases.
(iv) The dashed-line curves in these figures represent the analytical results of the damping rate $-\gamma_{0c}/\omega_{r}$ evaluated by Eq. (\ref{gammacwd}). These curves show that for the weak damping (i.e. $\gamma_{0c} \ll \omega_r$) and in the long wavelength limit (i.e. ${k_zR}\ll 1$) the oscillation frequency is not affected by the presence of the transitional layer. This is also confirmed by our numerical results. (vi) For a given $l/R$, the minimum value of the damping time to period ratio $\tau_D/T=2 \pi /|\gamma_{0c}|$
decreases with increasing $v_{zi}/v_{si}$. For instance, for the case
where $l/R=0.1$ and $k_zR=1$, the value of $\tau_D/T$ for $v_{zi}/v_{si}=0.8$ changes by $\sim 95\%$ less than the case where there
is no flow. So, the relation between the damping rate
(time) and the flow is of interest. Several researcher obtained similar results for the sausage modes in photospheric conditions. Yu et al. showed that for
$l/R = 0.1$ the minimum value of the damping time to period ratio is $\tau_D/T=14.11$ \cite{Yu2017b} and \cite{Sadeghi} showed that for
$l/R = 0.1$ the minimum value of the damping time to period ratio is $\tau_D/T=10.2$ for twist parameter $B_{\phi i}/B_{zi}=0.3$, while our results show that the minimum value of the damping time to period ratio forhigh upflow is much lower. vii) For $k_zR\rightarrow 0$, we see
that the damping rate go to zero for
finite values of the flow parameter, and it is an agreement with analytical relation Eq. (\ref{limitsu}).\\
 The right panels in Figs. \ref{5sl0.1} and \ref{5sl0.4} we plot the phase speed (or normalized frequency) $v/v_{si}\equiv\omega_r/\omega_{si}$, normalized Doppler Shifted $\Omega/\omega_{si}$ and the damping rate $\gamma_{0c}/\omega_r$ of the slow surface sausage modes for backward wave versus $k_z R$ for various flow parameters $v_{zi}/v_{si}=(10^{-5}, 0.1, 0.2, 0.3)$ and various thickness of the inhomogeneous layer $l/R=(0.1,0.2)$. The figures show that (i) the magnitude of the phase velocity decreases with increasing flow. (ii) The magnitude of the  Doppler shifted phase speed increases with increasing flow. (iii) The maximum value of $\gamma_{0c}/\omega_r$ decreases, and it corresponds to smaller $k_zR$ when $v_{zi}/v_{si}$ increases. (vi) For a given $l/R$, the minimum value of $\tau_D/T$
increases with increasing $v_{zi}/v_{si}$. For instance, for the case
where $l/R=0.1$ and $k_zR=1$, the value of $\tau_D/T$ for $v_{zi}/v_{si}=0.3$ changes by $\sim 238\%$ more than the case where there
is no flow. Due to the fact that at high flow parameters for backward waves, Doppler shifted frequencies out of the resonant region, so it is plotted up to a flow parameters of 0.3.\\
Figures \ref{5skz0.5} and \ref{5skz4} show the variations of the phase speed (or normalized frequency) $v/v_{si}\equiv\omega_r/\omega_{si}$, phase Doppler Shifted $\Omega/\omega_{si}$ and the damping rate $-\gamma_{0c}/\omega_r$ ($\gamma_{0c}/\omega_r$) of the slow surface sausage modes for forward and backward wave versus the inhomogeneous layer $(l/R)$ for various flow parameters and $k_zR=(0.5,2)$. The left panels of figures \ref{5skz0.5} and \ref{5skz4} show that for forward waves  and various flow parameters $v_{zi}/v_{si}=(10^{-5}, 0.2, 0.4, 0.6, 0.8)$ (i) the frequency increases with increasing the flow $(v_{zi}/v_{si})$. (ii) With increasing $l/R$ for $k_zR\ll 1$, the Doppler shifted frequency increases, but for $k_zR\gg 1$ the Doppler shifted frequency reaches a peak value then tends to $v_{ci}/v_{si}$. (iii) For $k_zR\ll 1$, the Doppler shifted frequency decrease when the flow increases, for $k_zR\gg 1$, when the Doppler shifted frequency reaches above $v_{ci}/v_{si}$, it decreases with increasing flow and tends to the value of $v_{ci}/v_{si}$. (iv) For a given $k_zR$, the damping rate values increases and the damping time to period ratio values decreases with increasing flow. For example, for $k_zR=2$ the minimum value of $\tau_D/T$ for $v_{zi}/v_{si}=0.8$ decreases $\sim 93\%$ with respect to the case where there is no flow.\\
The right panels of figures \ref{5skz0.5} and \ref{5skz4} show the variations of the phase speed (or normalized frequency) $v/v_{si}\equiv\omega_r/\omega_{si}$, Doppler shifted phase speed $\Omega/\omega_{si}$ and the damping rate $\gamma_{0c}/\omega_r$ of the slow surface sausage modes for backward wave versus the inhomogeneous layer $(l/R)$ for various flow parameters $v_{zi}/v_{si}=(10^{-5},0.1,0.2,0.3)$ and $k_zR=(0.5,4)$. The figures show that (i) the magnitude of the phase velocity increases with increasing flow. (ii) With increasing $l/R$ for $k_zR\ll 1$, the Doppler shifted frequency decreases, but for $k_zR\gg 1$ the Doppler shifted frequency reaches a minimum value then tends to $-v_{ci}/v_{si}$. (iii) For $k_zR\ll 1$, the magnitude of the Doppler shifted frequency increase when the flow increases. For $k_zR\gg 1$, the Doppler shifted frequency reaches $-v_{ci}/v_{si}$, it decreases with increasing flow and tends to the value of $-v_{ci}/v_{si}$. (iv) For a given $k_zR$, the values of the damping rate decrease and the values of the damping time to period ratio increase with increasing flow. For example, for $k_zR=2$ the minimum value of $\tau_D/T$ for $v_{zi}/v_{si}=0.3$ increases $\sim 278\%$ than the case where there is no flow.\\
We plot the results for kink waves in Figs. \ref{5kl0.1} and \ref{5kl0.4}. Same as the case of sausage modes in Figs. \ref{5sl0.1} and \ref{5sl0.4} for the forward wave (the left panels) the maximum value of $-\gamma_{1c}/\omega_r$ increases, and for low flow parameter correspond to smaller $k_zR$ when $v_{zi}/v_{si}$ increases but for high flow parameter correspond to larger $k_zR$ when $v_{zi}/v_{si}$ increases. For a given $l/R$, the minimum value of $\tau_D/T$ increases with increasing $v_{zi}/v_{si}$. For instance, for the case where $l/R=0.2$ the minimum value of $\tau_D/T$ for $v_{zi}/v_{si}=0.8$ changes by $\sim 57\%$ less than the case where there
is no flow. Yu et al. \cite{Yu2017b} showed that for
$l/R = 0.2$, a minimum value of $\tau_D/T$ is about 18.8 but our result
gives value about 2.8. It is now that Soler et al. \cite{Soler2009} have obtained this number about 1000 for $l/R = 0.2$. Also for the backward wave (the right panels) the maximum value of $\gamma_{1c}/\omega_r$ increases, and its position moves to smaller $k_zR$ when $v_{zi}/v_{si}$ increases.\\
Figures \ref{5kkz0.5}-\ref{5kkz4} are similar to Figs. \ref{5skz0.5}-\ref{5skz4} but for kink modes. The results show that the effect of flow on the slow resonance absorption of sausage and kink modes is almost the same. The effect of the slow resonance in the presence of flow on the
wave damping is significant under photospheric conditions.\\
It should be noted that for the case there is no flow, the results are similar to the results of \cite{Yu2017b}. When the flow is very small (i.e $v_{zi}/v_{si}=10^{-5}$) the results overlap with the no flow case.\\
Figure \ref{taudperiodvz} shows the minimum value of damping time to period ratio ($\tau_D/T$) for the forward wave of the slow surface sausage (solid line) and kink (dashed line) modes versus upflow velocity ($v_{zi}/v_{si}$). This figure shows that when the upflow velocity increases, the minimum value of damping time to period ratio can be considerably reduced. For instance, for the upflow velocity value $v_{zi}/v_{si}=0.87$, the damping time to period ratio of the surface sausage mode will reach about 0.30. This confirms that the resonant absorption in the presence of flow can be considered as an effective mechanism to justify the rapid damping of slow surface sausage mode observed by \cite{Grant2015}. Note that for all the results indicated in Fig. \ref{taudperiodvz}, the longitudinal wave number is in the observational range i.e. $k_zR\leqslant 5$. In the observational range, the minimum number of oscillations increases slightly for large values of $v_{zi}/v_{si}$.

\section{Conclusions}\label{con}
 In this paper  we studied the effect of the flow parameter on the frequencies, the damping rates in slow continuum of slow sausage and kink waves in magnetic flux tubes under solar photospheric (or magnetic pore) conditions. We considered a straight cylindrical flux tube with tree region inside, annulus and outside in which the linear density, squared magnetic field (linear pressure) and  linear flow profiles are considered in the annulus region or transitional layer. In addition, we numerically solved the dispersion relation and obtained the phase speed (or normalized frequency) $v/v_{si} \equiv \omega_{r}/\omega_{si}$, the normalized Doppler shifted frequency, the damping rate $\gamma_{mc}/\omega_{r}$, and the damping time to period ratio $\tau_D/T$ of the slow surface sausage and kink modes for forward and backward waves under photospheric (magnetic pore) conditions. Our results show that:\\
\begin{itemize}

\item For forward waves, the frequency and the damping rate increase when the flow
parameter increases but for backward waves, the frequencies and the damping rate decreases when the flow
parameter increases.
\item For forward waves, the damping time to period ratio decreases when the flow
parameter increase  but for backward waves, the damping time to period ratio increase when the flow parameter increases.
\item  For a given $l/R$, the Doppler shifted frequency, approach $\Omega/\omega_{si}\rightarrow v_{ci}/v_{si}$ for forward waves and  approach $\Omega/\omega_{si}\rightarrow -v_{ci}/v_{si}$ for backward waves and
$\gamma_{mc}/\omega_{r}\rightarrow 0$ for both forward and backward waves, in the long and short-wavelength limit.

\item For a given $k_zR$, the maximum value of $\gamma_{mc}/\omega_{r}$ (or
minimum value of $\tau_D/T$) increases (or decreases) for forward waves and decreases (increases) for backward waves.

\item For the case where $l/R=0.1$, the minimum value of $\tau_D/T$ for $v_{zi}/v_{si}=0.6$, for instance, changes $\sim89\%$
less for forward sausage waves and for backward sausage waves the minimum value of $\tau_D/T$ for $v_{zi}/v_{si}=0.3$, changes $\sim204\%$ more with respect to the case where there is no flow. Also, for kink mode changes $\sim83\%$ less for forward waves and $\sim272\%$
more for backward waves with respect to the case where there is no flow.
 According to these results, it can be said that the flow has a significant effect on the resonant
absorption of the slow surface sausage and kink modes in magnetic
flux tubes under magnetic pore conditions.

\item For the case of $l/R=0.1$ and $v_{zi}/v_{si}=0.87$, the damping time to period ratio of the surface sausage mode can reach $\tau_D/T=0.30$. For comparison, for a static tube (no flow) with $l/R= 0.1$, \cite{Yu2017b} obtained $\tau_D/T = 14.11$. This confirms that the resonant absorption in the presence of plasma flow can justify the extremely rapid damping of the slow surface sausage mode observed by \cite{Grant2015}.

\end{itemize}
\begin{appendices}
\section{Weak damping rate in long wavelength limit for the sausage mode}\label{A}
For the sausage mode $m=0$, we have
\begin{dmath}\label{limb}
Q_0=\frac{I_{0}^'\left(x\right)K_{0}\left(y\right)}{I_{0}\left(x\right)K_{0}^'\left(y\right)}=-\frac{I_{1}\left(x\right)K_{0}\left(y\right)}{I_{0}\left(x\right)K_{1}\left(y\right)}\approx \frac{xy\left(\ln(y/2)+\gamma_e\right)}{2},
\end{dmath}
\begin{dmath}
G_0=\frac{K_{0}(y)}{K_{0}^'(y)}=\frac{K_{0}(y)}{-K_{1}(y)}\approx \frac{-\ln(y/2)-\gamma_e}{-1/y}=y\left(\ln(y/2)+\gamma_e\right),
\end{dmath}
\begin{dmath}
P_0=\left(\frac{I''_0(x)}{I_0(x)}-\frac{{I'_0(x)}^2}{I_0(x)^2}\right)\frac{K_0(y)}{K_0'(y)}\approx y\left(\frac{1}{2}-\frac{3y^2}{16}\right)\left(\ln(y/2)+\gamma_e\right),
\end{dmath}
\begin{dmath}\label{limb1}
S_0=\left(1-\frac{K''_0(y) K_0(y)}{{K'_0(y)}^2}\right)\frac{I'_0(x)}{I_0(x)}\approx \frac{x}{2}\left(1+\ln(y/2)+\gamma_e\right).
\end{dmath}
Inserting Equations (\ref{limb})-(\ref{limb1}) into Equation (\ref{di11}) yields
\begin{align}
T_0=\left(\Omega_e^2-\omega_{Ae}^2\right)\frac{k_{ri}}{k_{re}}&\Bigg(\frac{xy\ln(y)\left(1-\frac{3y^2}{16}\right)(\Omega_i^2-2\omega_{ci}^2)\Omega_i^3}{(\omega_{si}^2-\Omega_i^2)(\omega_{Ai}^2-\Omega_i^2)(\Omega_i^2-\omega_{ci}^2)}\nonumber\\
&+\frac{ xy(\Omega_e^2-2\omega_{ce}^2)\Omega_e^3}{2(\omega_{se}^2-\Omega_e^2)(\omega_{Ae}^2-\Omega_e^2)(\Omega_e^2-\omega_{ce}^2) }\Bigg),
\end{align}
where $\ln(y/2)+\gamma_e=\ln(y)$. In the limit $k_zR<<1$ $(\Omega_{i}\approx \omega_{ci})$ above relation becomes singular. To avoid singularity, we need to evaluate the quantity $\alpha$. To this aim, following  \cite{Yu2017b} we first replace $\Omega^2_{i}=\omega_{ci}^2-\alpha$ into Eq. (\ref{k2}) and get
\begin{equation}\label{kix}
k_{ri}^2\simeq \frac{k_z^2}{\alpha} \frac{\left(\omega_{ci}^2-\omega_{si}^2\right)\left(\omega_{ci}^2-\omega_{Ai}^2\right)}{\left(\omega_{Ai}^2+\omega_{si}^2\right)}
=\frac{k_z^2}{\alpha}\frac{\omega_{ci}^6}{\omega_{si}^2\omega_{Ai}^2},
\end{equation}
where we have used the definition $\omega_c^2\equiv \frac{\omega_s^2\omega_A^2}{\omega_s^2+\omega_A^2}$ in obtaining the second equality of the above relation. In the next, the dispersion relation (\ref{dispersionrelation}) in long wavelength limit (${k_zR}\ll 1$) reads
\begin{equation}\label{disp}
\rho_i \left(\Omega_{i}^2-\omega_{Ai}^2\right)-\frac{k_{ri}}{k_{re}}\rho_e \left(\Omega_{e}^2-\omega_{Ae}^2\right)\frac{xy\ln(y)}{2}=0.
\end{equation}
Now, replacing $k_{ri}^2$ from Eq. (\ref{kix}) into (\ref{disp}), the quantity $\alpha$ can be obtained as follows
\begin{equation}\label{ki}
\alpha=\frac{\chi}{2}\frac{\omega_{ci}^4}{\omega_{Ai}^4}\left(\Omega_{e}^2-\omega_{Ae}^2\right)k_z^2R^2 \ln(k_zR),
\end{equation}
where $\Omega_{e}=\omega_{ci}+k_z\left(v_{zi}-v_{ze}\right)$ and
\begin{equation}\label{kii}
k_{ri}^2=\frac{-2 \omega_{Ai}^2\omega_{ci}^2}{\chi\omega_{si}^2\left(\Omega_{e}^2-\omega_{Ae}^2\right) R^2 \ln(k_zR)},
\end{equation}
\begin{align}\label{Tm0}
T_0=&\Bigg(\frac{x^2\ln(y)\omega_{ci}^5\left(\Omega_{e}^2-\omega_{Ae}^2\right)}{(\omega_{si}^2-\omega_{ci}^2)(\omega_{Ai}^2-\omega_{ci}^2)\alpha}\nonumber\\
&-\frac{3x^4\ln(y)\omega_{ci}^5\left(\Omega_{e}^2-\omega_{Ae}^2\right)}{16(\omega_{si}^2-\omega_{ci}^2)(\omega_{Ai}^2-\omega_{ci}^2)\alpha}\nonumber\\
&+\frac{ x^2(\Omega_{e}^2-2\omega_{ce}^2)\Omega_{e}^3}{2(\omega_{se}^2-\Omega_{e}^2)(\omega_{Ae}^2-\Omega_{e}^2)(\Omega_{e}^2-\omega_{ce}^2) }\Bigg).
\end{align}
Now, replacing Eqs. (\ref{ki}) and (\ref{kii}) into Eq. (\ref{Tm0}) we obtain
\begin{align}
T_0=&\Bigg(-\frac{4 \omega_{Ai}^6}{\chi^2 \omega_{ci}\omega_{si}^2 \left(\Omega_{e}^2-\omega_{Ae}^2\right) k_z^2 R^2 }\nonumber\\
&-\frac{3 \omega_{Ai}^8 \omega_{ci} \ln(y)}{2\chi^3 \omega_{si}^4 \left(\Omega_{e}^2-\omega_{Ae}^2\right)^2 k_z^2 R^2 \ln^2(k_zR)}\nonumber\\
&-\frac{ \omega_{ci}^4 \omega_{Ai}^2(\Omega_{e}^2-2\omega_{ce}^2)\Omega_{e}^3}{\chi \omega_{si}^2(\omega_{se}^2-\Omega_{e}^2)\left(\Omega_{e}^2-\omega_{Ae}^2\right) \ln(k_zR) }\Bigg),
\end{align}
and Finally we reach
\begin{align}\label{Tm00}
T_0=-\frac{3 \omega_{Ai}^8 \omega_{ci}^2+8 \chi \omega_{si}^2 \omega_{Ai}^6 \left(\Omega_{e}^2-\omega_{Ae}^2\right)\ln(k_zR)}{2\chi^3 \omega_{ci} \omega_{si}^4 \left(\Omega_{e}^2-\omega_{Ae}^2\right)^2 k_z^2 R^2 \ln^2(k_zR)},
\end{align}
substituting Eq. (\ref{Tm00}) in (\ref{gammacwd}) we have
\begin{align}
\gamma_{0c}=-\frac{\frac{ \pi \rho_e k_z^2}{k_{re} }\frac{\Sign~\Omega}{\rho_c|\Delta_c|}\Big|_{r=r_c}\left(\frac{v_{s}^2}{v_{A}^2+v_{s}^2}\right)^2 (\omega_{ci}^2-\omega_{Ai}^2)(\Omega_e^2-\omega_{Ae}^2) k_zR}{ \chi \frac{3 \omega_{Ai}^8 \omega_{ci}^2+8 \chi \omega_{si}^2 \omega_{Ai}^6 \left(\Omega_{e}^2-\omega_{Ae}^2\right)\ln(k_zR)}{2\chi^3 \omega_{ci} \omega_{si}^4 \left(\Omega_{e}^2-\omega_{Ae}^2\right)^2 k_z^2 R^2 \ln^2(k_zR)}},
\end{align}
and after some algebra we get
\begin{align}\label{gammacwd211211}
\gamma_{0c}=\frac{ 2\pi \chi^3 \Sign~\Omega}{|\Delta_c| R}\left[\frac{ \omega_{ci}^7\omega_{si}^2\left(\Omega_{e}^2-\omega_{Ae}^2\right)^3 }{3\omega_{Ai}^{10} \omega_{ci}^2+8\chi \omega_{Ai}^8 \omega_{si}^2\left(\Omega_{e}^2-\omega_{Ae}^2\right)\ln(k_zR)}\right](k_zR)^4\ln^3(k_zR).
\end{align}
\section{Weak damping rate in long wavelength limit for the kink mode}\label{B}
For the kink mode $m=1$, we have
\begin{dmath}\label{limb2}
Q_1=\frac{I_{1}^'\left(x\right)K_{1}\left(y\right)}{I_{1}\left(x\right)K_{1}^'\left(y\right)}=-\frac{K_{1}\left(y\right)\left(I_{0}\left(x\right)+I_{2}\left(x\right)\right)}{I_{1}\left(x\right)\left(K_{0}\left(y\right)+K_{2}\left(y\right)\right)}\approx -\left(\frac{y}{x}+\frac{xy}{4}\right),
\end{dmath}
\begin{dmath}
G_1=\frac{K_{1}(y)}{K_{1}^'(y)}=\frac{-2K_{1}(y)}{K_{0}(y)+K_{2}(y)}\approx \frac{\frac{1}{y}-\frac{1}{4}+\frac{1}{2}\left(ln(y/2)+\gamma_e\right)}{-\frac{1}{y^2}+\frac{1}{4}+\frac{1}{2}\left(ln(y/2)+\gamma_e\right)}=-y,
\end{dmath}
\begin{dmath}
P_1=\left(\frac{I''_1(x)}{I_1(x)}-\frac{{I'_1(x)}^2}{I_1(x)^2}\right)\frac{K_1(y)}{K_1'(y)}\approx -y\left(\frac{1}{4}-\frac{1}{x^2}\right),
\end{dmath}
\begin{dmath}\label{limb12}
S_1=\left(1-\frac{K''_1(y) K_1(y)}{{K'_1(y)}^2}\right)\frac{I'_1(x)}{I_1(x)}\approx -\frac{1+\left(1+3\ln(y)\right)y^2}{x}.
\end{dmath}
Inserting Eqs. (\ref{limb2})-(\ref{limb12}) into (\ref{di11}) yields
\begin{align}
T_1=\left(\Omega_e^2-\omega_{Ae}^2\right)\frac{k_{ri}}{k_{re}}&\Bigg(\frac{\left(-\left(\frac{y}{x}+\frac{xy}{4}\right)-xy\left(\frac{1}{4}-\frac{1}{x^2}\right)\right)(\Omega_i^2-2\omega_{ci}^2)\Omega_i^3}{(\omega_{si}^2-\Omega_i^2)(\omega_{Ai}^2-\Omega_i^2)(\Omega_i^2-\omega_{ci}^2)}\nonumber\\
&-\frac{ \left(-\left(\frac{y}{x}+\frac{xy}{4}\right)+y\frac{1+\left(1+3\ln(y)\right)y^2}{x}\right)(\Omega_e^2-2\omega_{ce}^2)\Omega_e^3}{(\omega_{se}^2-\Omega_e^2)(\omega_{Ae}^2-\Omega_e^2)(\Omega_e^2-\omega_{ce}^2) }\Bigg).
\end{align}
For $\Omega^2_{i}=\omega_{ci}^2-\alpha$, we obtain
\begin{align}\label{Tm1}
T_1=\left(\Omega_e^2-\omega_{Ae}^2\right)&\Bigg(\frac{-x^2\omega_{ci}^5}{2(\omega_{si}^2-\Omega_i^2)(\omega_{Ai}^2-\Omega_i^2)\alpha}\nonumber\\
&-\frac{ \left(-\frac{x^2}{4}+\left(1+3\ln(y)\right)y^2\right)(\Omega_e^2-2\omega_{ce}^2)\Omega_e^3}{(\omega_{se}^2-\Omega_e^2)(\omega_{Ae}^2-\Omega_e^2)(\Omega_e^2-\omega_{ce}^2) }\Bigg).
\end{align}
In the next, the dispersion relation (\ref{dispersionrelation}) in long wavelength limit (${k_zR}\ll 1$) for $m=1$ reads
\begin{equation}\label{disp1}
\rho_i \left(\omega_{ci}^2-\omega_{Ai}^2\right)+\rho_e \left(\Omega_{e}^2-\omega_{Ae}^2\right)\left[1+\frac{x^2}{4}\right]=0,
\end{equation}
now, replacing $k_{ri}^2$ from Eq. (\ref{kix}) into (\ref{disp}), the quantity $\alpha$ can be obtained as follows
\begin{equation}\label{ki1}
\alpha=\frac{\chi}{4}\frac{\omega_{ci}^6\left(\Omega_{e}^2-\omega_{Ae}^2\right)}{\omega_{ci}^2\omega_{Ai}^4-\chi \omega_{si}^2 \omega_{Ai}^2 \left(\Omega_{e}^2-\omega_{Ae}^2\right)}k_z^2R^2,
\end{equation}
\begin{equation}\label{kii1}
k_{ri}^2=\frac{4}{\chi}\frac{ \omega_{ci}^2\omega_{Ai}^2-\chi \omega_{si}^2 \left(\Omega_{e}^2-\omega_{Ae}^2\right) }{\omega_{si}^2 \left(\Omega_{e}^2-\omega_{Ae}^2\right) R^2},
\end{equation}
putting Eqs. (\ref{ki1}) and (\ref{kii1}) in (\ref{Tm1}) and keeping only the sentence proportional to the sentence $\frac{1}{k_z^2R^2}$ we  obtain
\begin{align}
T_1=\frac{8 \omega_{si}^2 \omega_{Ai}^2 \left(\Omega_{e}^2-\omega_{Ae}^2\right)}{\omega_{ci}^5 k_z^2 R^2}\left(\frac{\omega_{ci}^2 \omega_{Ai}^2}{\chi \omega_{si}^2 \left(\Omega_{e}^2-\omega_{Ae}^2\right)}-1\right)^2.
\end{align}
In the following with the help of Eq. (\ref{gammacwd}), we get
\begin{align}
\gamma_{1c}=-\frac{\frac{ \pi \rho_e k_z^2}{k_{re} }\frac{\Sign~\Omega}{\rho_c|\Delta_c|}\Big|_{r=r_c}\left(\frac{v_{s}^2}{v_{A}^2+v_{s}^2}\right)^2 (\omega_{ci}^2-\omega_{Ai}^2)(\Omega_e^2-\omega_{Ae}^2) k_zR}{ \chi \frac{8 \omega_{si}^2 \omega_{Ai}^2 \left(\Omega_{e}^2-\omega_{Ae}^2\right)}{\omega_{ci}^5 k_z^2 R^2}\left(\frac{\omega_{ci}^2 \omega_{Ai}^2}{\chi \omega_{si}^2 \left(\Omega_{e}^2-\omega_{Ae}^2\right)}-1\right)^2},
\end{align}
this can be simplified as
\begin{align}
\gamma_{1c}=-\frac{ \pi \chi^2 \Sign~\Omega}{8|\Delta_c| R}\frac{ \omega_{ci}^{11}\left(\Omega_e^2-\omega_{Ae}^2\right)^2}{\omega_{Ai}^{4}\left(\omega_{ci}^2\omega_{Ai}^2 -\chi\omega_{si}^2 \left(\Omega_e^2-\omega_{Ae}^2\right)\right)^2}(k_zR)^4.
\end{align}
\end{appendices}

\newpage
\begin{figure}
	\begin{subfigure}[b]{0.45\textwidth}
		\centering
		\includegraphics[width=\textwidth]{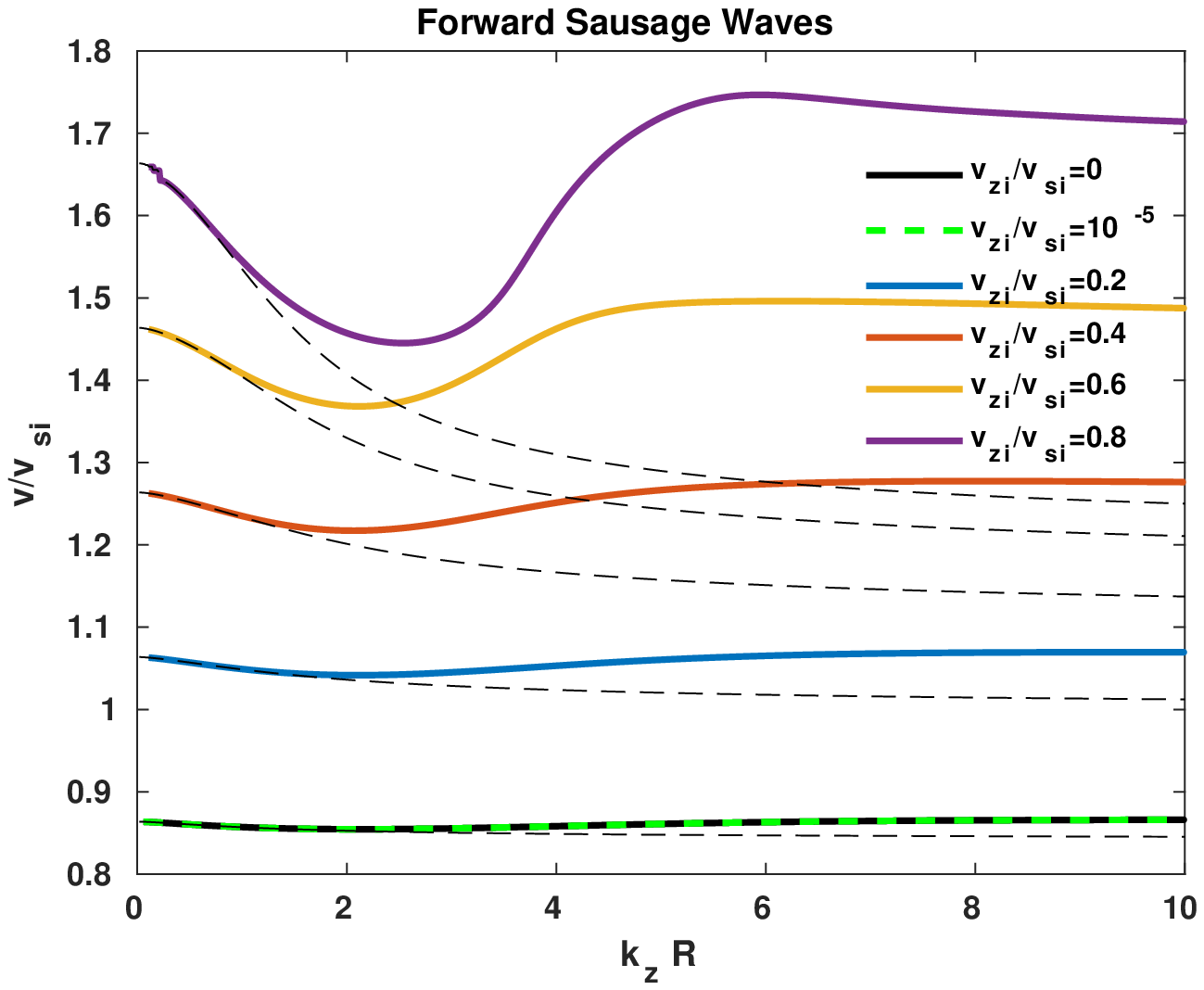}
		\caption{}
		\label{5sl0.1a}
	\end{subfigure}
\hfill
\begin{subfigure}[b]{0.45\textwidth}
	\centering
	\includegraphics[width=\textwidth]{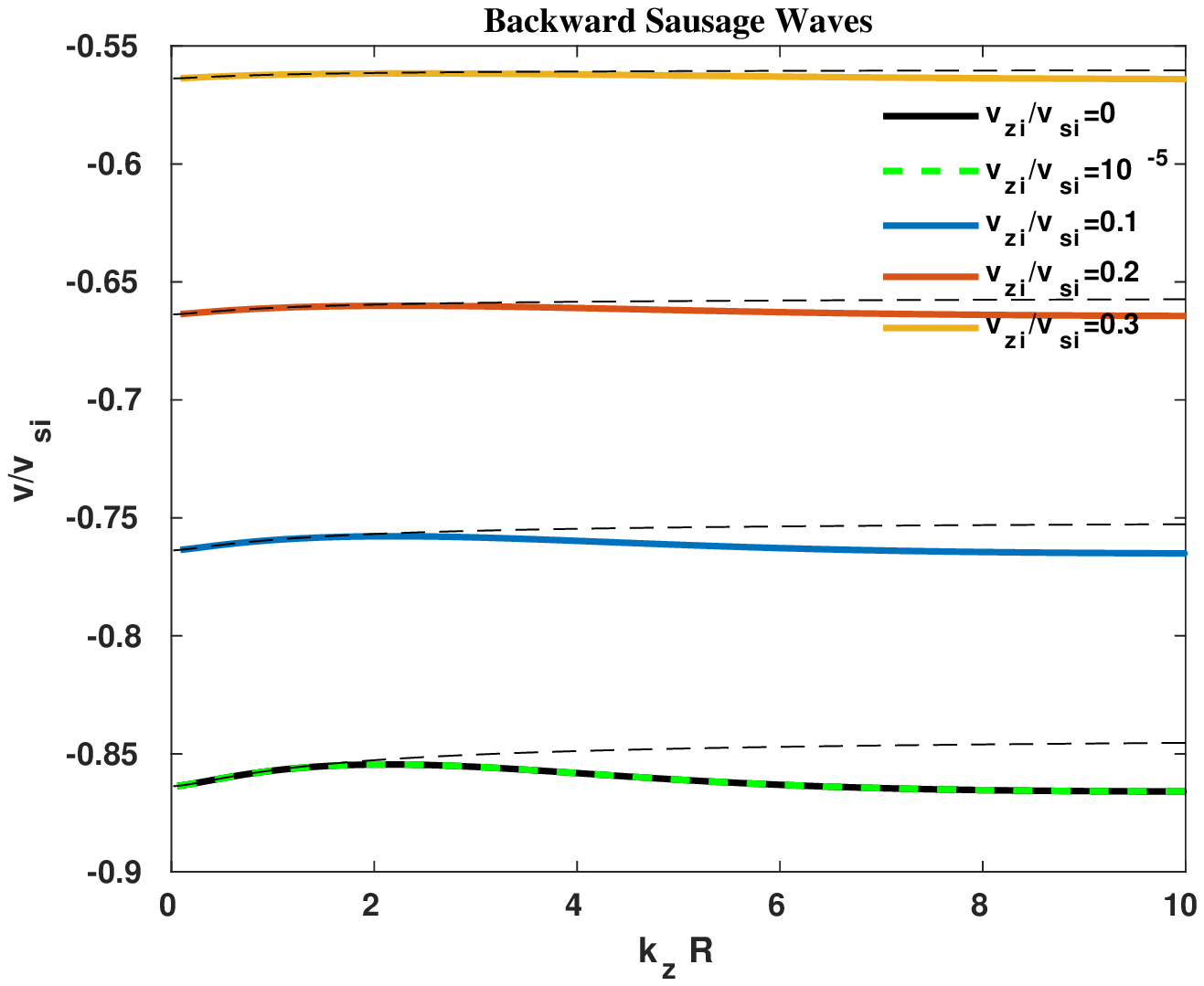}
	\caption{}
	\label{5sl0.1b}
\end{subfigure}
	\vfill
	\begin{subfigure}[b]{0.45\textwidth}
		\centering
		\includegraphics[width=\textwidth]{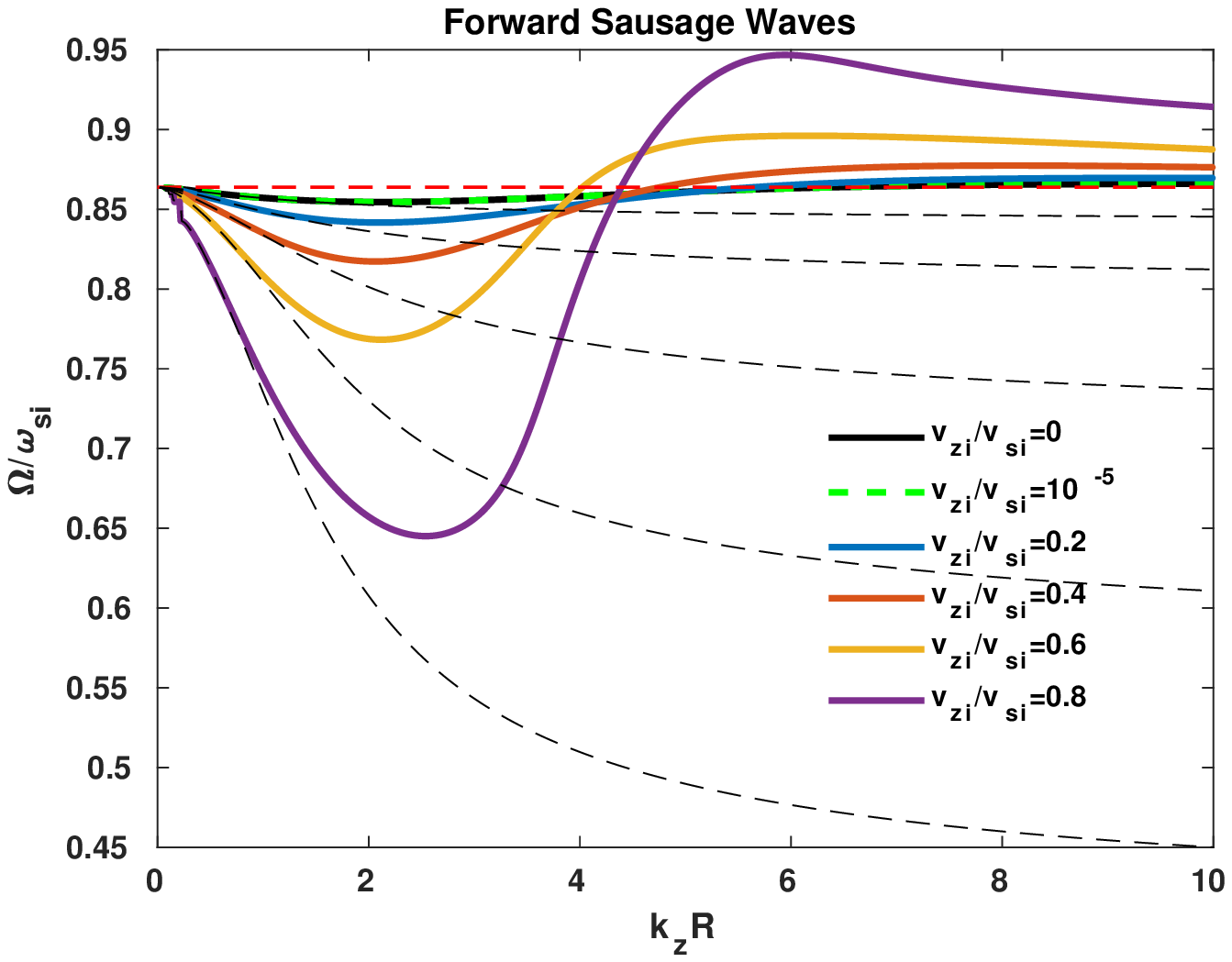}
		\caption{}
		\label{5sl0.1c}
	\end{subfigure}
	\hfill
	\begin{subfigure}[b]{0.45\textwidth}
		\centering
		\includegraphics[width=\textwidth]{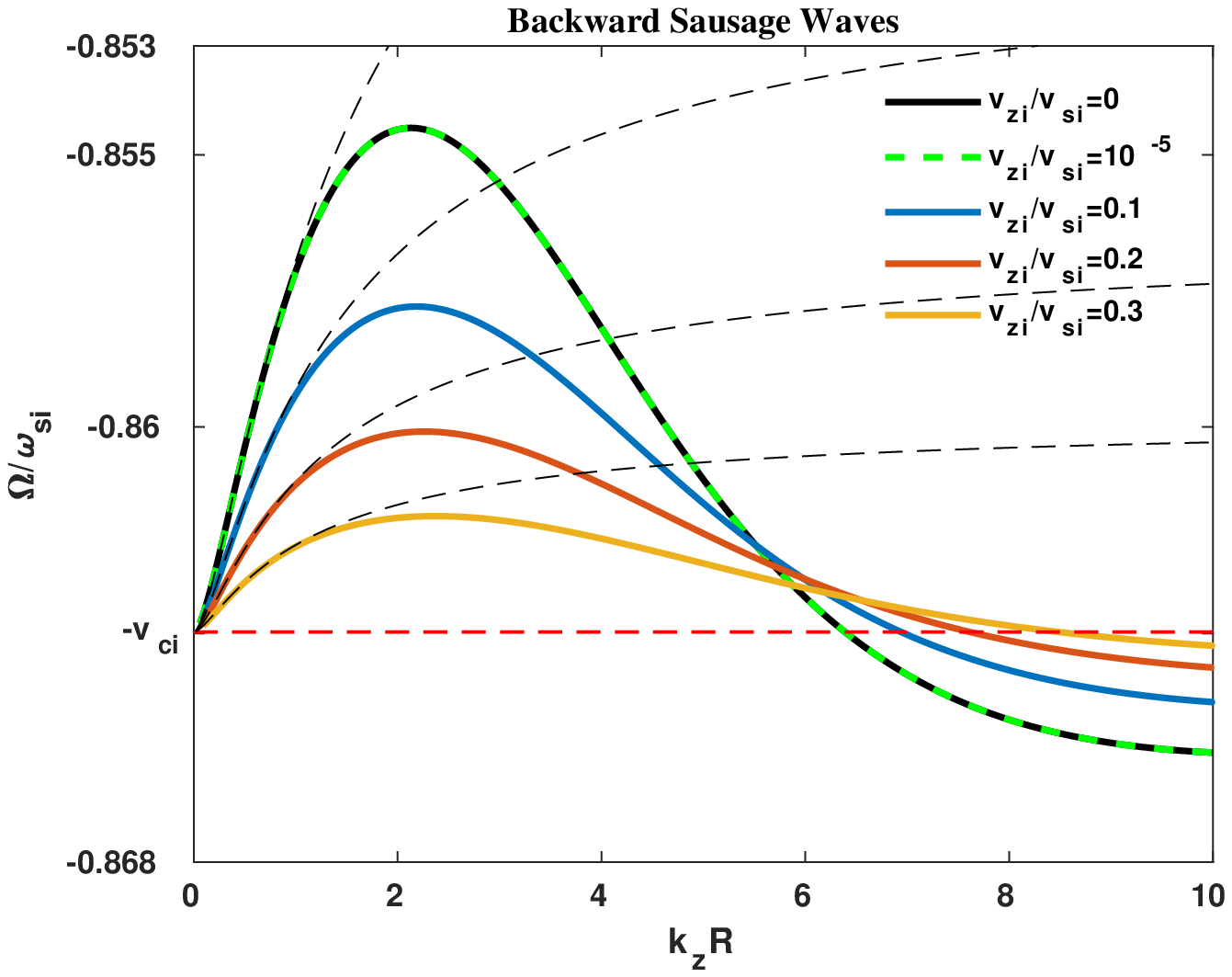}
		\caption{}
		\label{5sl0.1d}
	\end{subfigure}
	\vfill
	\begin{subfigure}[b]{0.45\textwidth}
		\centering
		\includegraphics[width=\textwidth]{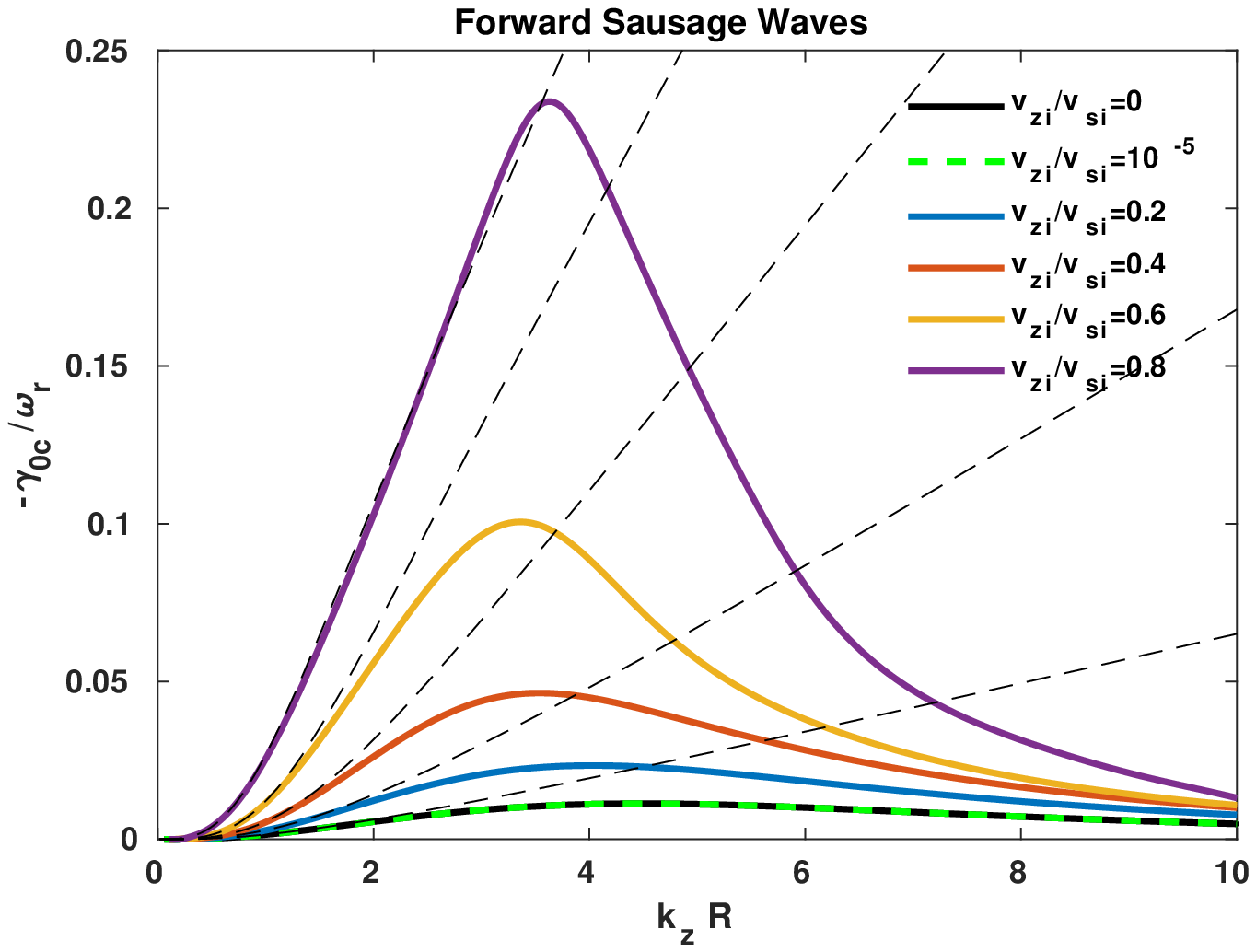}
		\caption{}
		\label{5sl0.1e}
	\end{subfigure}
	\hfill
	\begin{subfigure}[b]{0.45\textwidth}
		\centering
		\includegraphics[width=\textwidth]{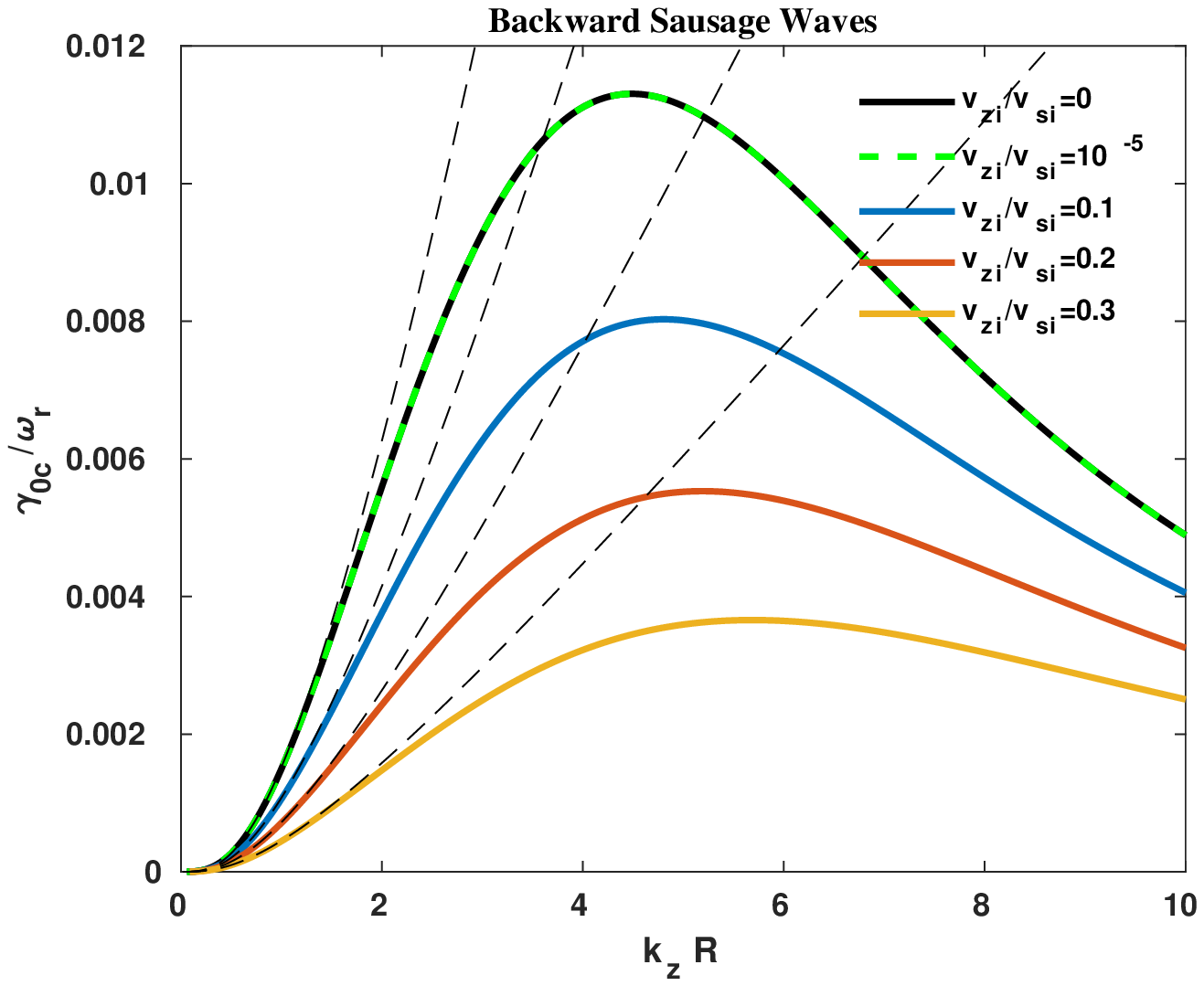}
		\caption{}
		\label{5sl0.1f}
	\end{subfigure}
	\caption{The left panels are for the forward sausage waves and the diagrams in (a), (b) and (c) represent the phase speed $v/v_{si}\equiv\omega_r/\omega_{si}$, the Doppler Shifted phase speed $\Omega/\omega_{si}$ and the damping rate $-\gamma_{0c}/\omega_r$ as functions of $k_zR$ for various values of plasma flow. The right panels are the same as the left panels for the backward sausage waves. For the damping rate the dashed curves represent the analytical solutions determined from Eq. (\ref{gammacwd}). The dashed curves in the other diagrams show the results obtained in the case of no boundary layer i.e. Eq. (\ref{dispersionrelation}). Other parameters of the tube are $l/R=0.1$, $v_{Ai}=12 ~{\rm km~s}^{-1}$, $v_{Ae}=0 ~{\rm km~s}^{-1}$ (i.e. $B_{ze}=0$), \textbf{$v_{ze}=0 ~{\rm km~s}^{-1}$}, $v_{si}=7 ~{\rm km~s}^{-1}$, $v_{se}=11.5 ~{\rm km~s}^{-1}$, $v_{ci}=6.0464~{\rm km~s}^{-1}(\simeq 0.8638~ v_{si})$ and $v_{ce}=0 ~{\rm km~s}^{-1}$.}
	\label{5sl0.1}
\end{figure}
\begin{figure}
	\begin{subfigure}[b]{0.45\textwidth}
		\centering
		\includegraphics[width=\textwidth]{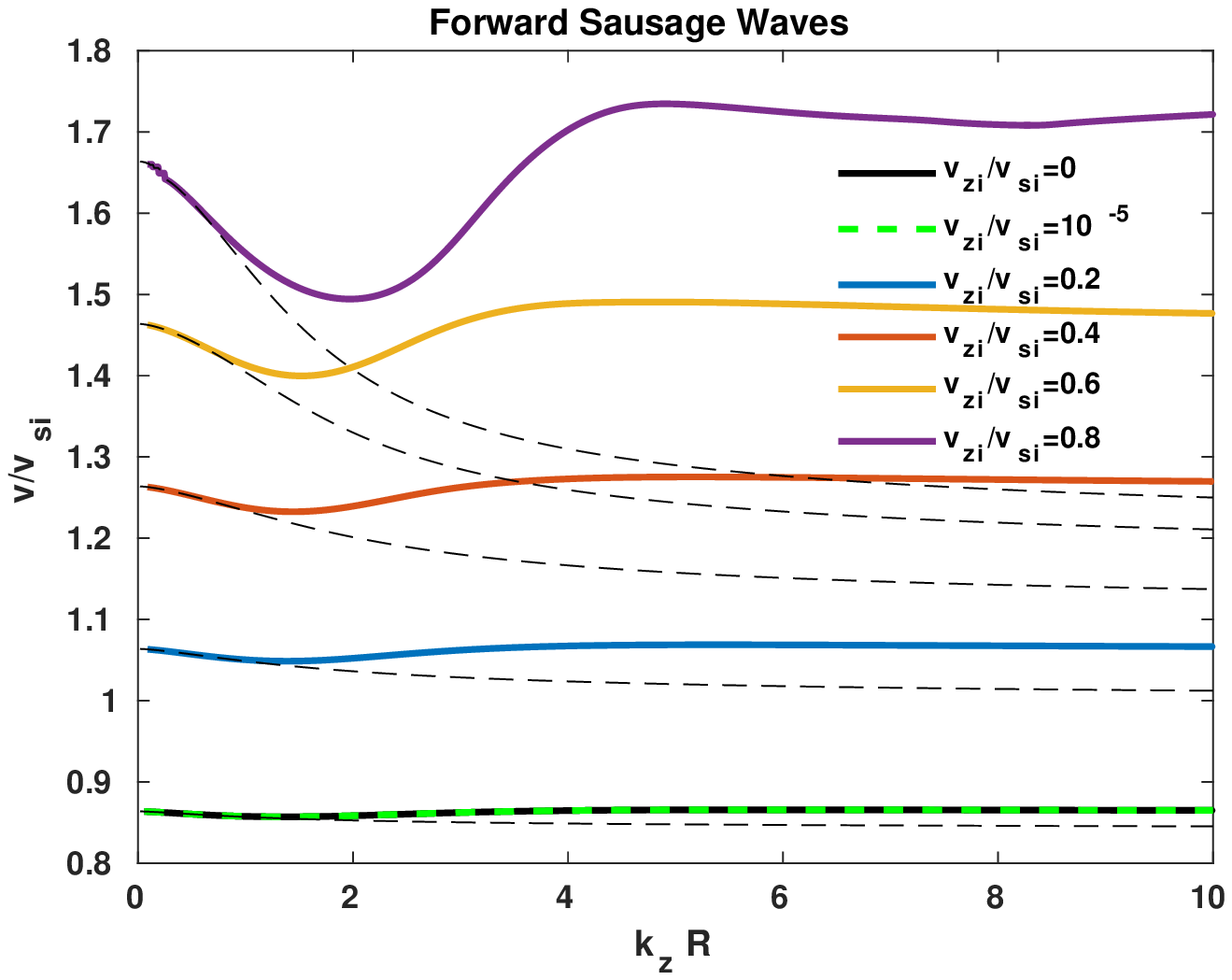}
		\caption{}
		\label{5sl0.4a}
	\end{subfigure}
\hfill
\begin{subfigure}[b]{0.45\textwidth}
	\centering
	\includegraphics[width=\textwidth]{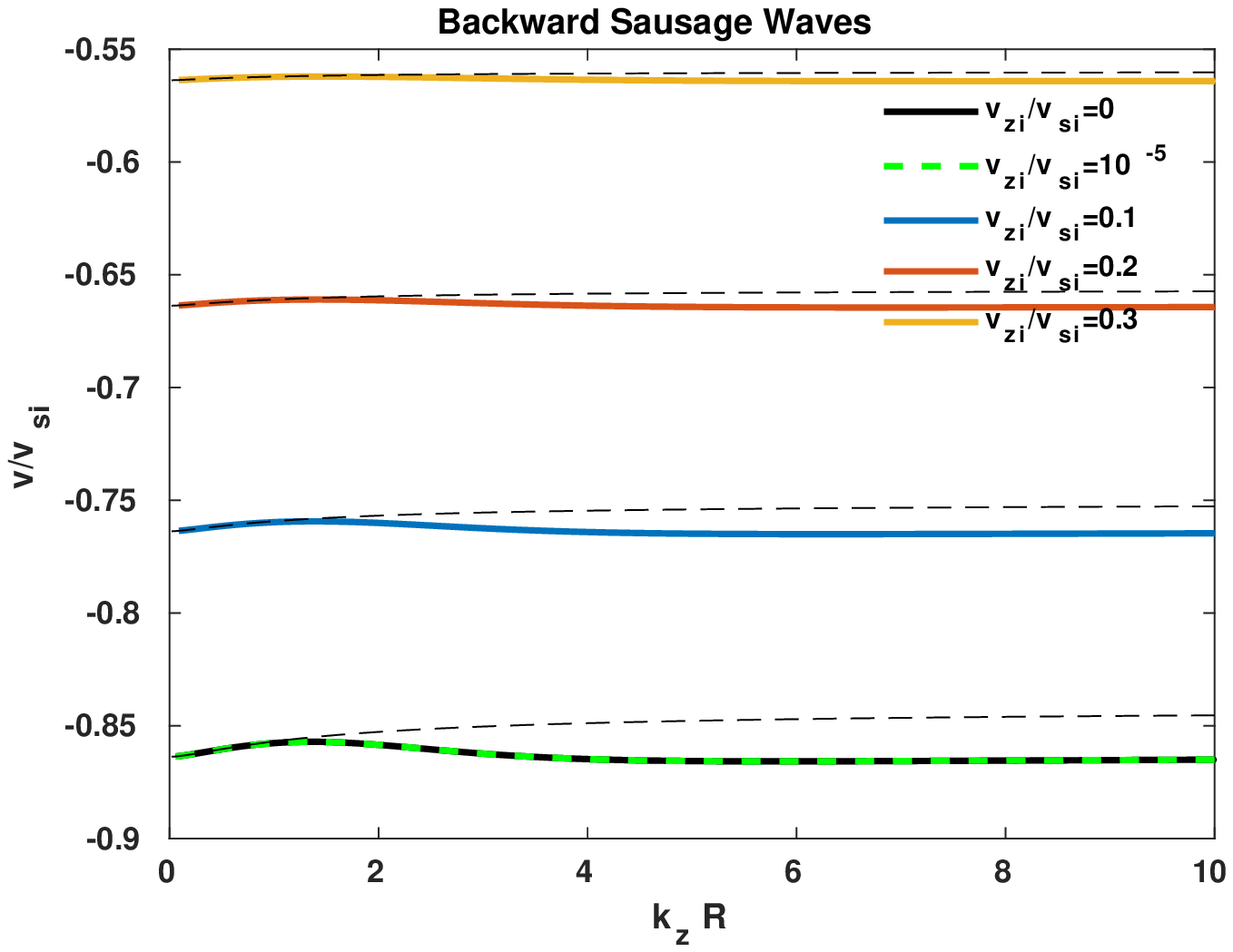}
	\caption{}
	\label{5sl0.4b}
\end{subfigure}
	\vfill
	\begin{subfigure}[b]{0.45\textwidth}
		\centering
		\includegraphics[width=\textwidth]{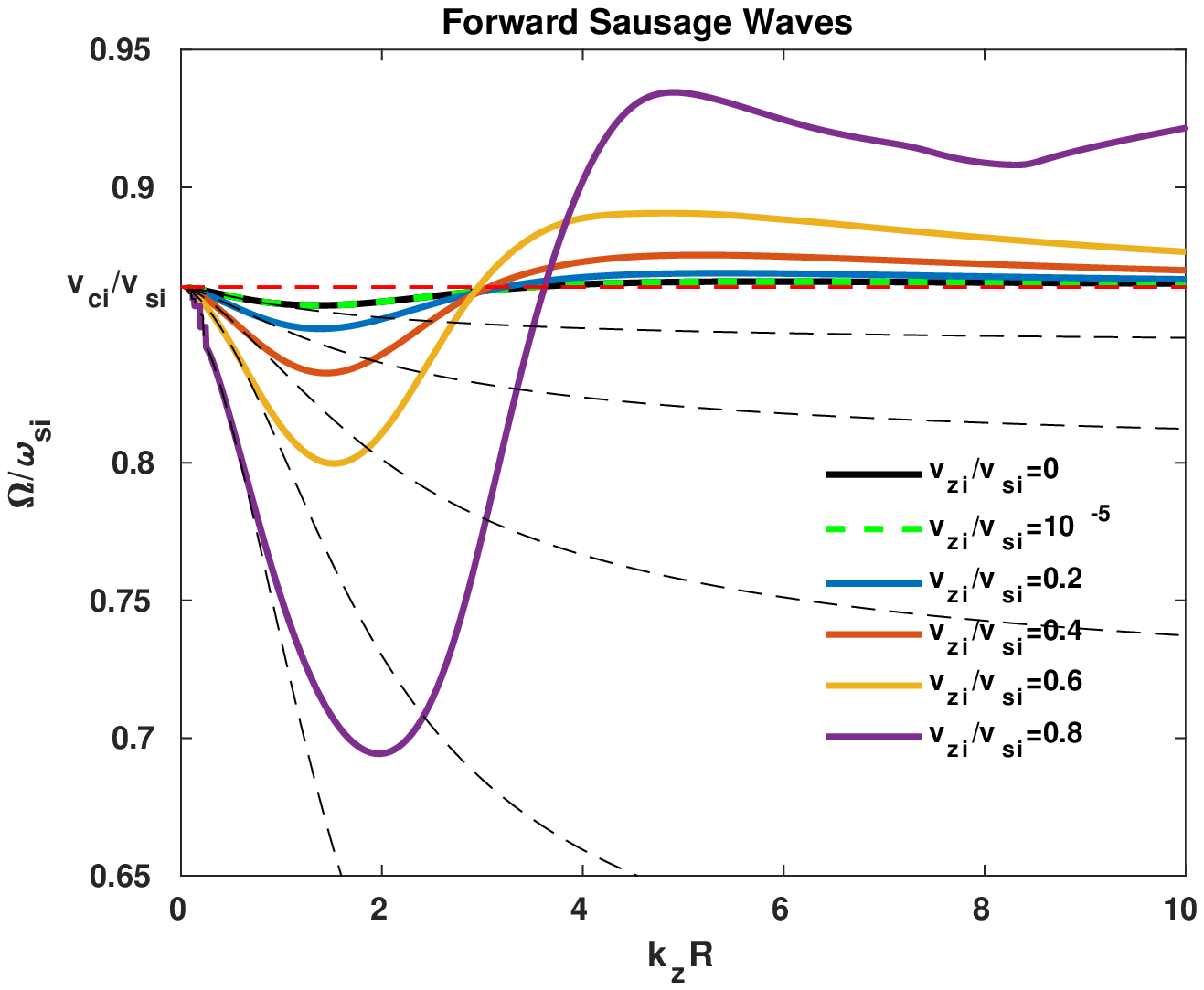}
		\caption{}
		\label{5sl0.4c}
	\end{subfigure}
	\hfill
	\begin{subfigure}[b]{0.45\textwidth}
		\centering
		\includegraphics[width=\textwidth]{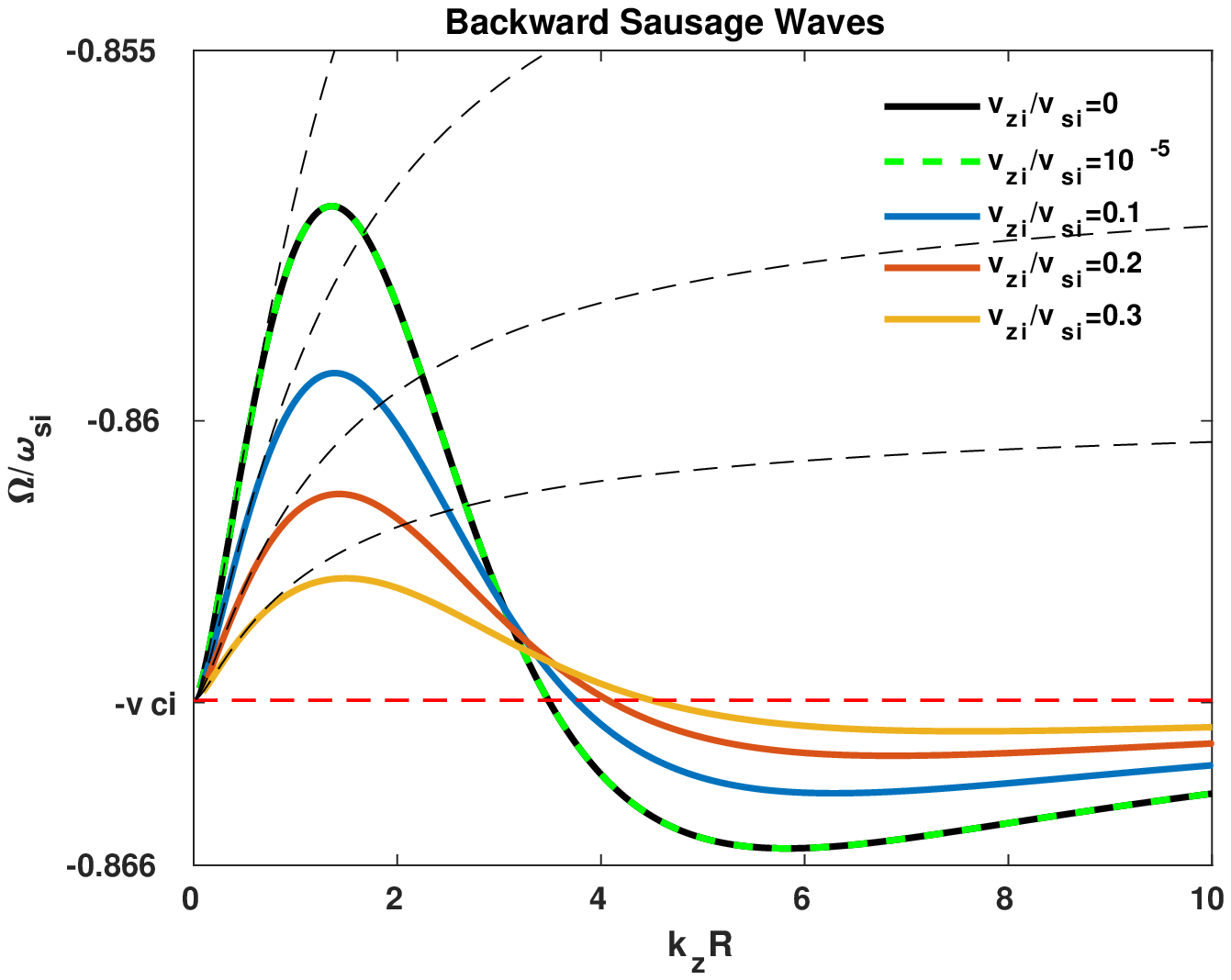}
		\caption{}
		\label{5sl0.4d}
	\end{subfigure}
	\vfill
	\begin{subfigure}[b]{0.45\textwidth}
		\centering
		\includegraphics[width=\textwidth]{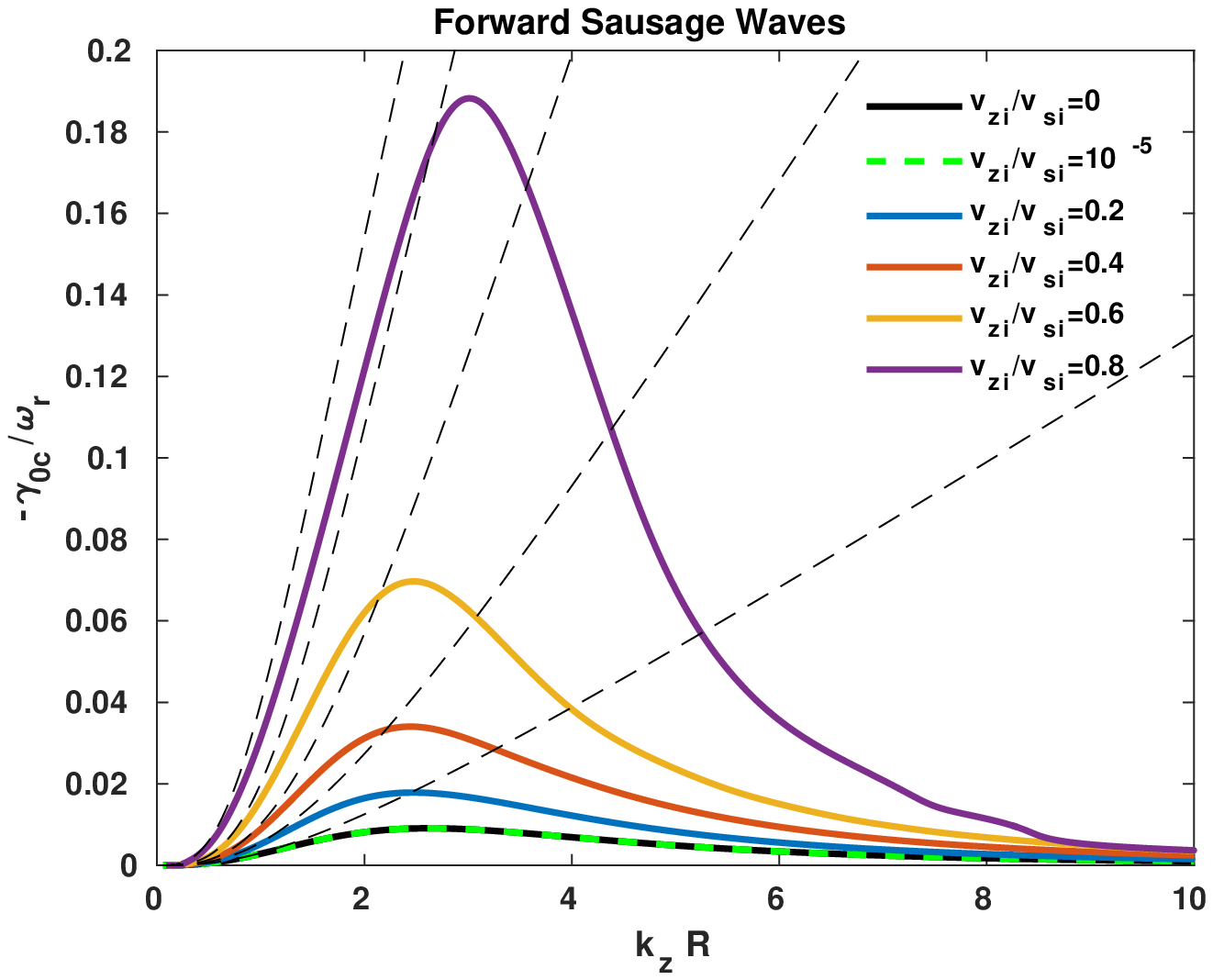}
		\caption{}
		\label{5sl0.4e}
	\end{subfigure}
	\hfill
	\begin{subfigure}[b]{0.45\textwidth}
		\centering
		\includegraphics[width=\textwidth]{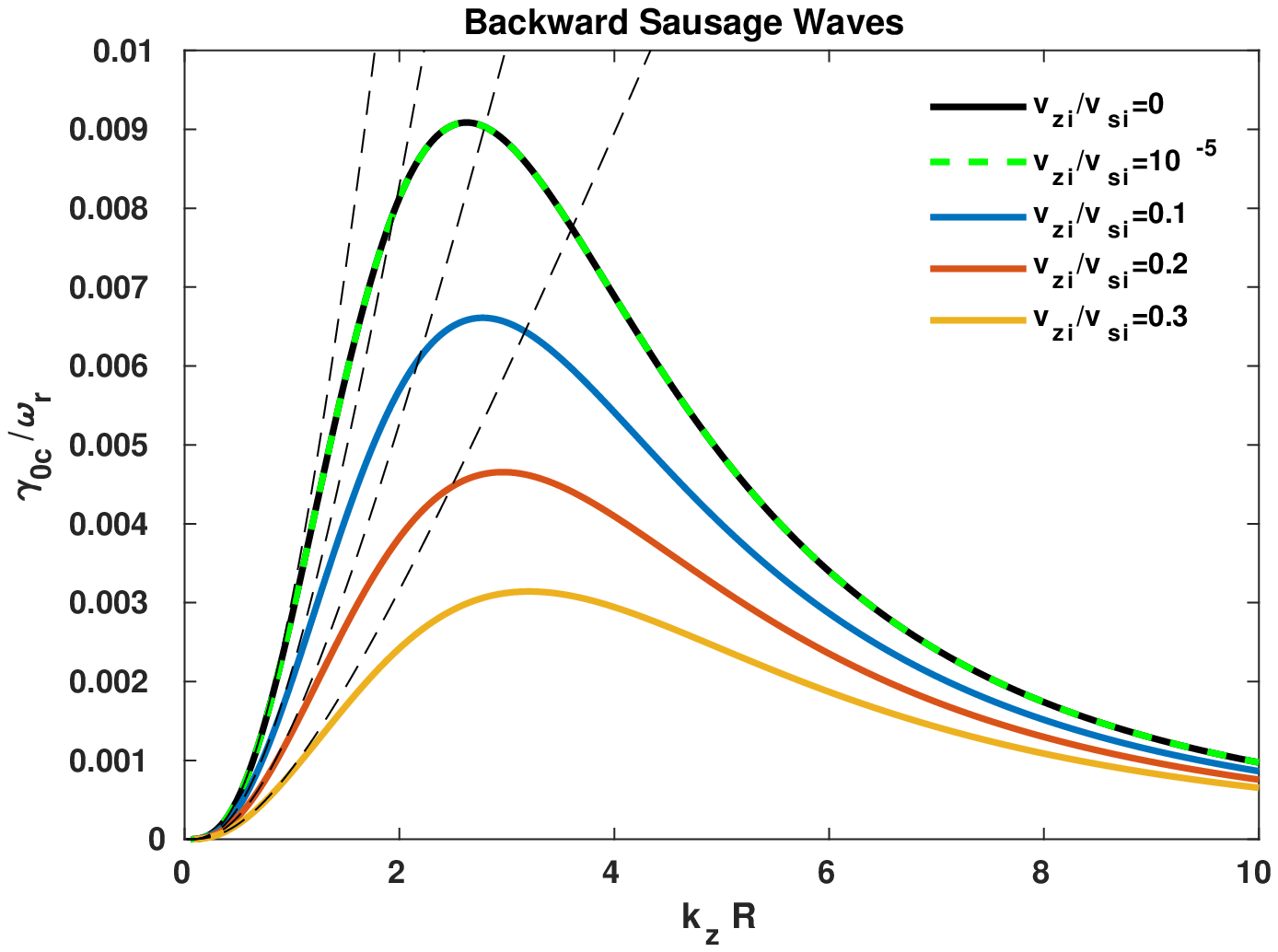}
		\caption{}
		\label{5sl0.4f}
	\end{subfigure}
	\caption{Same as Fig. \ref{5sl0.1}  , but for $l/R=0.2$.}
	\label{5sl0.4}
\end{figure}
\begin{figure}
	\begin{subfigure}[b]{0.45\textwidth}
		\centering
		\includegraphics[width=\textwidth]{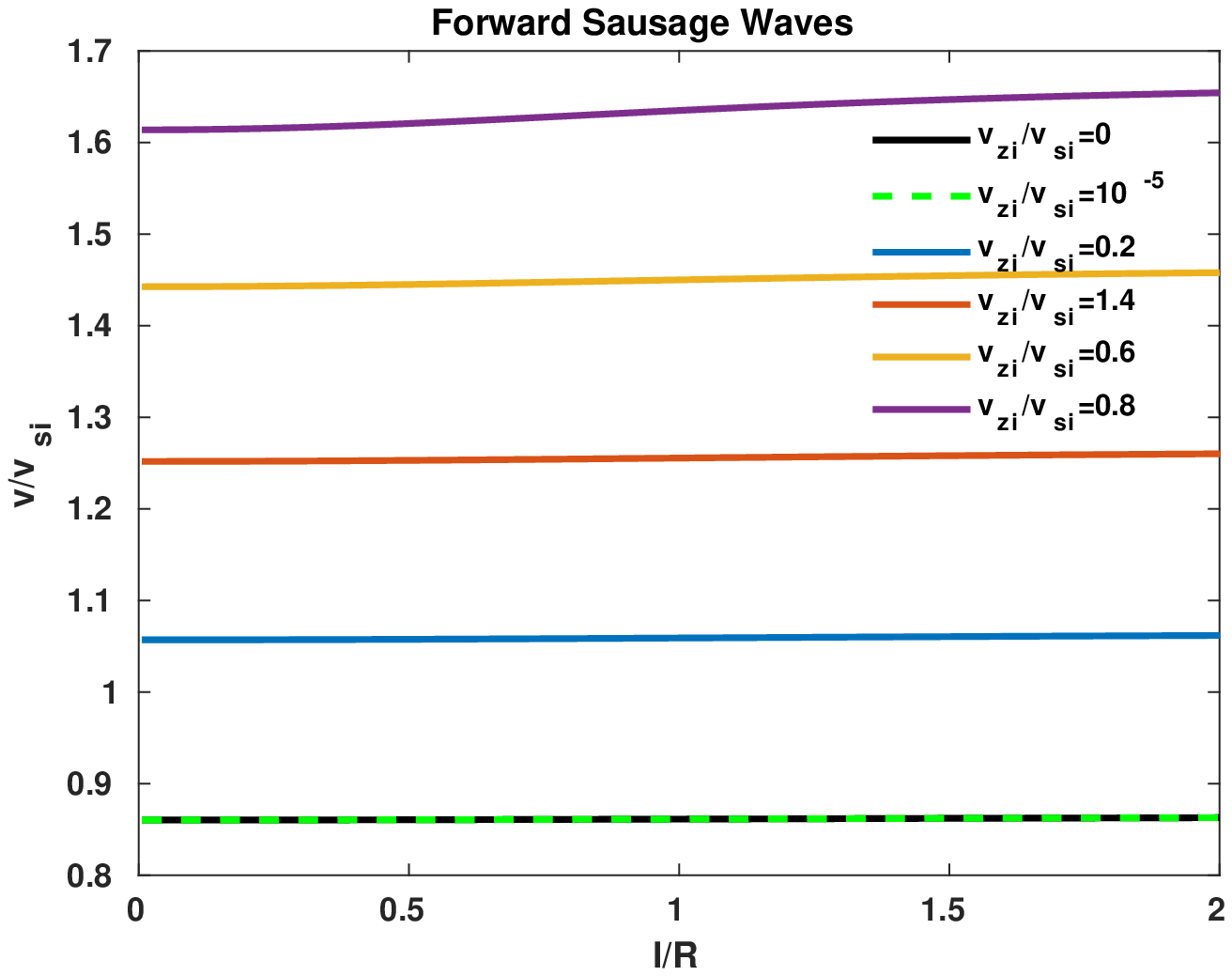}
		\caption{}
		\label{5skz0.5a}
	\end{subfigure}
\hfill
\begin{subfigure}[b]{0.45\textwidth}
	\centering
	\includegraphics[width=\textwidth]{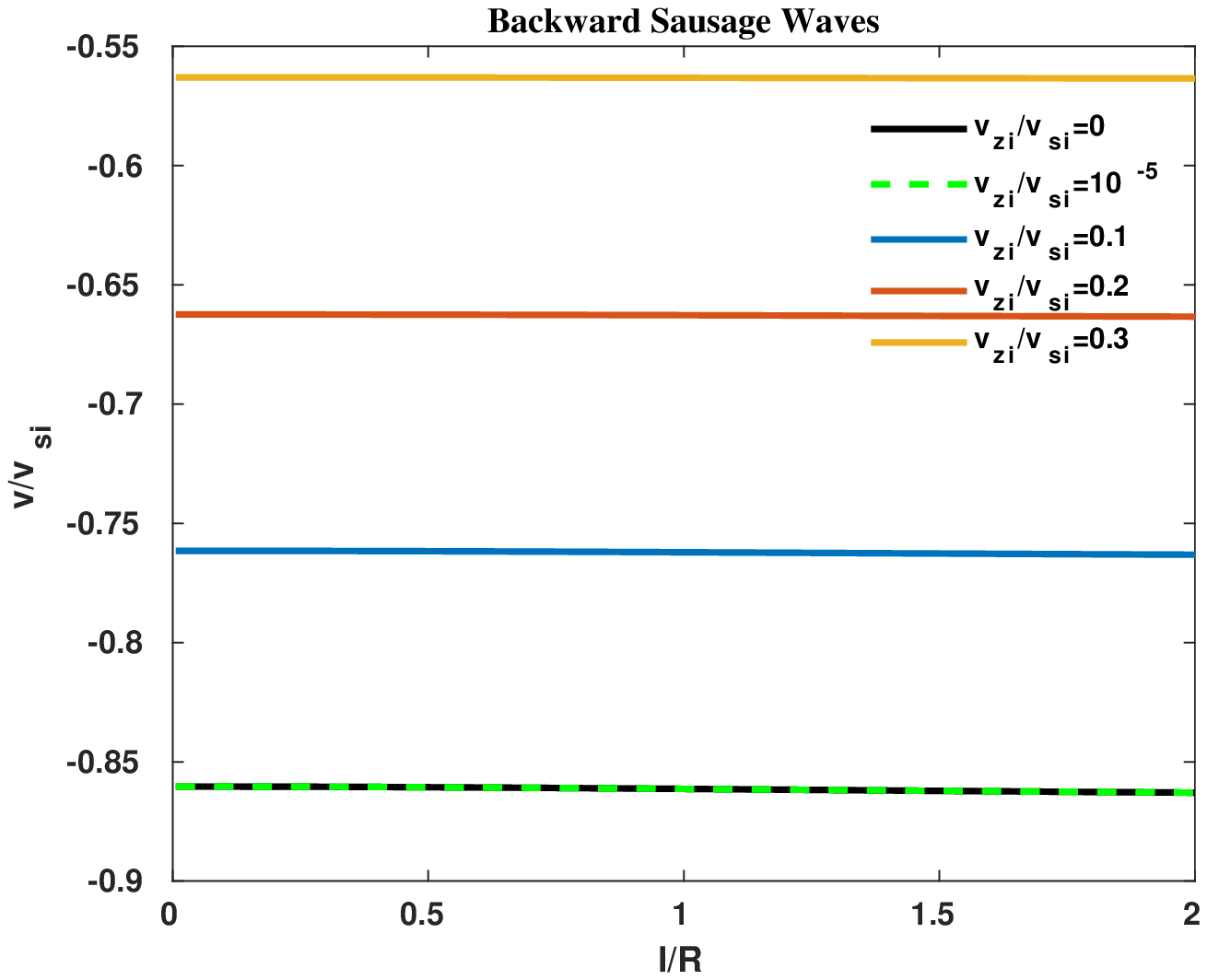}
	\caption{}
	\label{5skz0.5b}
\end{subfigure}
	\vfill
	\begin{subfigure}[b]{0.45\textwidth}
		\centering
		\includegraphics[width=\textwidth]{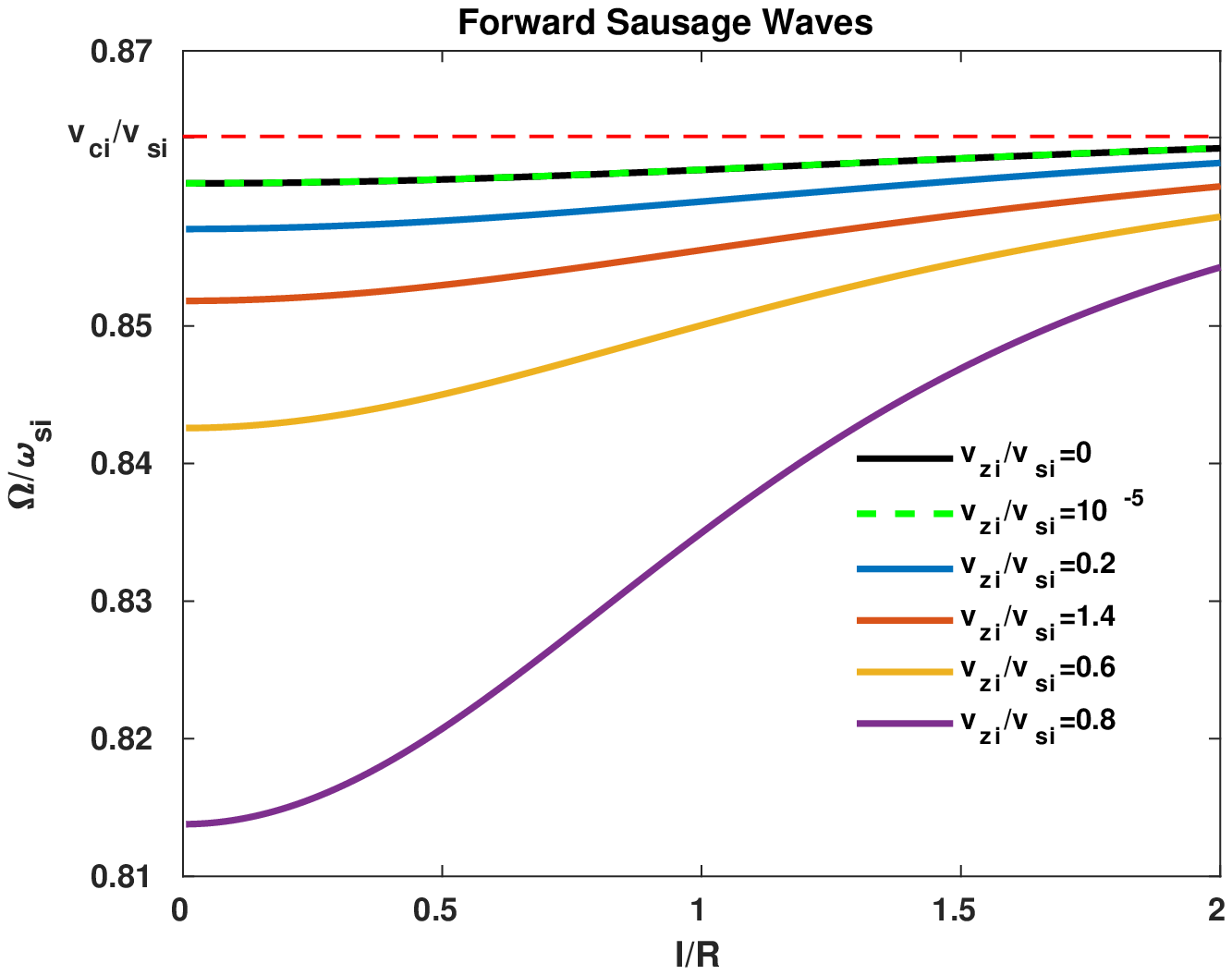}
		\caption{}
		\label{5skz0.5c}
	\end{subfigure}
	\hfill
	\begin{subfigure}[b]{0.45\textwidth}
		\centering
		\includegraphics[width=\textwidth]{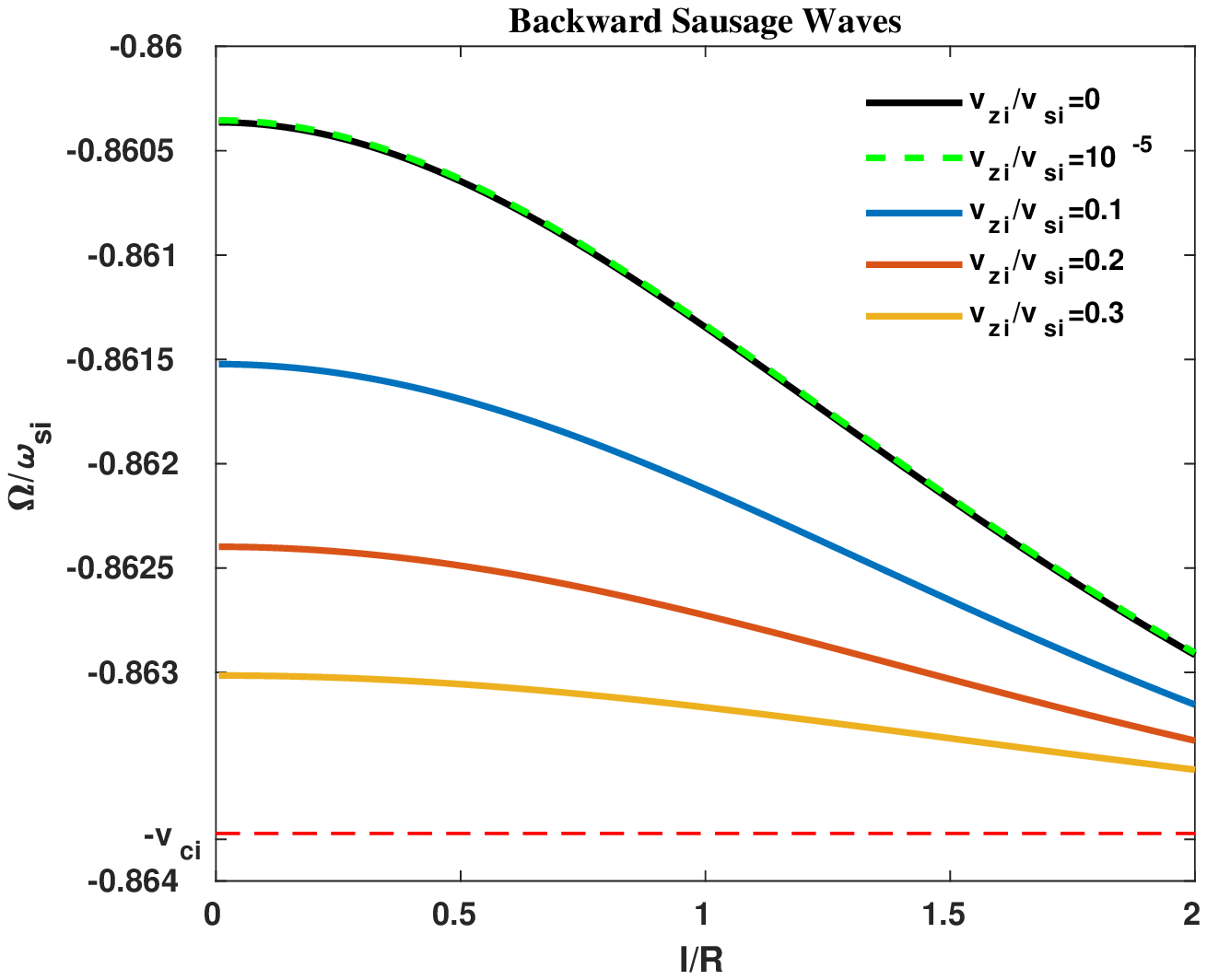}
		\caption{}
		\label{5skz0.5d}
	\end{subfigure}
	\vfill
	\begin{subfigure}[b]{0.45\textwidth}
		\centering
		\includegraphics[width=\textwidth]{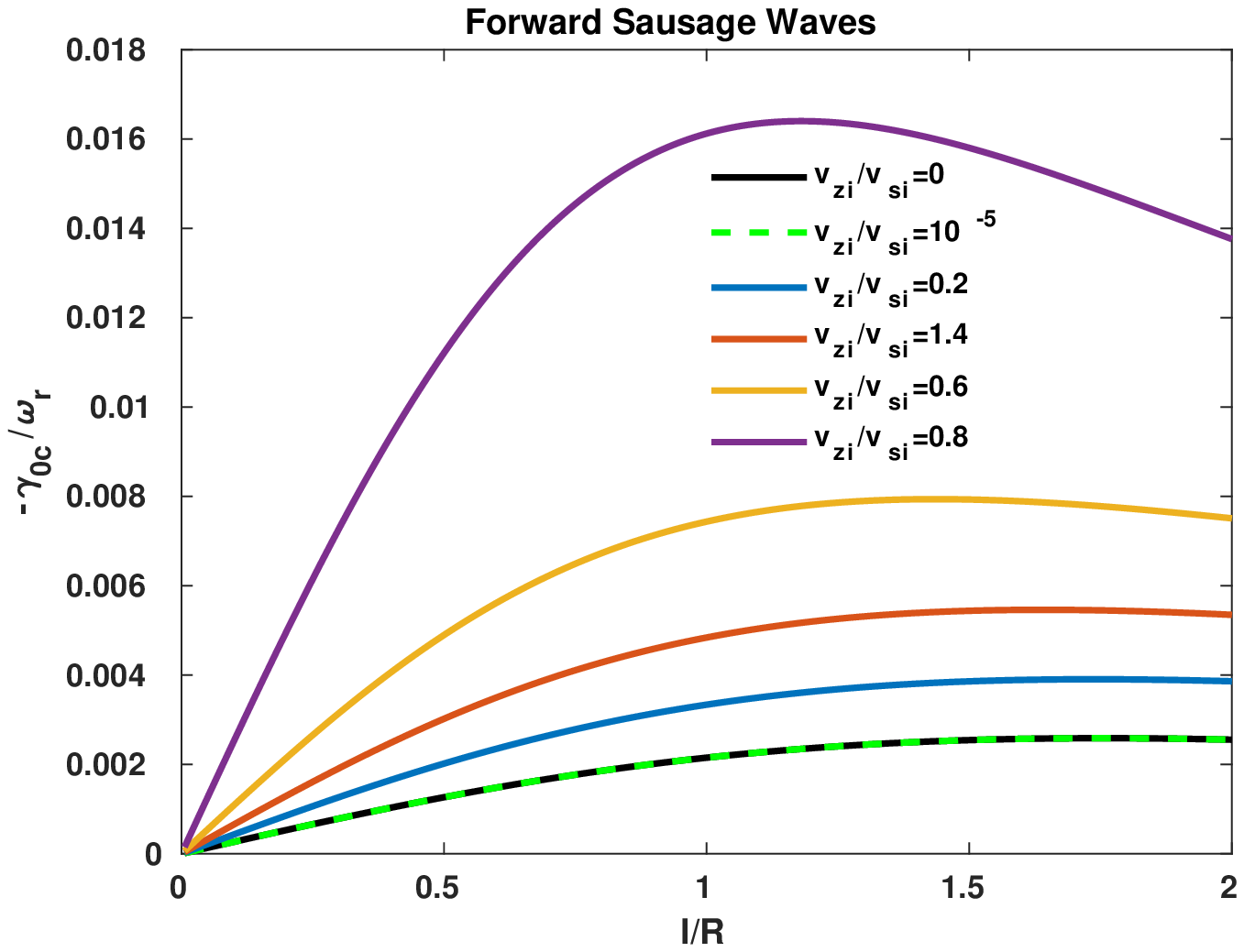}
		\caption{}
		\label{5skz0.5f}
	\end{subfigure}
	\hfill
	\begin{subfigure}[b]{0.45\textwidth}
		\centering
		\includegraphics[width=\textwidth]{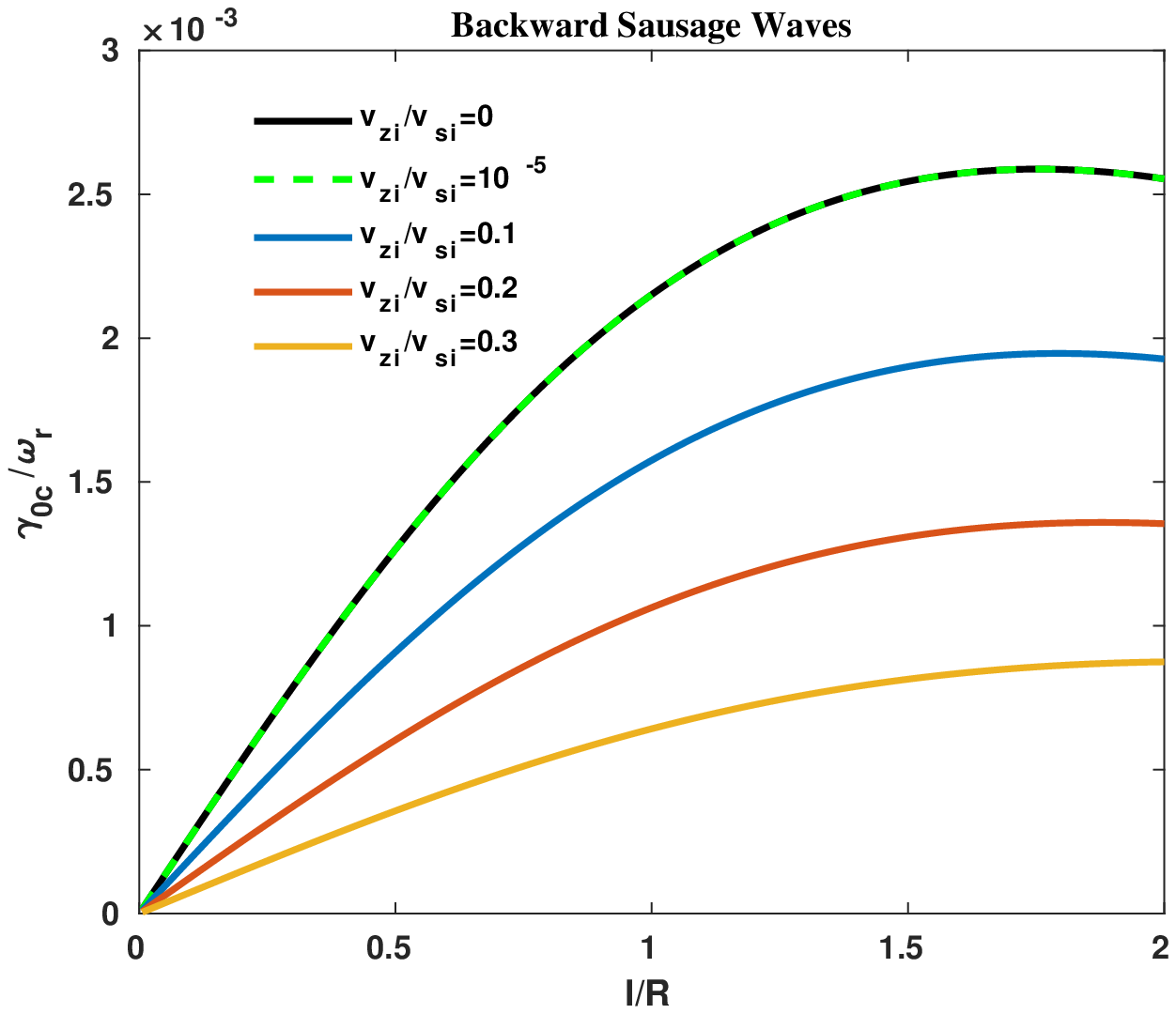}
		\caption{}
		\label{5skz0.5e}
	\end{subfigure}
	\caption{The left panels are for the forward sausage waves and the diagrams in (a), (b) and (c) represent the phase speed $v/v_{si}\equiv\omega_r/\omega_{si}$, the Doppler Shifted phase speed $\Omega/\omega_{si}$ and the damping rate $-\gamma_{0c}/\omega_r$ as functions of $l/R$ for various values of plasma flow. The right panels are the same as the left panels for the backward sausage waves. For all panels we have assumed $k_zR=0.5$, other parameters are the same as Fig. \ref{5sl0.1}.}
	\label{5skz0.5}
\end{figure}
\begin{figure}
	\begin{subfigure}[b]{0.45\textwidth}
		\centering
		\includegraphics[width=\textwidth]{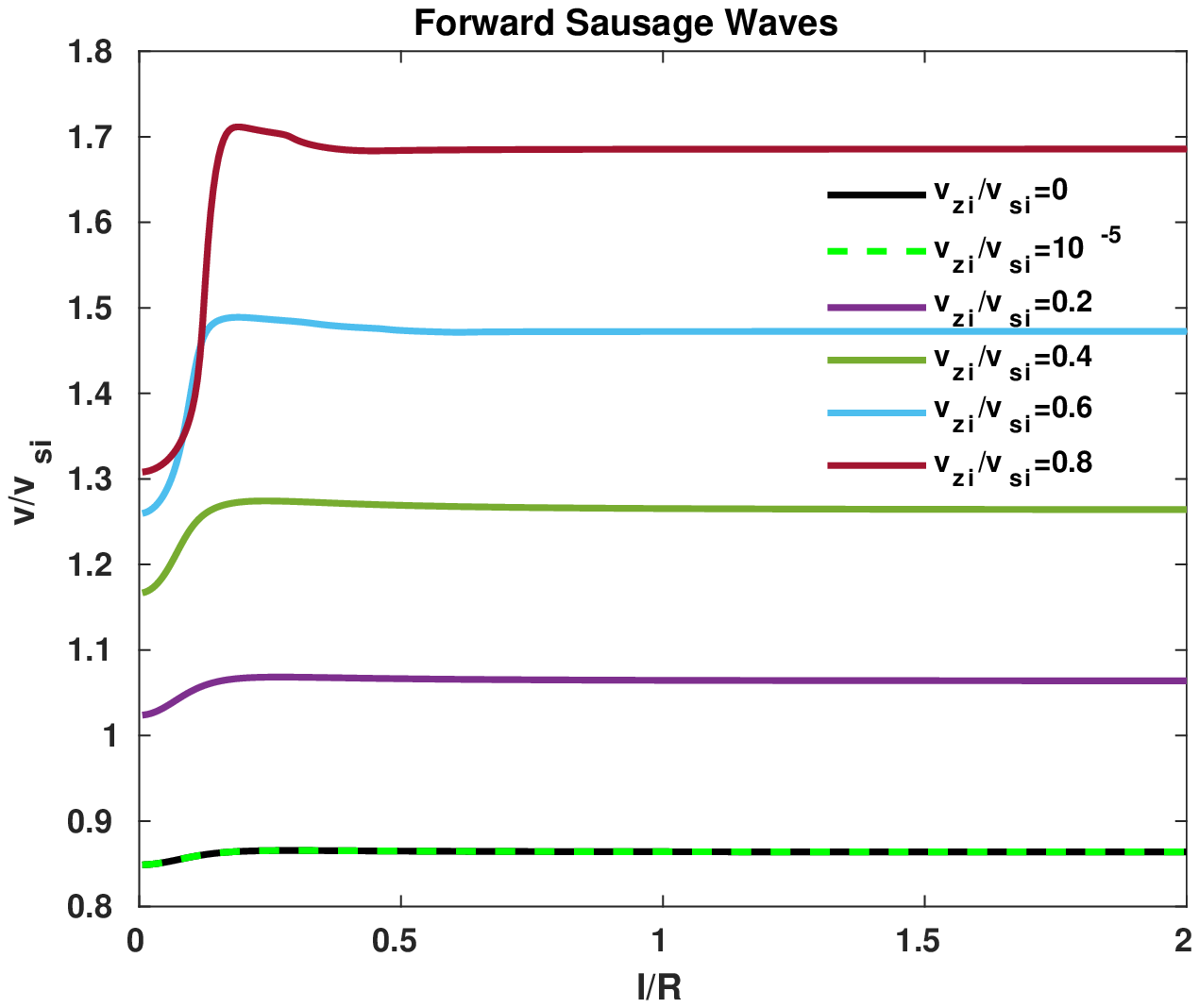}
		\caption{}
		\label{5skz4a}
	\end{subfigure}
\hfill
\begin{subfigure}[b]{0.45\textwidth}
	\centering
	\includegraphics[width=\textwidth]{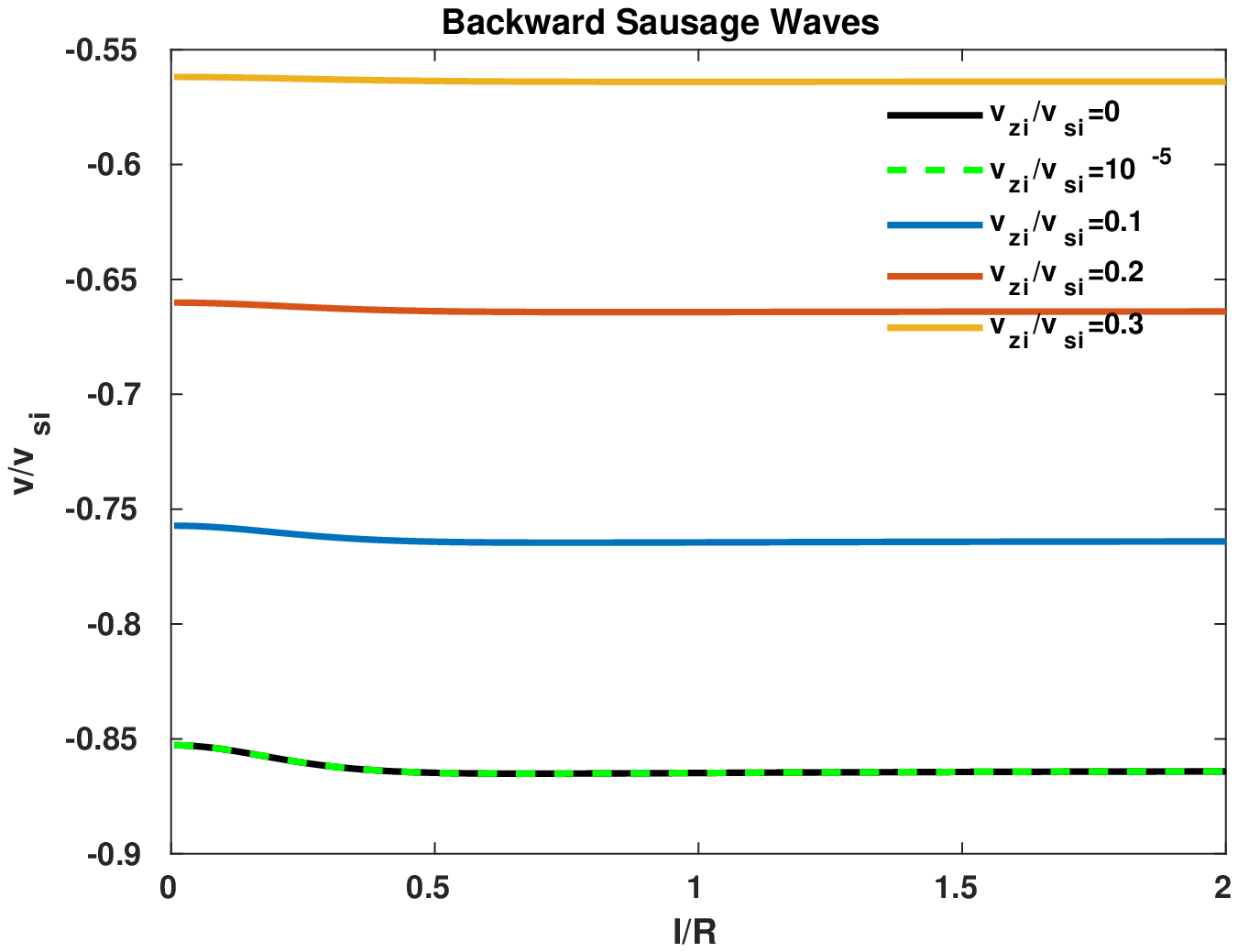}
	\caption{}
	\label{5skz4b}
\end{subfigure}
	\vfill
	\begin{subfigure}[b]{0.45\textwidth}
		\centering
		\includegraphics[width=\textwidth]{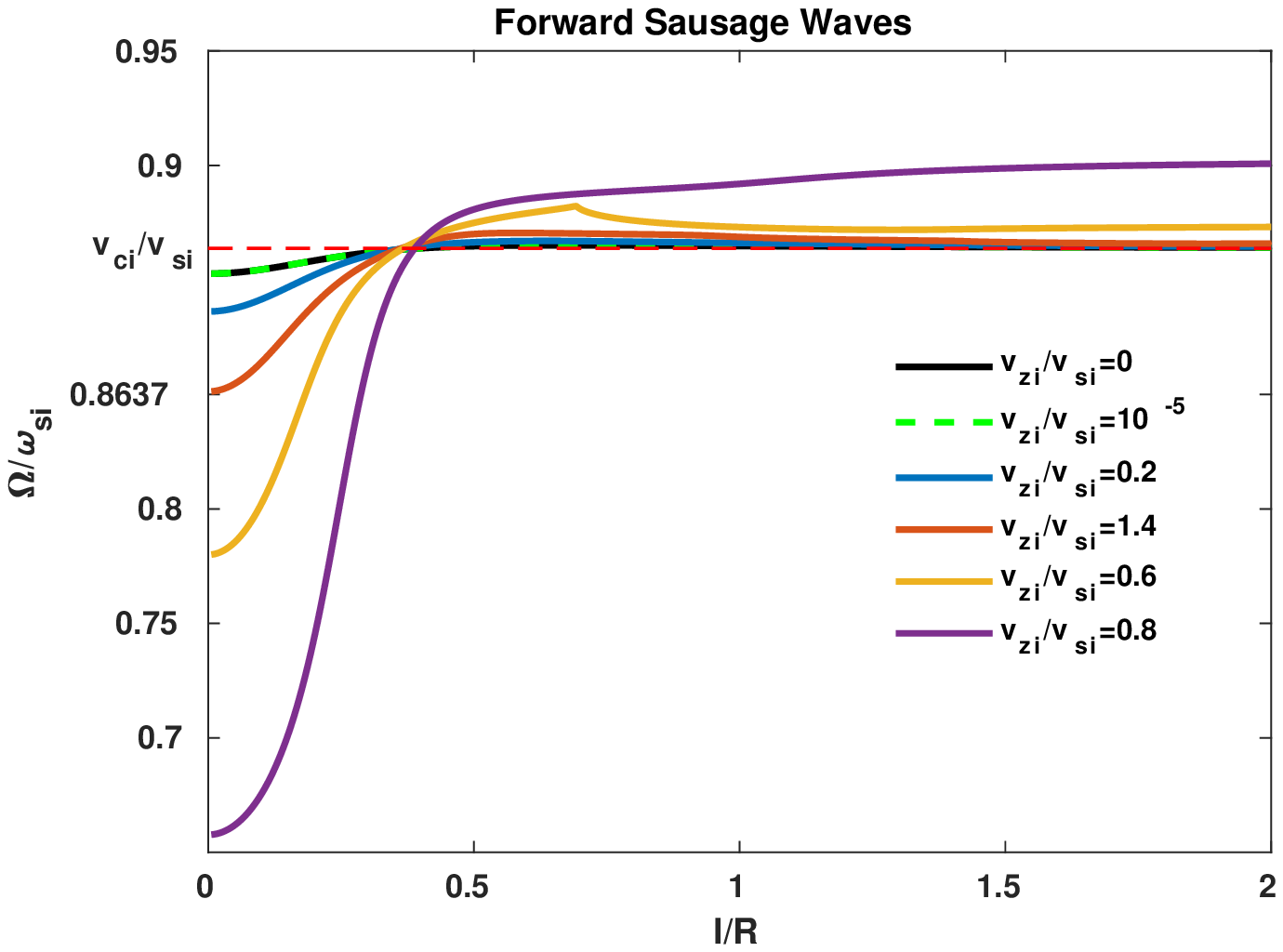}
		\caption{}
		\label{5skz4c}
	\end{subfigure}
	\hfill
	\begin{subfigure}[b]{0.45\textwidth}
		\centering
		\includegraphics[width=\textwidth]{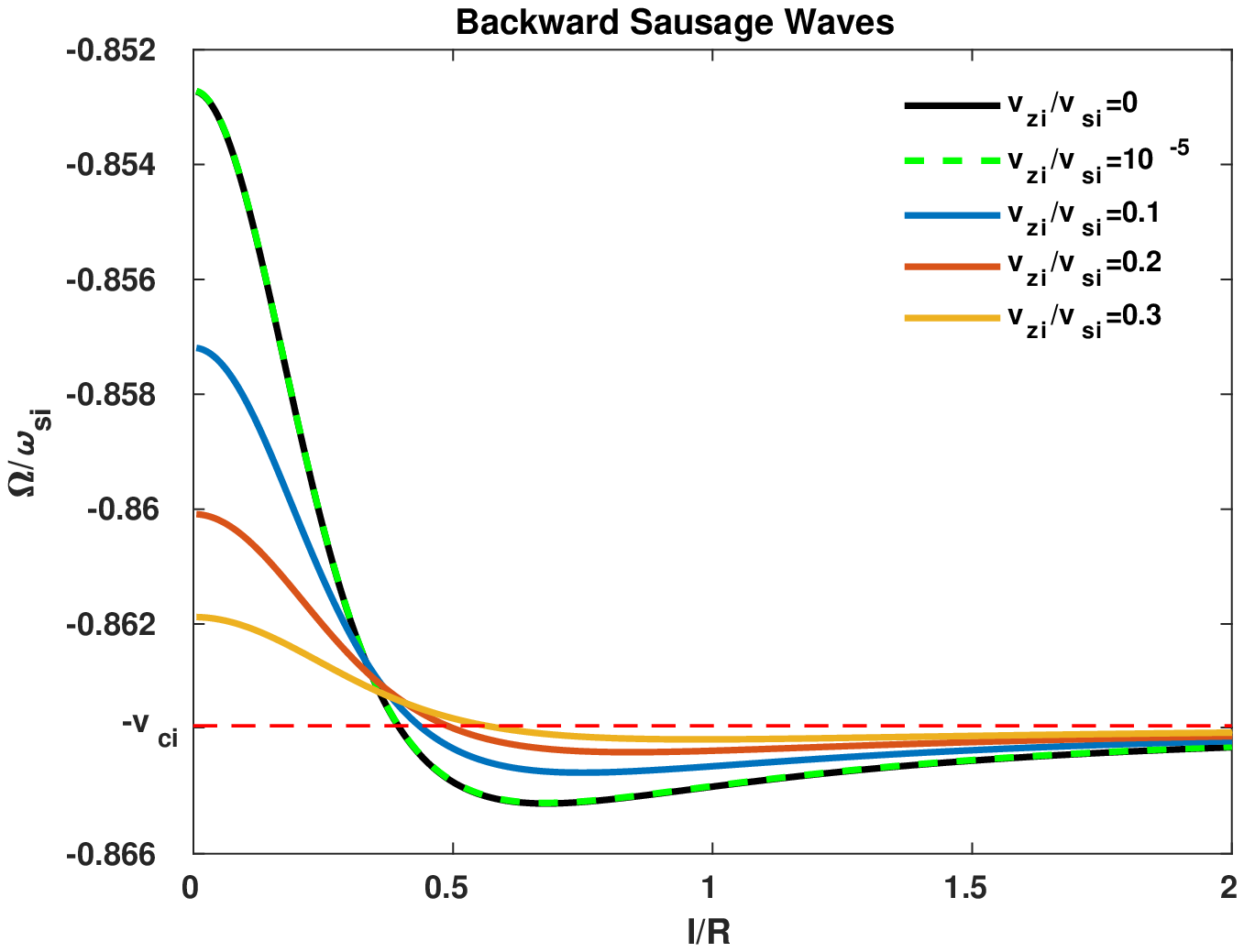}
		\caption{}
		\label{5skz4d}
	\end{subfigure}
	\vfill
	\begin{subfigure}[b]{0.45\textwidth}
		\centering
		\includegraphics[width=\textwidth]{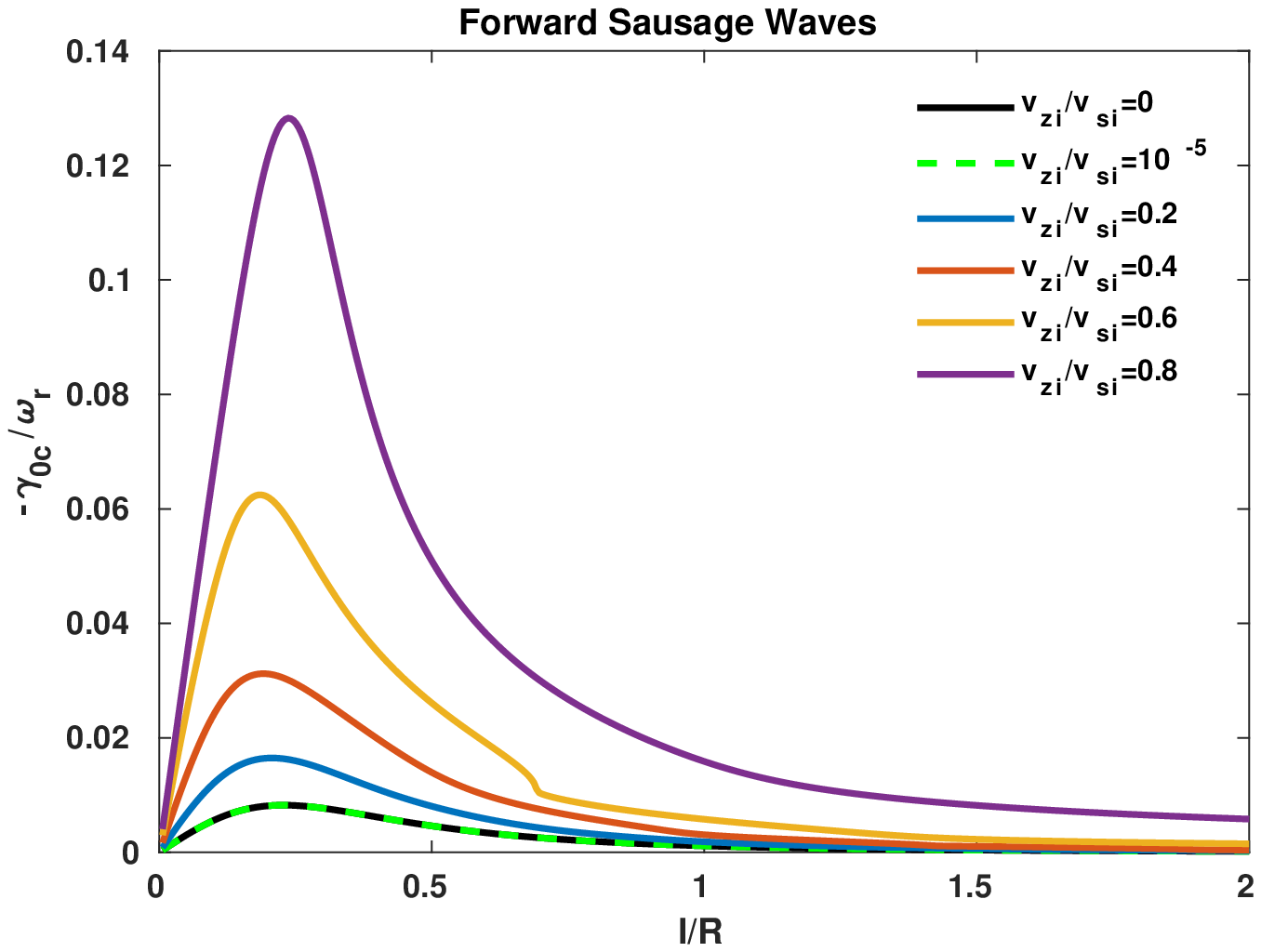}
		\caption{}
		\label{5skz4e}
	\end{subfigure}
	\hfill
	\begin{subfigure}[b]{0.45\textwidth}
		\centering
		\includegraphics[width=\textwidth]{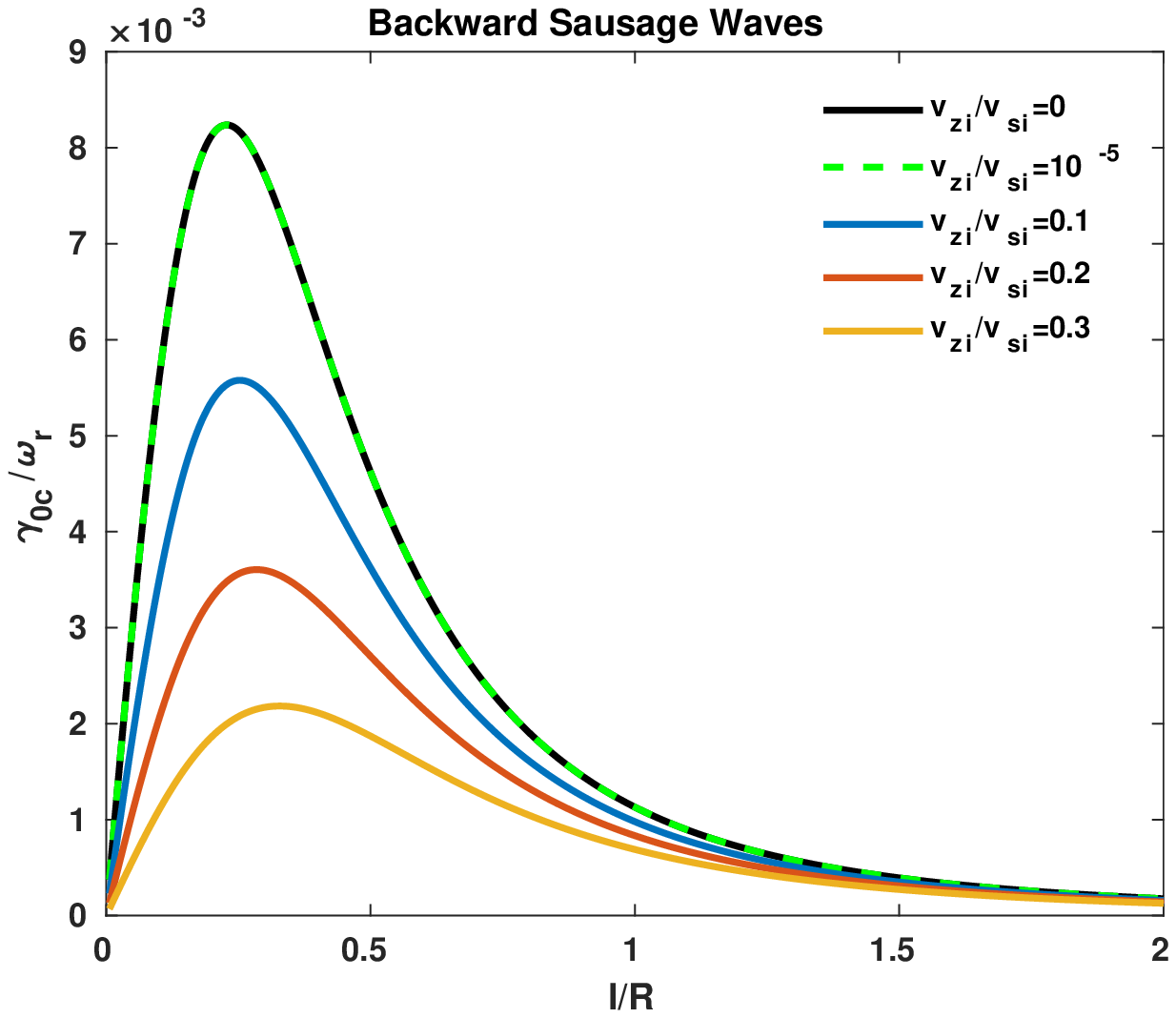}
		\caption{}
		\label{5slkz4f}
	\end{subfigure}
	\caption{Same as Fig. \ref{5skz0.5}  , but for $k_zR=2$.}
	\label{5skz4}
\end{figure}
\begin{figure}
	\begin{subfigure}[b]{0.45\textwidth}
		\centering
		\includegraphics[width=\textwidth]{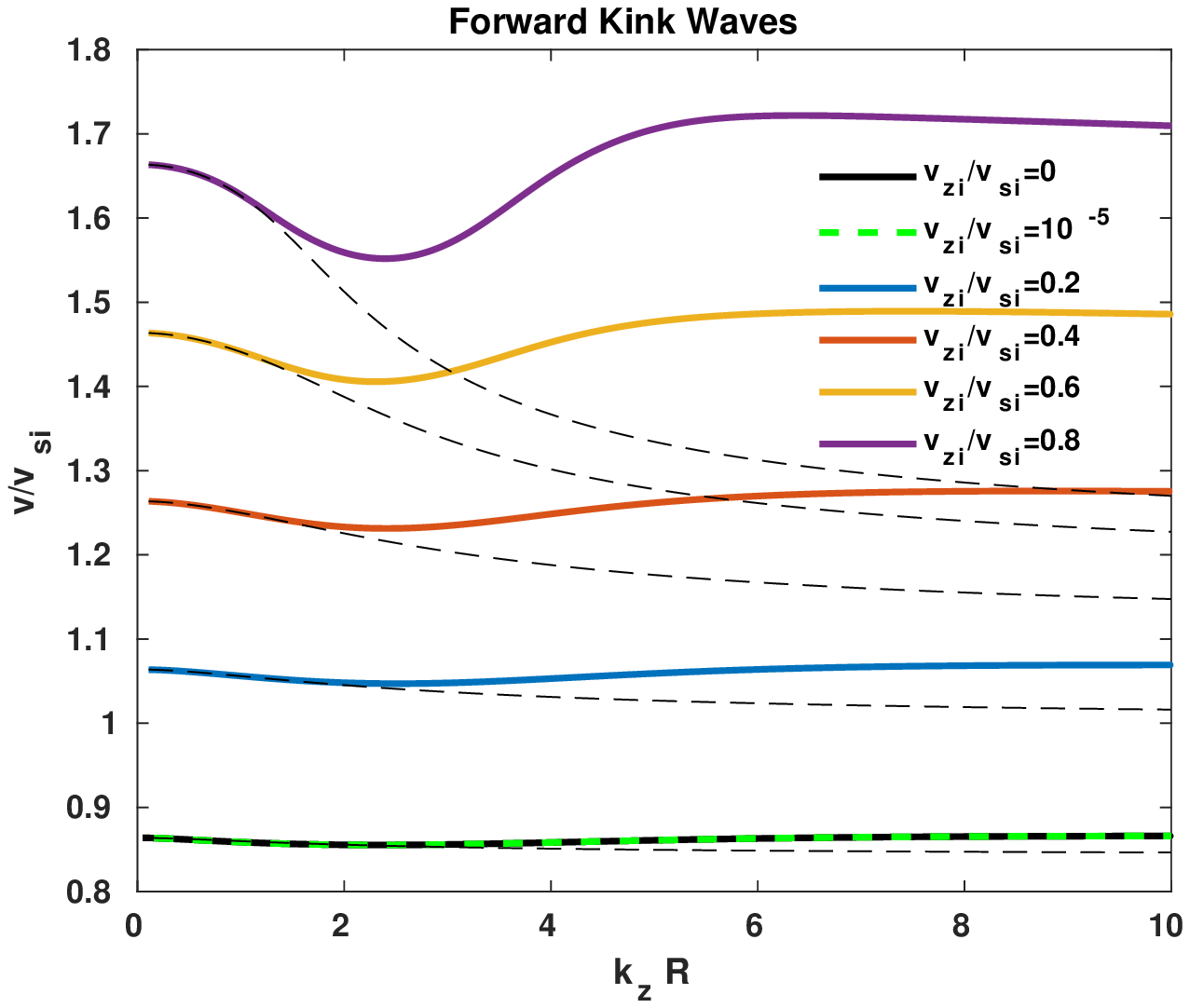}
		\caption{}
		\label{5kl0.1a}
	\end{subfigure}
\hfill
\begin{subfigure}[b]{0.45\textwidth}
	\centering
	\includegraphics[width=\textwidth]{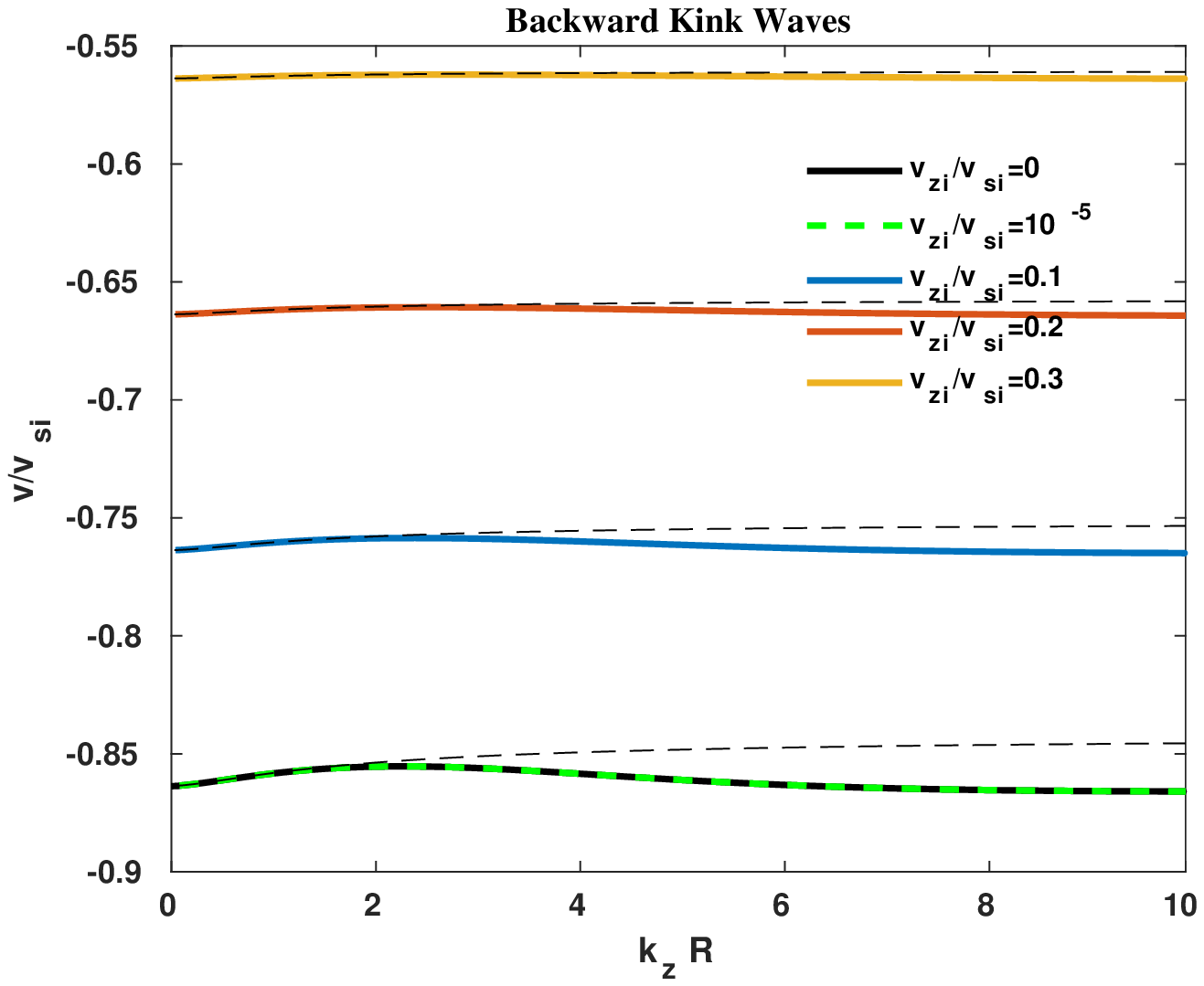}
	\caption{}
	\label{5kl0.1b}
\end{subfigure}
	\vfill
	\begin{subfigure}[b]{0.45\textwidth}
		\centering
		\includegraphics[width=\textwidth]{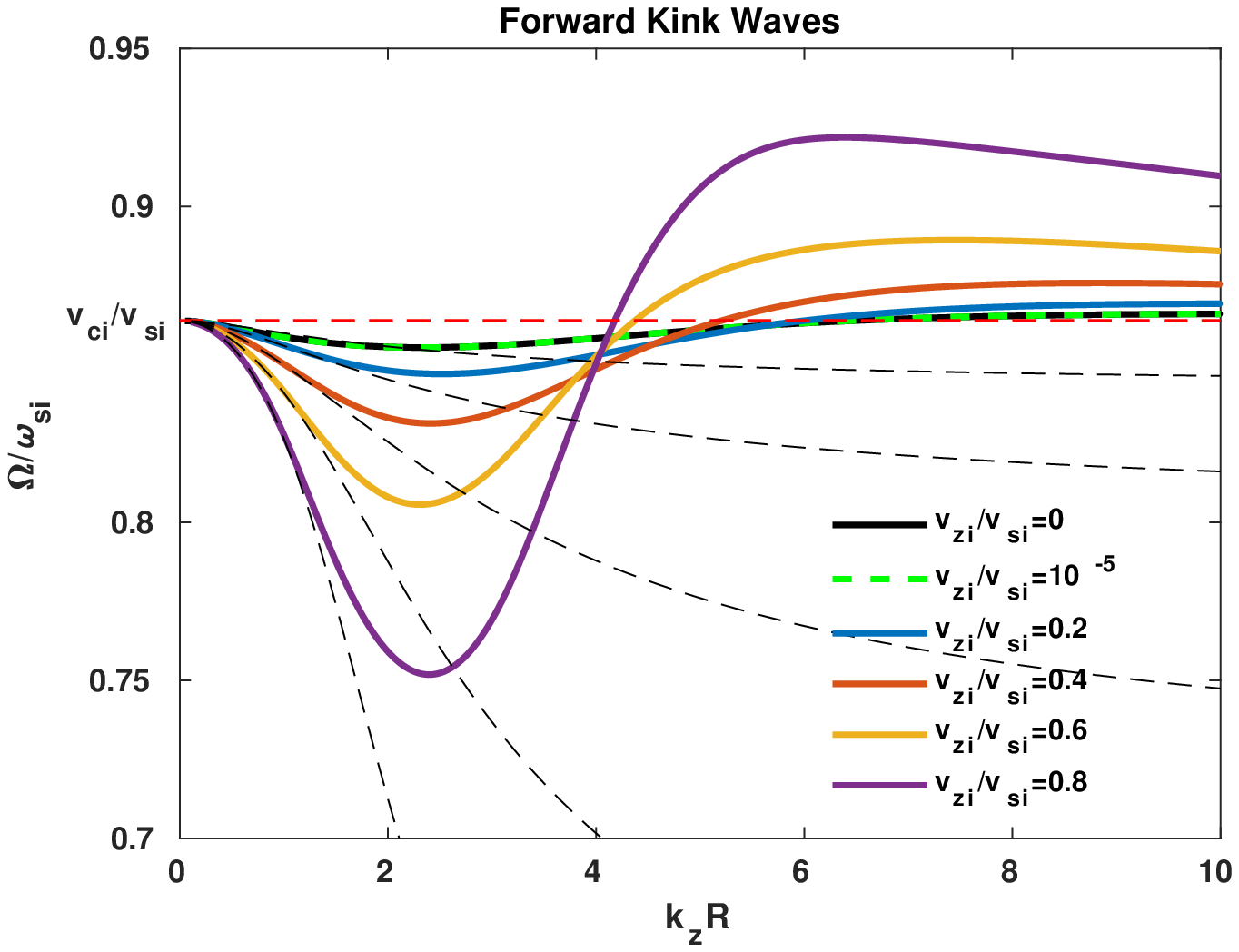}
		\caption{}
		\label{5kl0.1c}
	\end{subfigure}
	\hfill
	\begin{subfigure}[b]{0.45\textwidth}
		\centering
		\includegraphics[width=\textwidth]{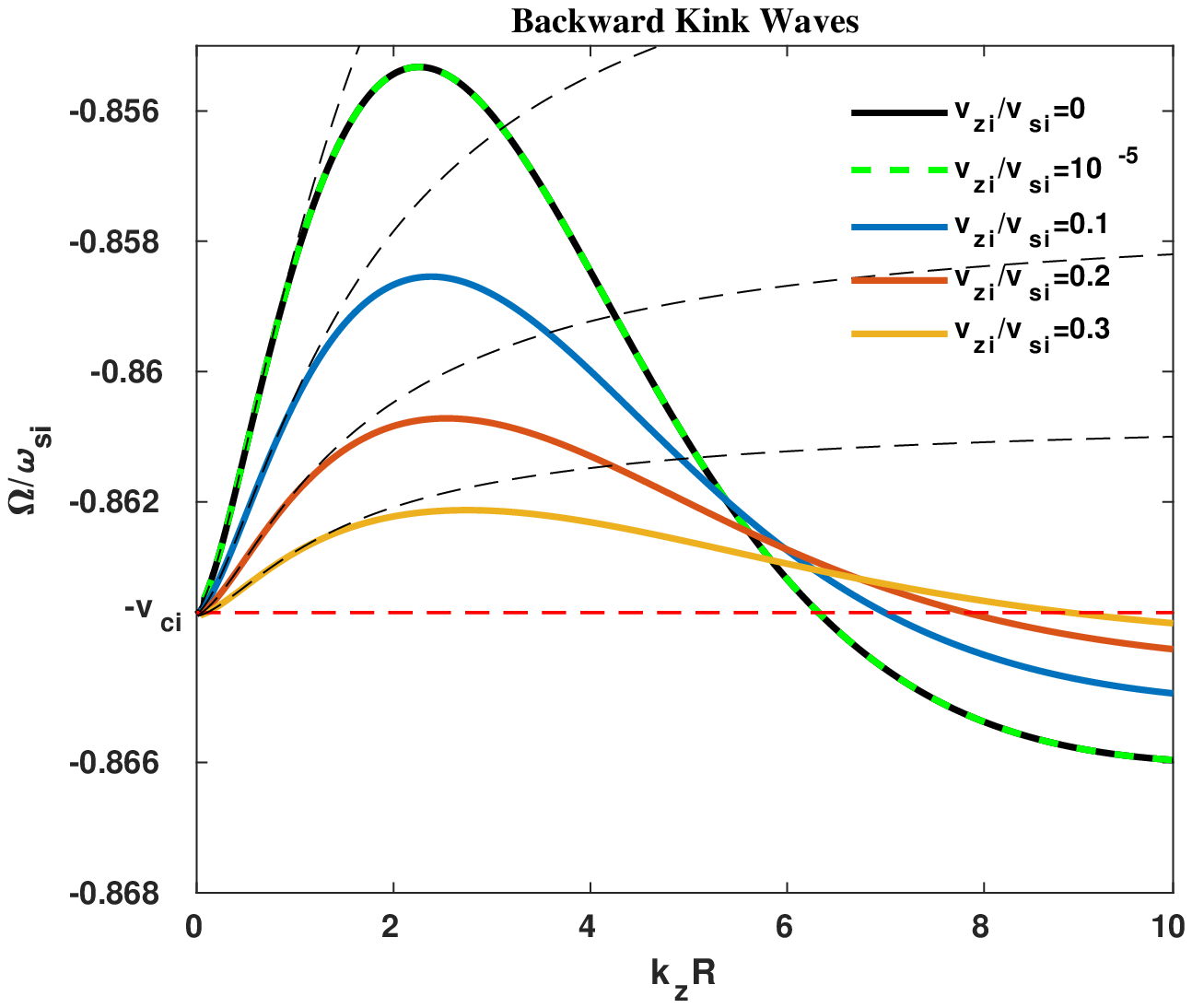}
		\caption{}
		\label{5kl0.1d}
	\end{subfigure}
	\vfill
	\begin{subfigure}[b]{0.45\textwidth}
		\centering
		\includegraphics[width=\textwidth]{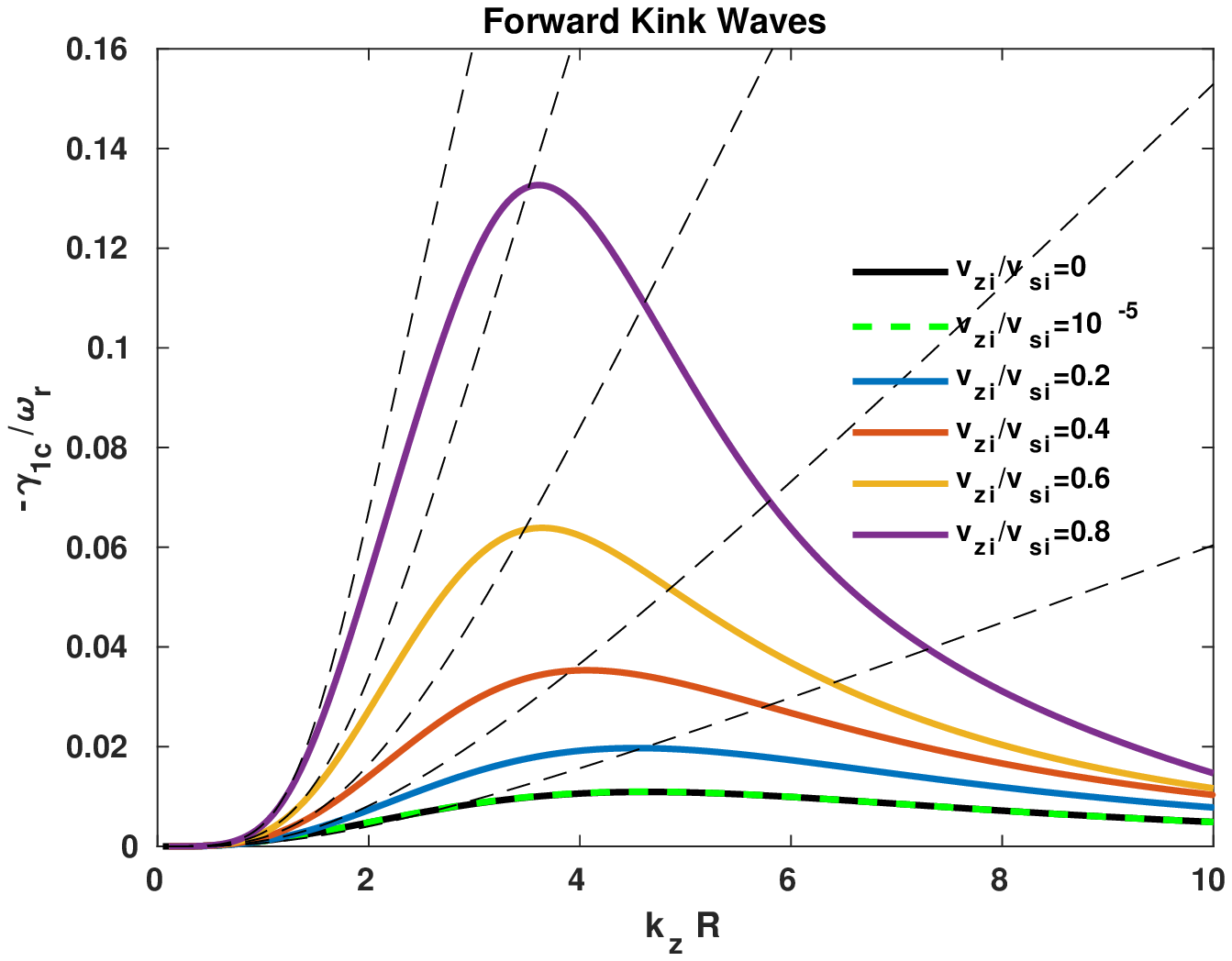}
		\caption{}
		\label{5kl0.1e}
	\end{subfigure}
	\hfill
	\begin{subfigure}[b]{0.45\textwidth}
		\centering
		\includegraphics[width=\textwidth]{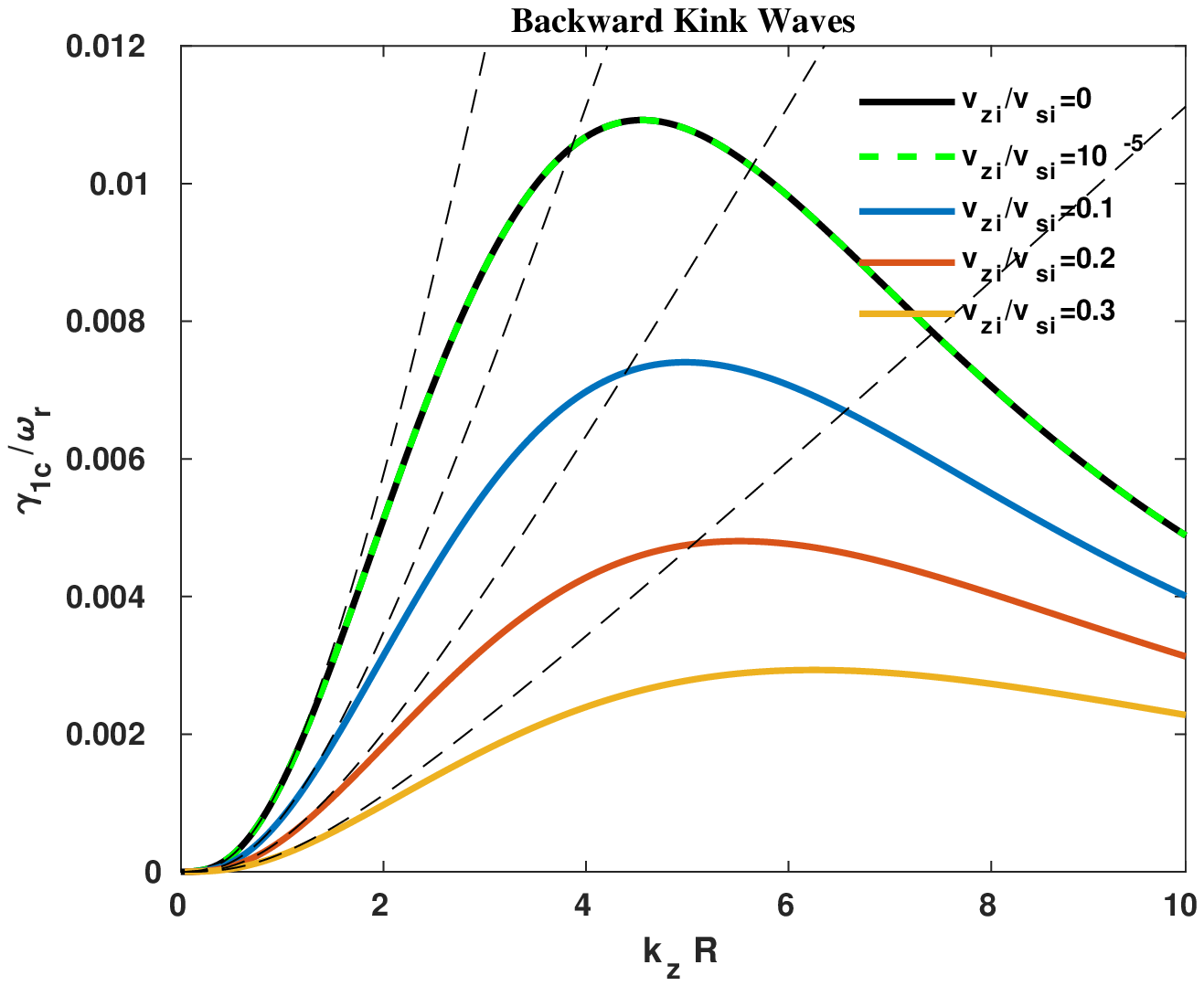}
		\caption{}
		\label{5kl0.1f}
	\end{subfigure}
	\caption{The left panels are for the forward kink waves and the diagrams in (a), (b) and (c) represent the phase speed $v/v_{si}\equiv\omega_r/\omega_{si}$, the Doppler Shifted phase speed $\Omega/\omega_{si}$ and the damping rate $-\gamma_{0c}/\omega_r$ as functions of $k_zR$ for various values of plasma flow. The right panels are the same as the left panels for the backward kink waves. For the damping rate the dashed curves represent the analytical solutions determined from Eq. (\ref{gammacwd}). The dashed curves in the other diagrams show the results obtained in the case of no boundary layer i.e. Eq. (\ref{dispersionrelation}). For all panels we have assumed $l/R=0.1$, other parameters are the same as Fig. \ref{5sl0.1}.}
	\label{5kl0.1}
\end{figure}
\begin{figure}
	\begin{subfigure}[b]{0.45\textwidth}
		\centering
		\includegraphics[width=\textwidth]{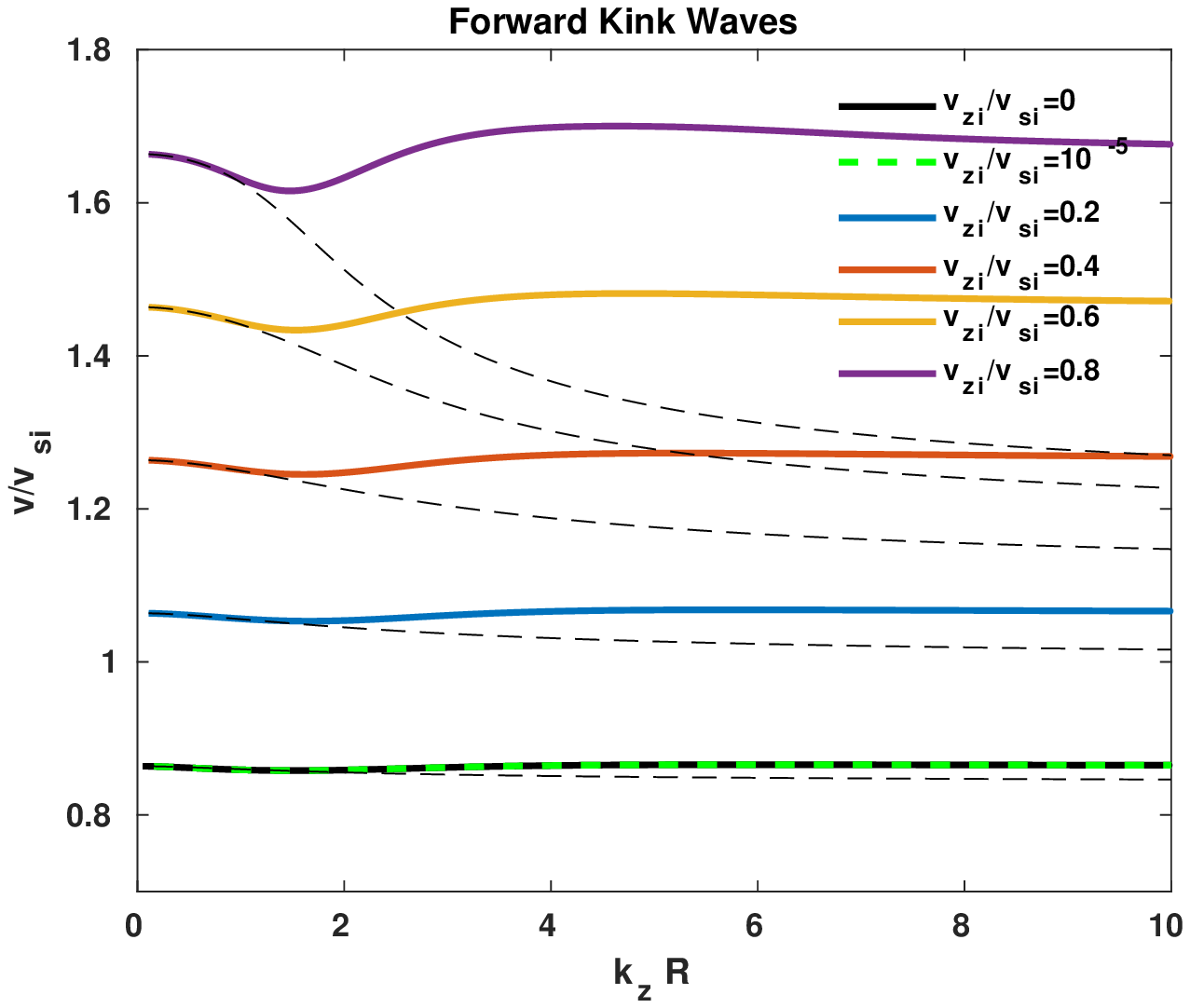}
		\caption{}
		\label{5kl0.4a}
	\end{subfigure}
\hfill
\begin{subfigure}[b]{0.45\textwidth}
	\centering
	\includegraphics[width=\textwidth]{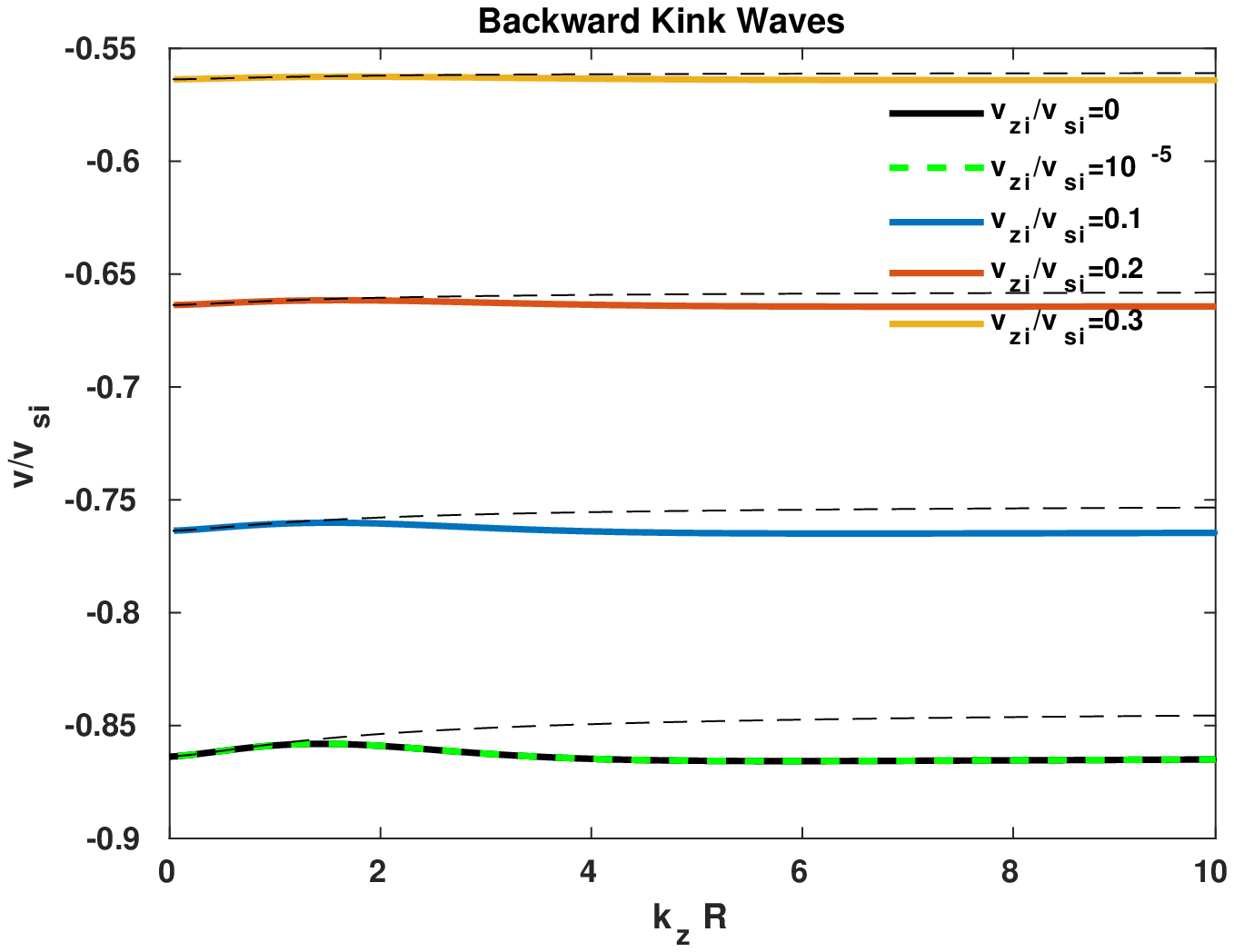}
	\caption{}
	\label{5kl0.4b}
\end{subfigure}
	\vfill
	\begin{subfigure}[b]{0.45\textwidth}
		\centering
		\includegraphics[width=\textwidth]{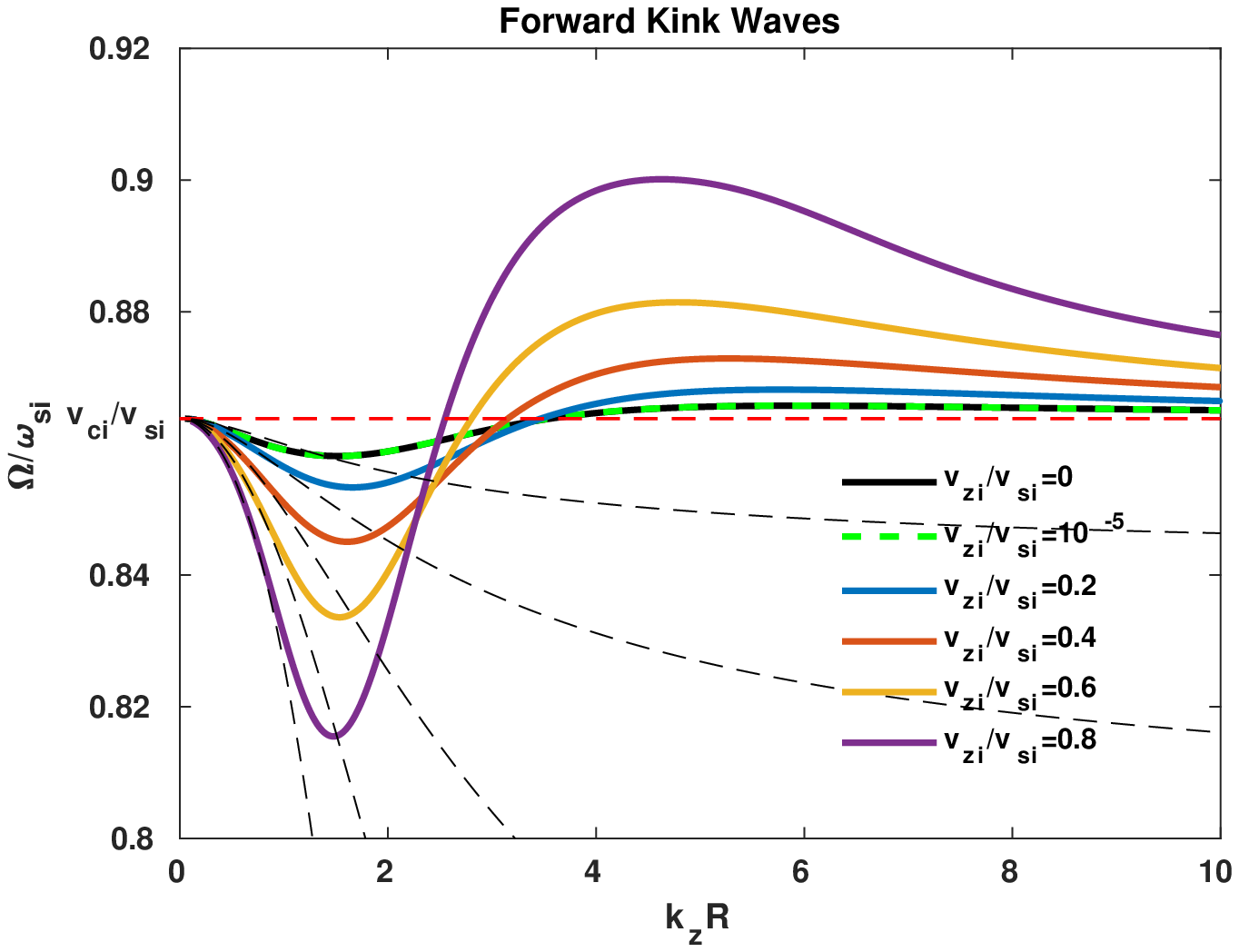}
		\caption{}
		\label{5kl0.4c}
	\end{subfigure}
	\hfill
	\begin{subfigure}[b]{0.45\textwidth}
		\centering
		\includegraphics[width=\textwidth]{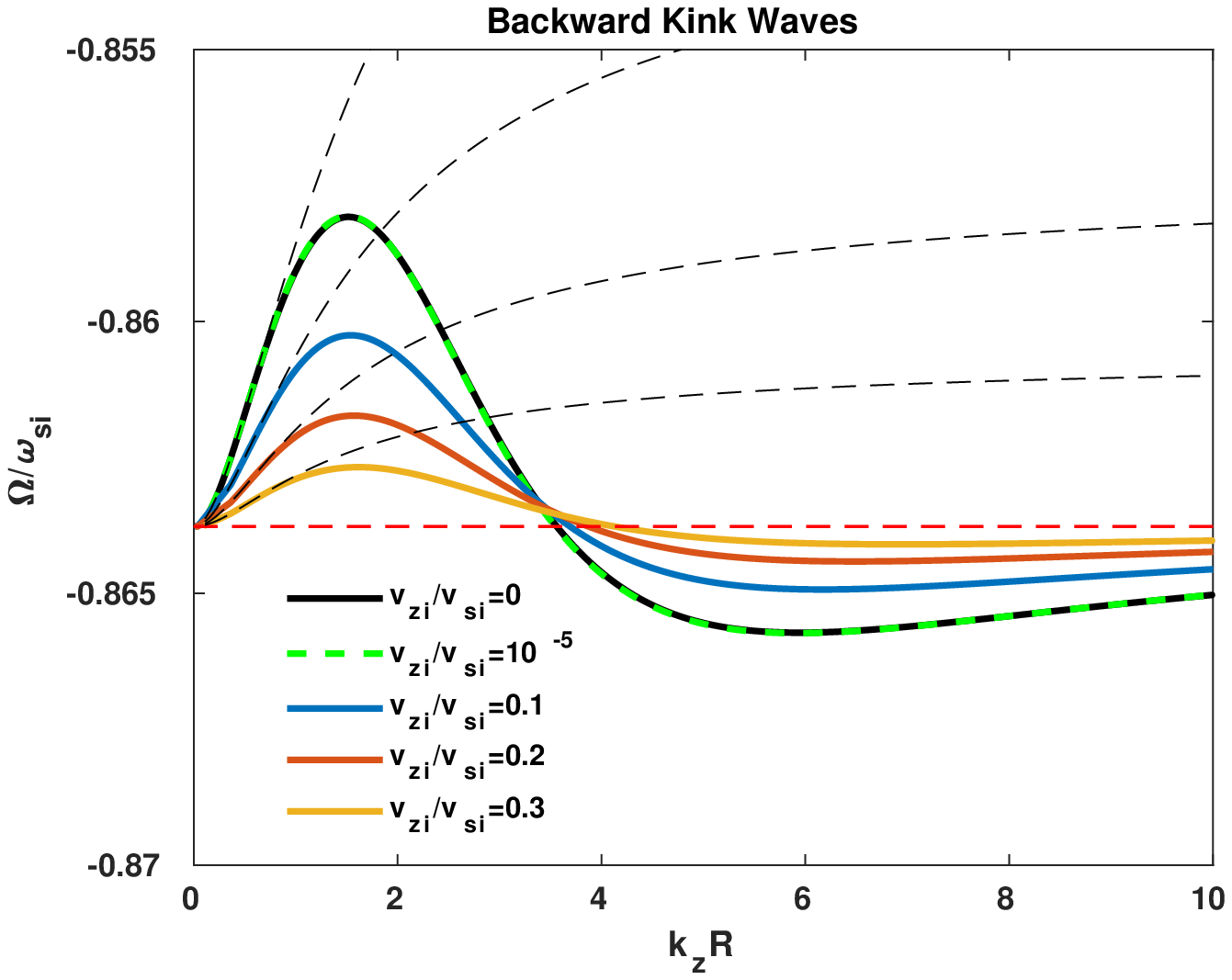}
		\caption{}
		\label{5kl0.4d}
	\end{subfigure}
	\vfill
	\begin{subfigure}[b]{0.45\textwidth}
		\centering
		\includegraphics[width=\textwidth]{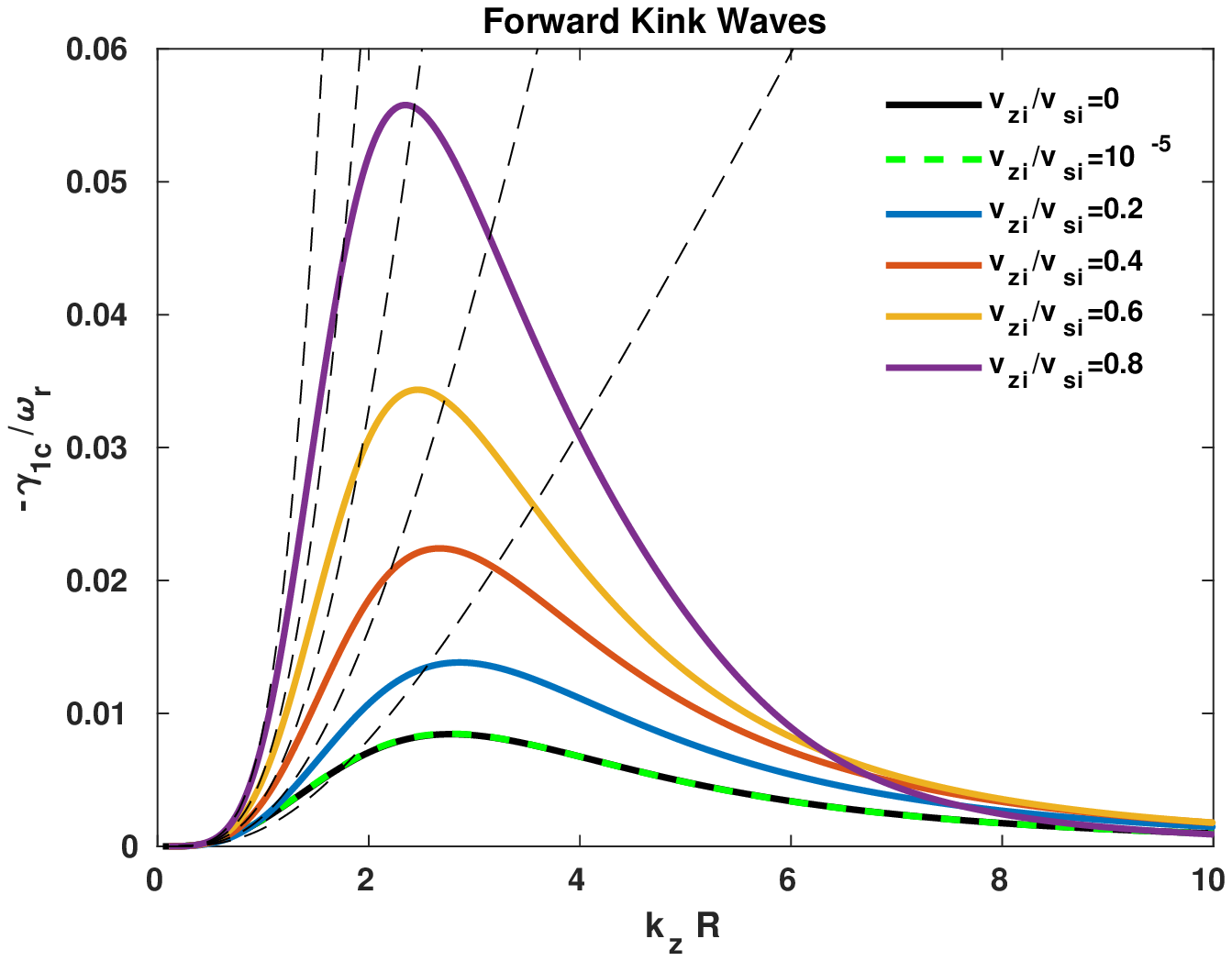}
		\caption{}
		\label{5kl0.4e}
	\end{subfigure}
	\hfill
	\begin{subfigure}[b]{0.45\textwidth}
		\centering
		\includegraphics[width=\textwidth]{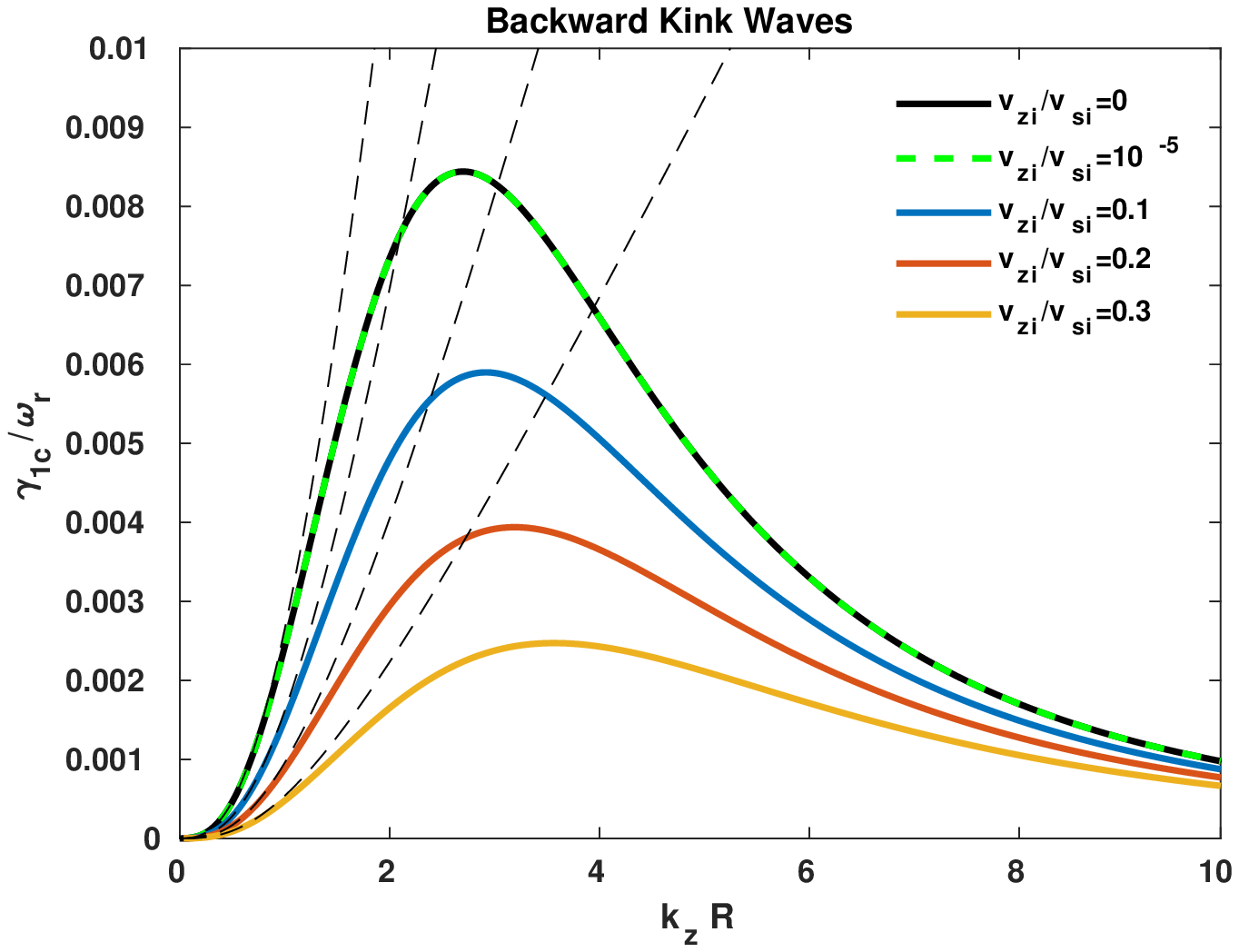}
		\caption{}
		\label{5kl0.4f}
	\end{subfigure}
	\caption{Same as Fig. \ref{5kl0.1}  , but for $l/R=0.2$}
	\label{5kl0.4}
\end{figure}
\begin{figure}
	\begin{subfigure}[b]{0.45\textwidth}
		\centering
		\includegraphics[width=\textwidth]{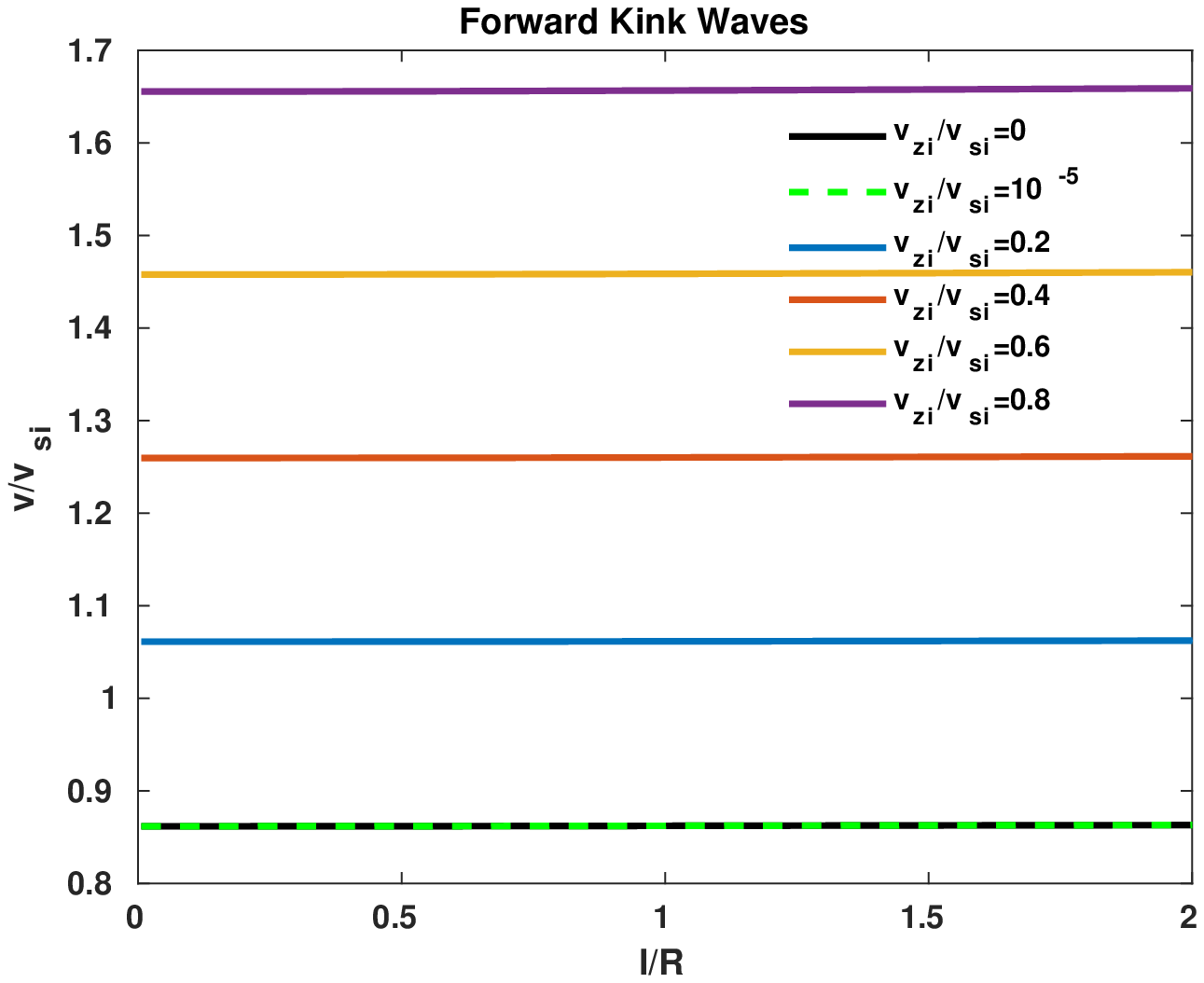}
		\caption{}
		\label{5kkz0.5a}
	\end{subfigure}
\hfill
\begin{subfigure}[b]{0.45\textwidth}
	\centering
	\includegraphics[width=\textwidth]{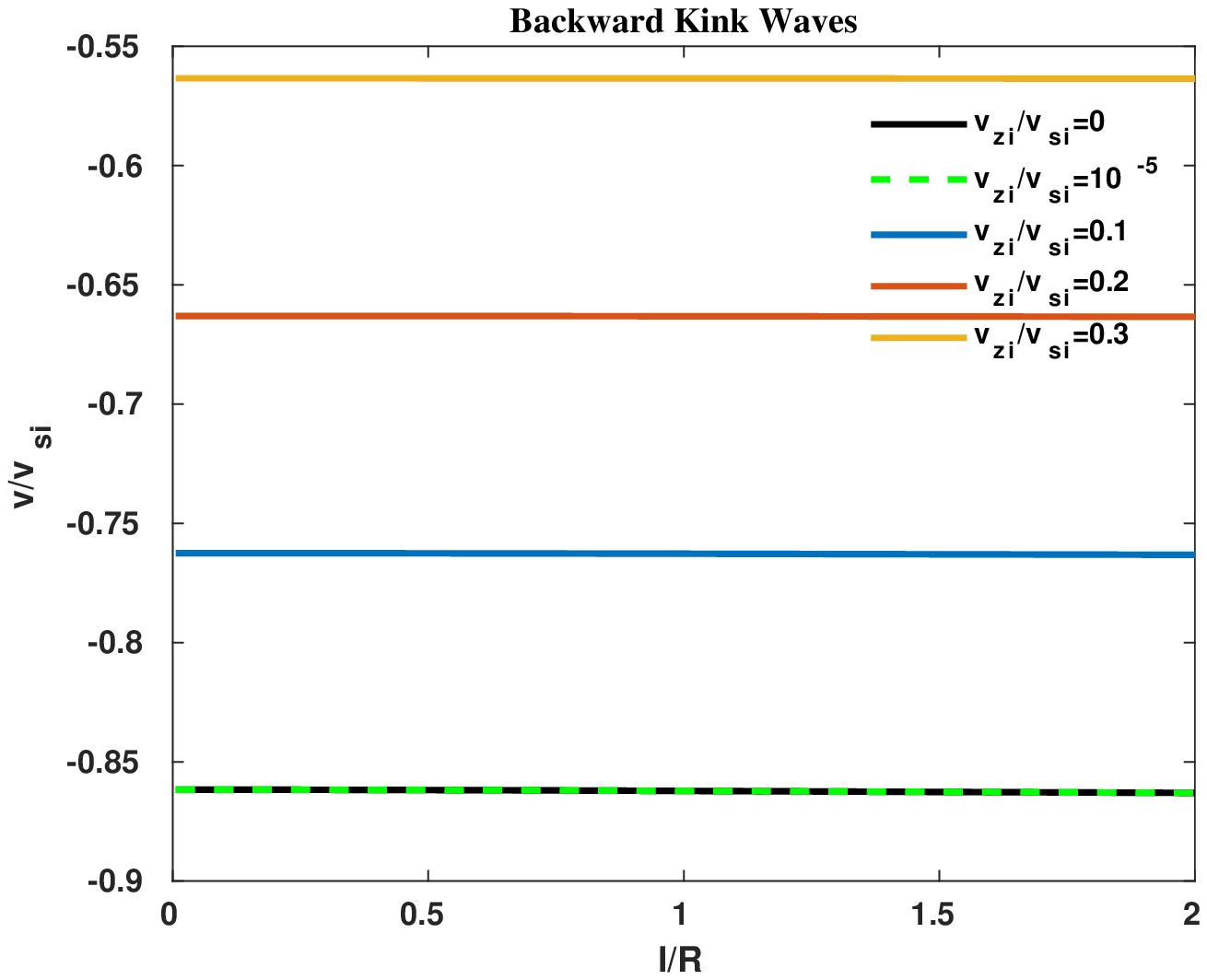}
	\caption{}
	\label{5kkz0.5b}
\end{subfigure}
	\vfill
	\begin{subfigure}[b]{0.45\textwidth}
		\centering
		\includegraphics[width=\textwidth]{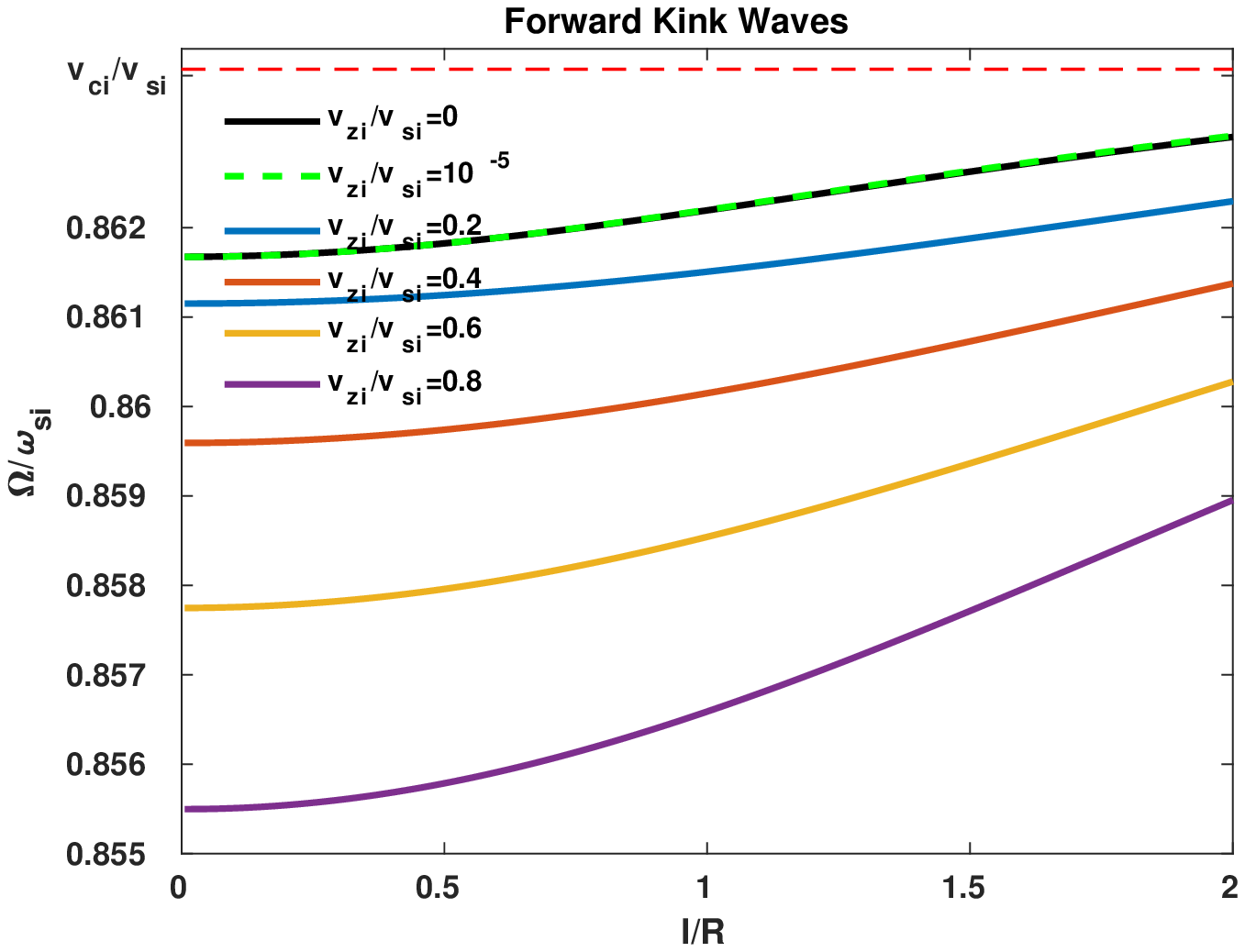}
		\caption{}
		\label{5kkz0.5c}
	\end{subfigure}
	\hfill
	\begin{subfigure}[b]{0.45\textwidth}
		\centering
		\includegraphics[width=\textwidth]{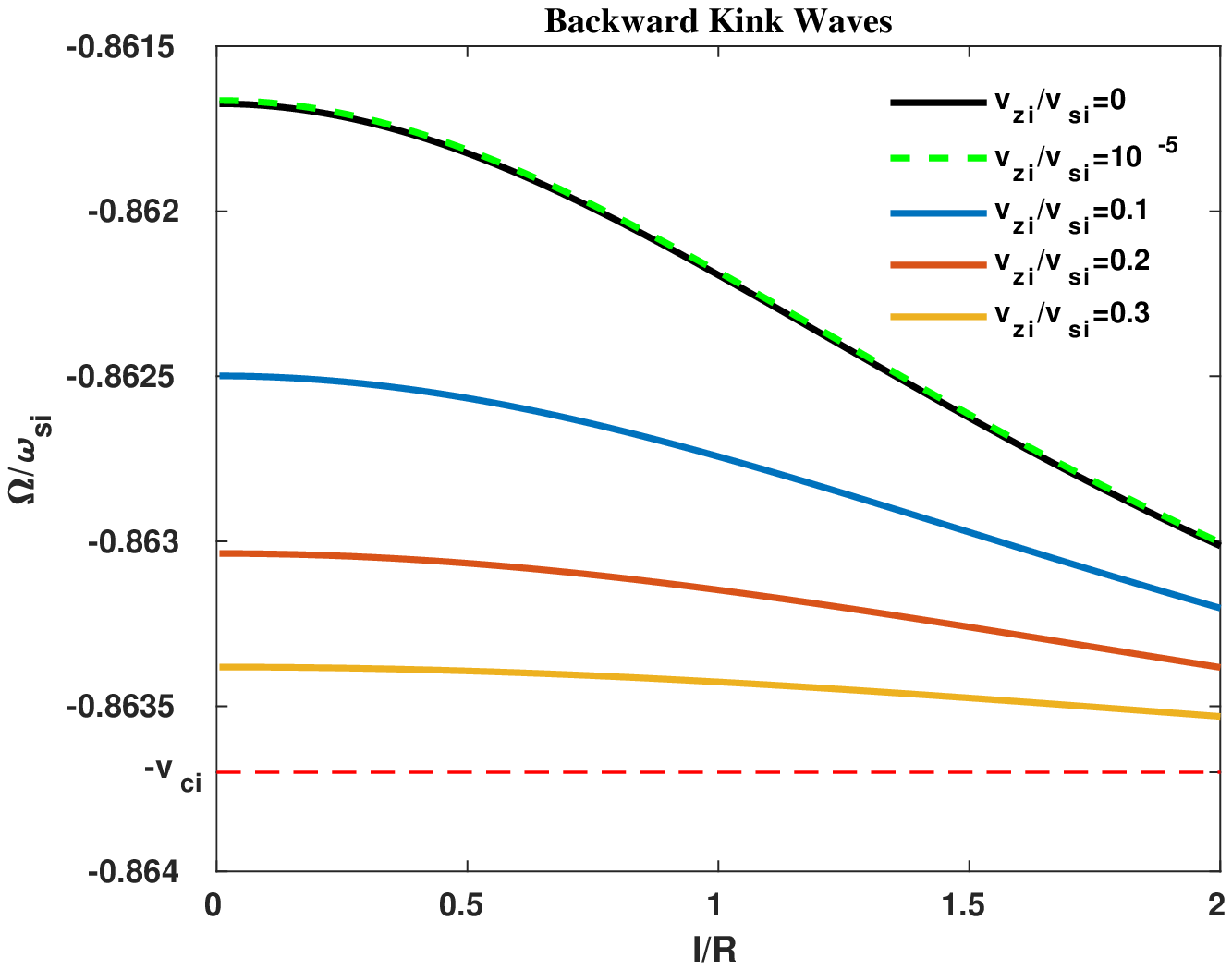}
		\caption{}
		\label{5kkz0.5d}
	\end{subfigure}
	\vfill
	\begin{subfigure}[b]{0.45\textwidth}
		\centering
		\includegraphics[width=\textwidth]{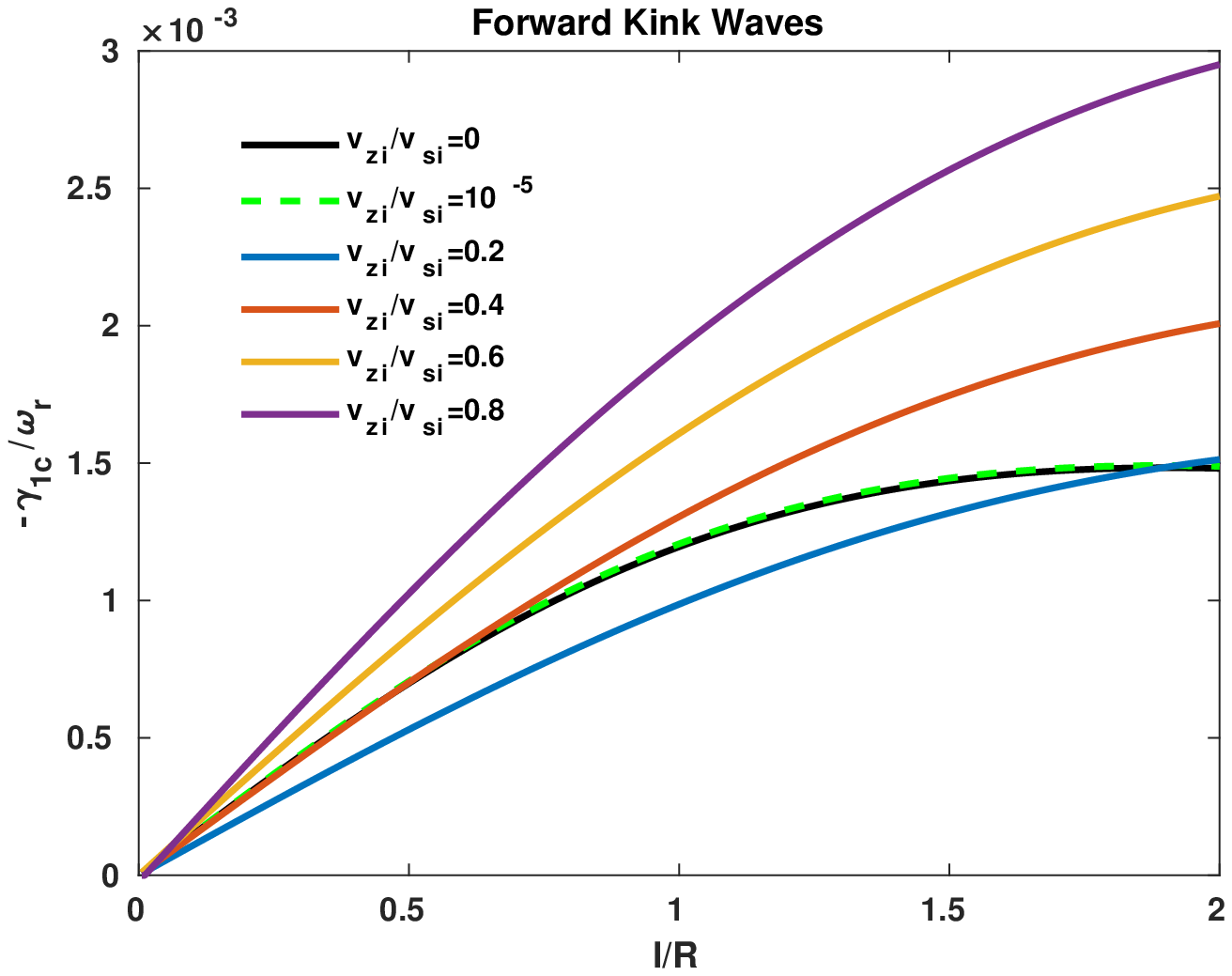}
		\caption{}
		\label{5kkz0.5e}
	\end{subfigure}
	\hfill
	\begin{subfigure}[b]{0.45\textwidth}
		\centering
		\includegraphics[width=\textwidth]{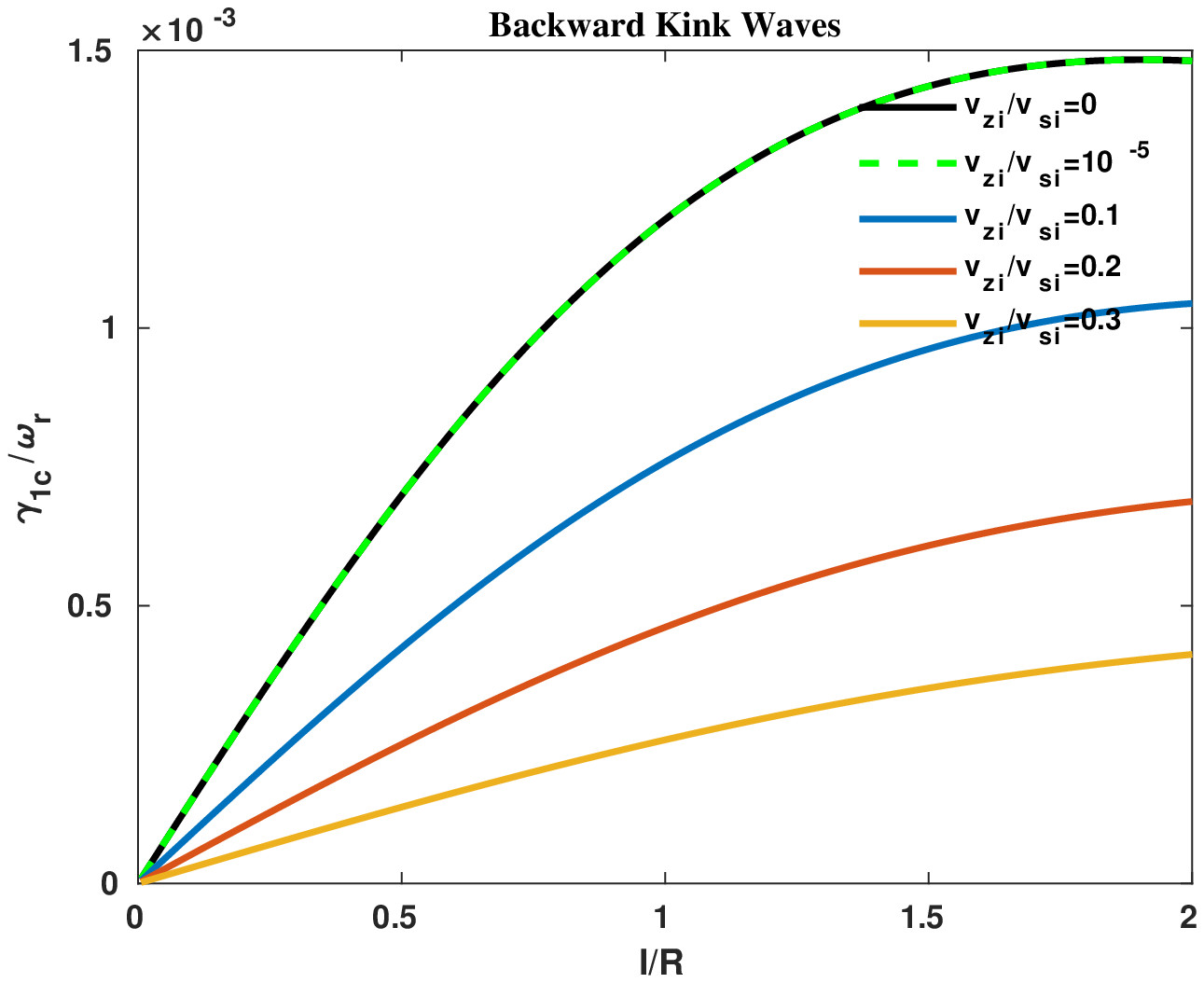}
		\caption{}
		\label{5kkz0.5f}
	\end{subfigure}
	\caption{The left panels are for the forward kink waves and the diagrams in (a), (b) and (c) represent the phase speed $v/v_{si}\equiv\omega_r/\omega_{si}$, the Doppler Shifted phase speed $\Omega/\omega_{si}$ and the damping rate $-\gamma_{0c}/\omega_r$ as functions of $l/R$ for various values of plasma flow. The right panels are the same as the left panels for the backward kink waves. For all panels we have assumed $k_zR=0.5$, other parameters are the same as Fig. \ref{5sl0.1}.}
	\label{5kkz0.5}
\end{figure}

\begin{figure}
	\begin{subfigure}[b]{0.45\textwidth}
		\centering
		\includegraphics[width=\textwidth]{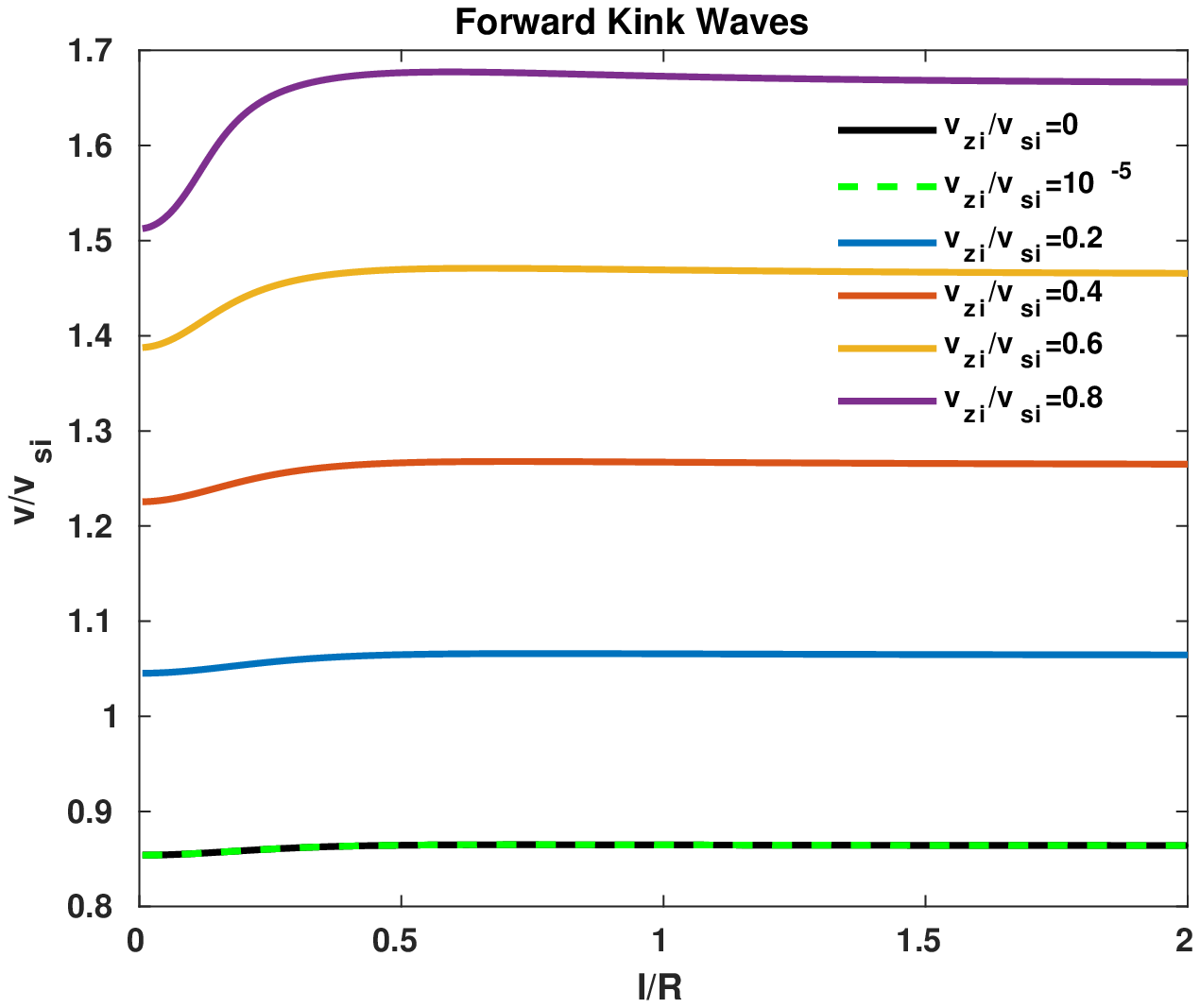}
		\caption{}
		\label{5kkz4a}
	\end{subfigure}
\hfill
\begin{subfigure}[b]{0.45\textwidth}
	\centering
	\includegraphics[width=\textwidth]{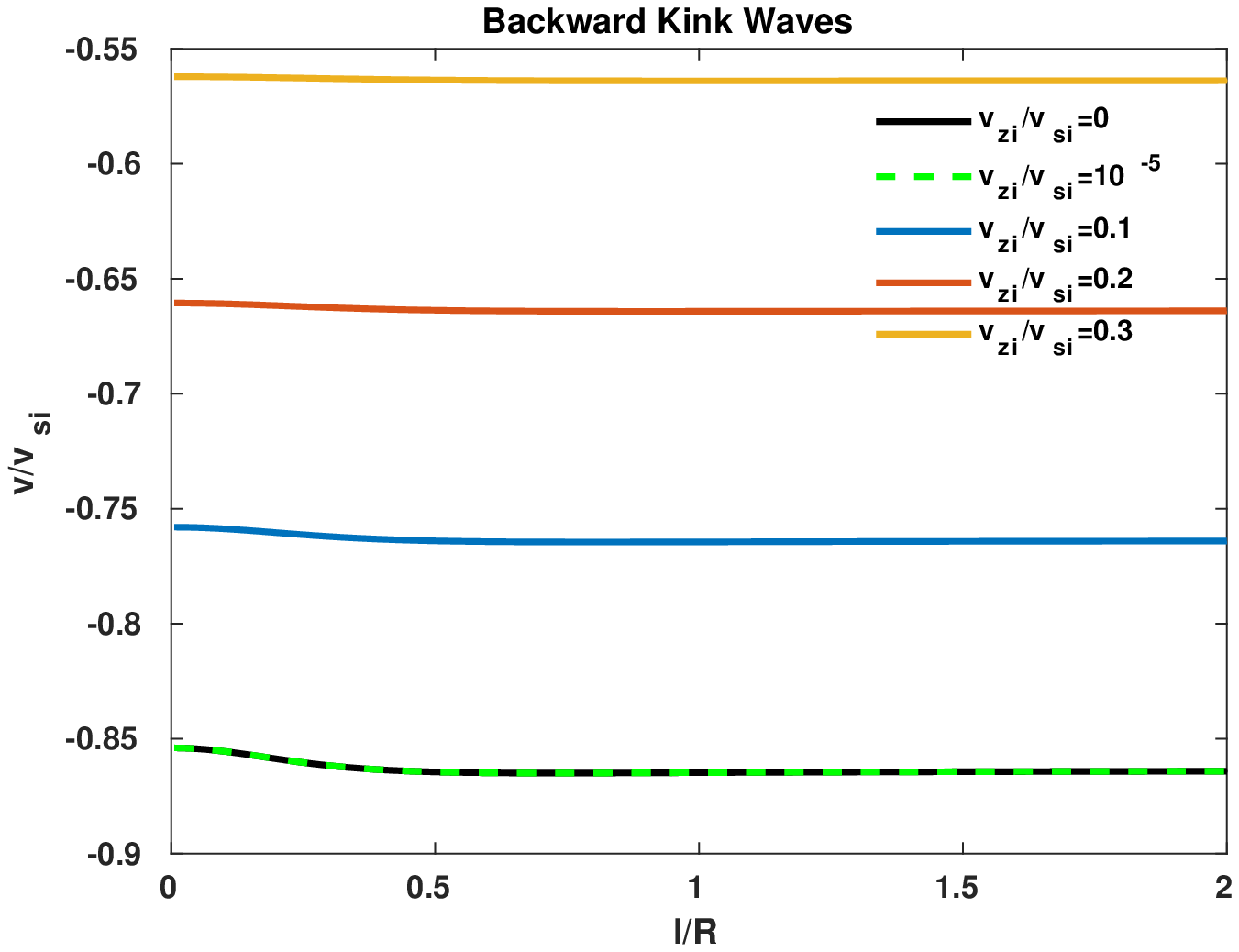}
	\caption{}
	\label{5kkz4b}
\end{subfigure}
	\vfill
	\begin{subfigure}[b]{0.45\textwidth}
		\centering
		\includegraphics[width=\textwidth]{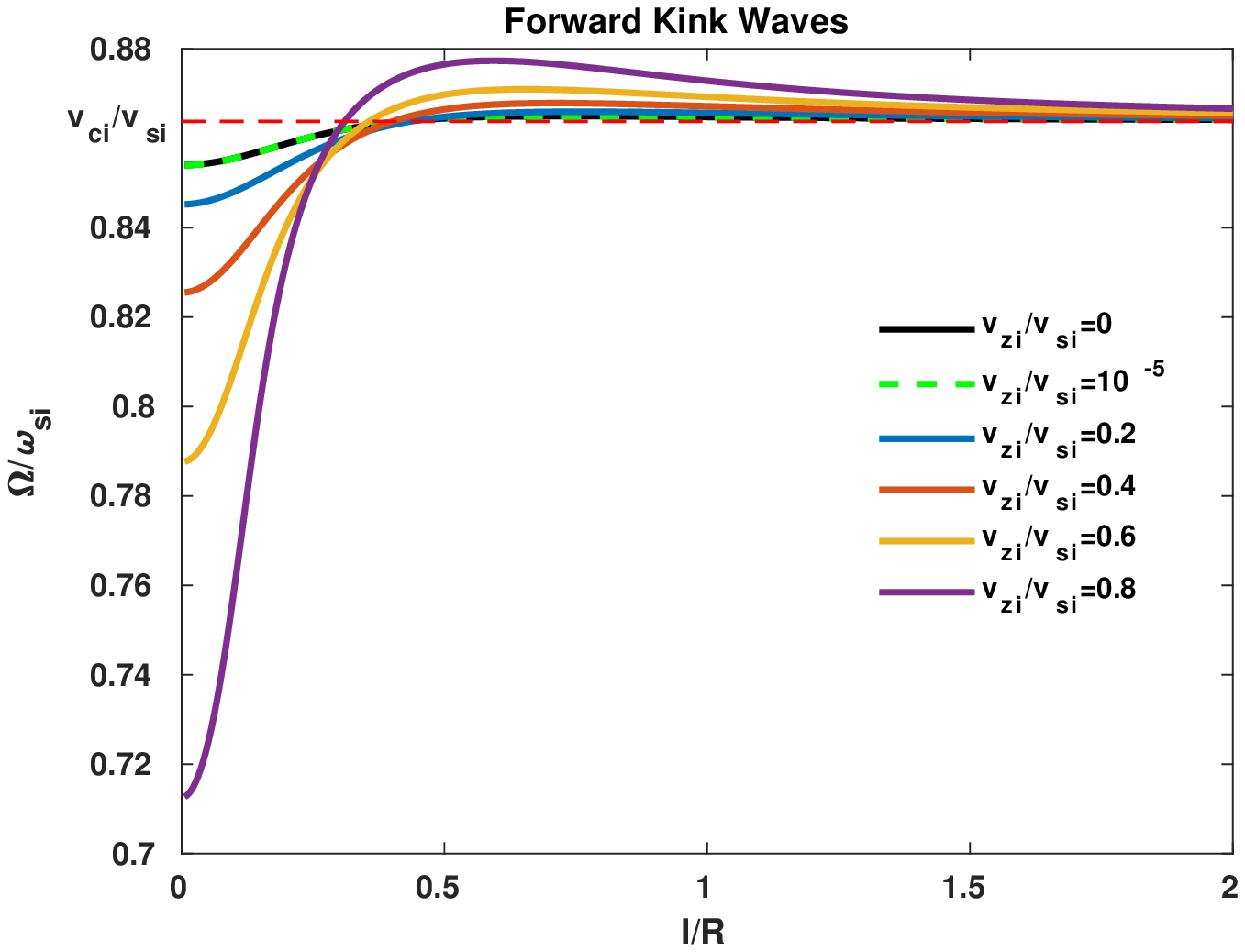}
		\caption{}
		\label{5kkz4c}
	\end{subfigure}
	\hfill
	\begin{subfigure}[b]{0.45\textwidth}
		\centering
		\includegraphics[width=\textwidth]{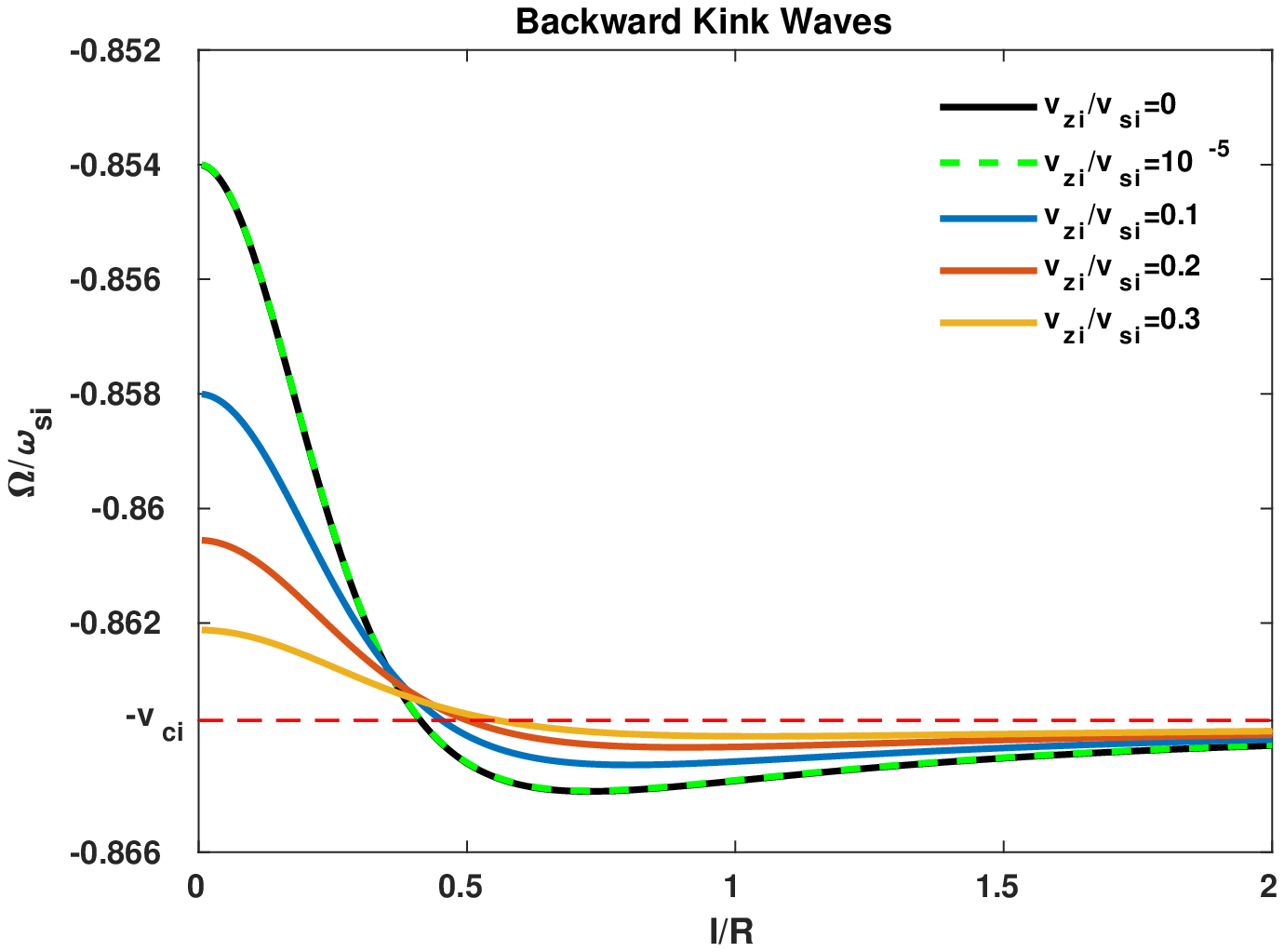}
		\caption{}
		\label{5kkz4d}
	\end{subfigure}
	\vfill
	\begin{subfigure}[b]{0.45\textwidth}
		\centering
		\includegraphics[width=\textwidth]{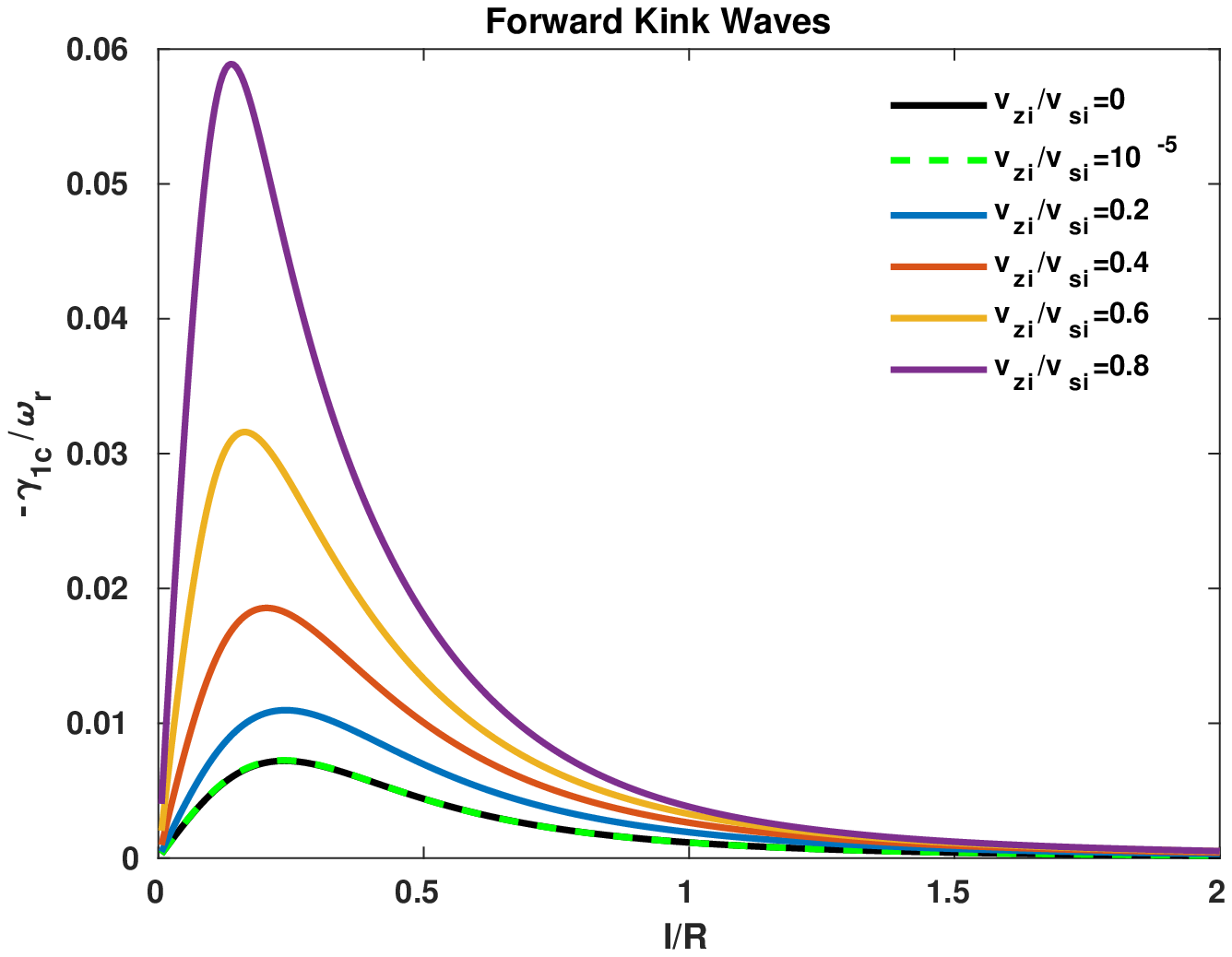}
		\caption{}
		\label{5kkz4e}
	\end{subfigure}
	\hfill
	\begin{subfigure}[b]{0.45\textwidth}
		\centering
		\includegraphics[width=\textwidth]{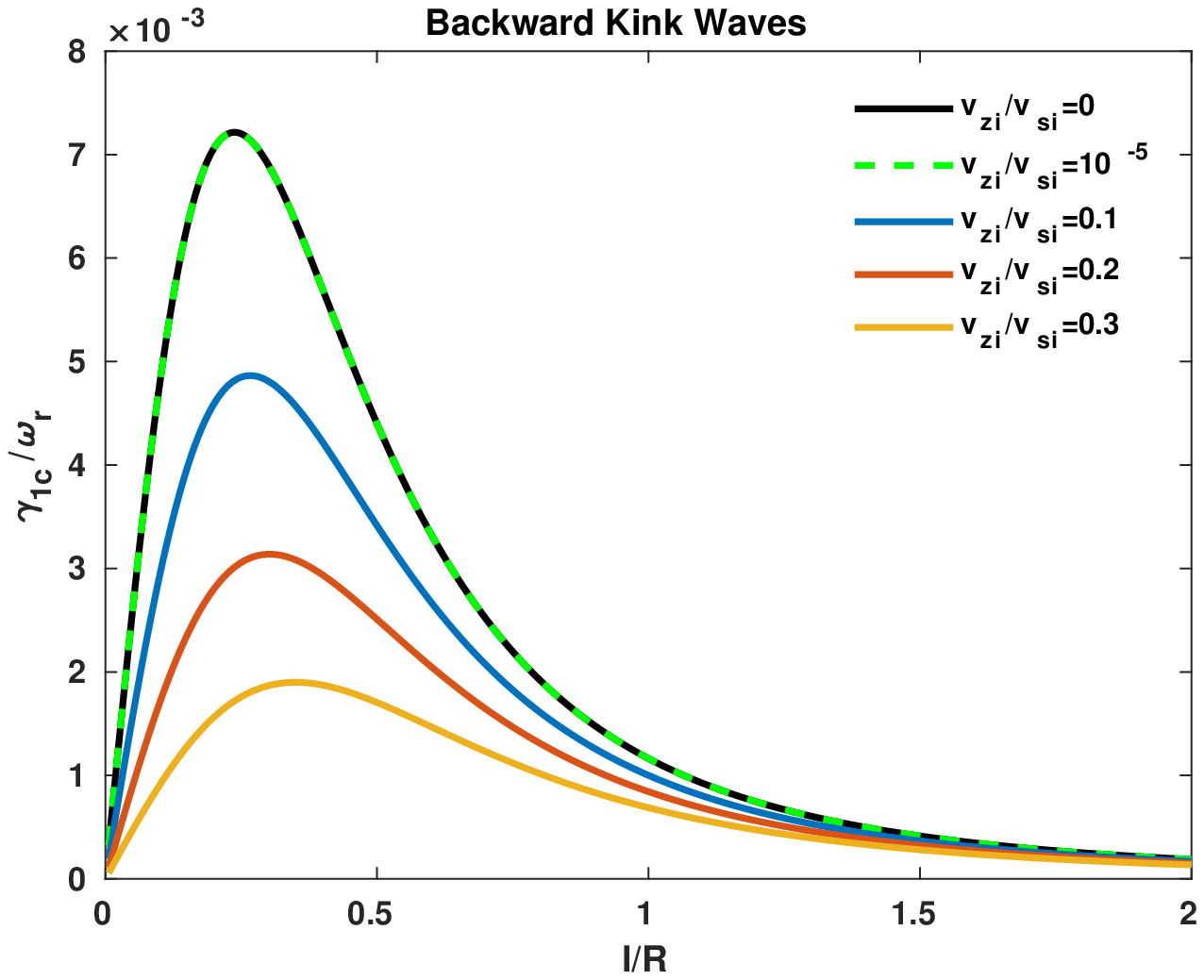}
		\caption{}
		\label{5kkz4f}
	\end{subfigure}
	\caption{Same as Fig. \ref{5kkz0.5}  , but for $k_zR=2$.}
	\label{5kkz4}
\end{figure}
\begin{figure}
 \centering\includegraphics[width=0.5\linewidth]{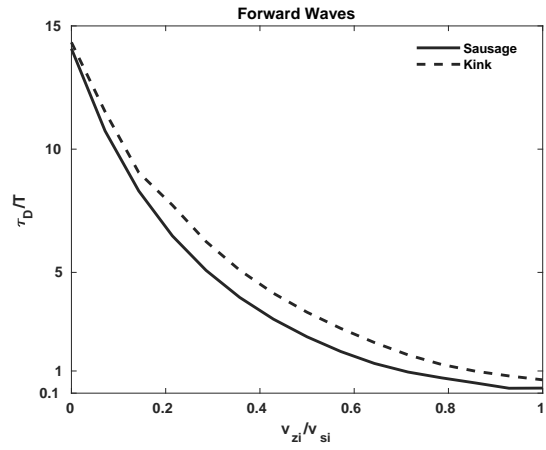}
 \caption{The minimum value of the damping time to period ratio ($\tau_D/T$) for the forward waves including the slow surface sausage (solid line) and kink (dashed line) modes versus upflow velocity ($v_{zi}/v_{si}$) for $l/R=0.1$. Here, we have $k_zR\leq 5$ and other parameters are the same as Fig. \ref{5sl0.1}.}
 \label{taudperiodvz}
\end{figure}
\end{document}